\documentclass[english,aps,prl,twocolumn,amsmath,amssymb,showpacs,superscriptaddress,notitlepage]{revtex4-2}

\usepackage[T1]{fontenc}
\usepackage{color}
\definecolor{page_backgroundcolor}{rgb}{1, 1, 1}
\pagecolor{page_backgroundcolor}
\usepackage{amsmath}
\usepackage{graphicx}
\usepackage{microtype}
\usepackage{bm}
\usepackage{makecell}
\usepackage[normalem]{ulem}
\usepackage{enumitem}
\usepackage[marginal]{footmisc}

\makeatletter
\usepackage{amssymb}
\usepackage{amsmath}
\usepackage{graphicx}
\usepackage[colorlinks=true,linkcolor=blue,anchorcolor=red,citecolor=blue,urlcolor=blue]{hyperref}
\usepackage[caption=false]{subfig}
\usepackage{lipsum}

\makeatother

\usepackage{babel}

\begin{document}
\global\long\def\figurename{Fig.}%

\title{Finite-momentum mixed singlet-triplet pairing in chiral antiferromagnets induced by even-parity spin texture}

\author{Song-Bo Zhang}
\email{songbozhang@ustc.edu.cn}
\thanks{equal contribution}
\address{Hefei National Laboratory, Hefei, Anhui, 230088, China}
\address{International Center for Quantum Design of Functional Materials (ICQD), University of Science and Technology of China, Hefei, Anhui 230026, China}

\author{Lun-Hui Hu}
\email{lunhui@zju.edu.cn}
\thanks{equal contribution}
\affiliation{Center for Correlated Matter and School of Physics, Zhejiang University, Hangzhou 310058, China}
\affiliation{Department of Applied Physics, Aalto University School of Science, FI-00076 Aalto, Finland}

\begin{abstract}
Non-relativistic spin-splitting in unconventional 
antiferromagnets has
garnered much attention for its promising spintronic applications and open fundamental questions. Here, we uncover a unique even-parity spin texture in chiral non-collinear antiferromagnets, exemplified using a kagome lattice. We consider two distinct types of electrons in the system: one with Schr\"odinger-like dispersion and the other exhibiting Dirac-like behavior. Remarkably, we show that, for both electron types, this spin texture induces an exotic coexistence of opposite-spin singlet and equal-spin triplet Cooper pairs with finite momentum when proximity-coupled to conventional superconductors. The triplet pairing arises from the intrinsic spin rotation of the antiferromagnet and does not require net magnetization or spin-orbit coupling. Moreover, we identify an unprecedented and tunable phase difference between singlet and triplet pairings, controllable through junction orientation. This mixed pairing state can be experimentally probed via damped oscillations in order parameters and 0-$\pi$ transitions in Josephson junctions. Additionally, we analyze the effect of out-of-plane spin canting, elucidating its role in generating spin-polarized supercurrents, and discuss Mn$_3$Ga and Mn$_3$Ge to test our predictions. 
\end{abstract}

\maketitle

{\color{blue}\emph{Introduction.}--}Non-relativistic spin splitting, independent of spin-orbit coupling (SOC), can arise in unconventional antiferromagnets or from Fermi-surface instabilities in strongly correlated systems~\cite{Chen2014Anomalous,nakatsuji2015large,wu2004prl,CJWu07PRB}. This phenomenon has been observed in antiferromagnets exhibiting large anomalous Hall effect, such as Mn$_3$Sn~\cite{Chen2014Anomalous,nakatsuji2015large}, and the recently discovered altermagnets~\cite{vsmejkal2020crystal,Naka19NC,hayami2019momentum,Yuan2020Giant,mazin2021prediction,ma2021multifunctional,ifmmode2022Beyond,ifmmode2022Emerging,Ahn19PRB}. Altermagnets belong to a new type of collinear compensated magnetic phase with anisotropic spin splitting. They have been found to exhibit many exotic properties and functionalities~\cite{shao2021spin,FengZX22NE,ifmmode2022Giant,HJLin25PRL}. In particular, the interplay between altermagnetism and superconductivity has been shown to engender a plethora of unconventional superconducting phenomena~\cite{SBZ2023arXiv,Sumita2023Fulde,chakraborty2023zero,Ouassou2023dc,Beenakker2023Phase,Cheng2024Orientation,sun2023Andreev,Papaj2023Andreev,wei2023gapless,Zhu2023Topological,Li2023Majorana,ghorashi2023altermagnetic,banerjee2024altermagnetic,HPSun2024arXiv,CLi2025submit}, including finite-momentum Cooper pairing~\cite{Fulde64PR,Larkin64JETP} with zero net magnetization~\cite{SBZ2023arXiv,Sumita2023Fulde,chakraborty2023zero}, 
peculiar Andreev reflections~\cite{sun2023Andreev,Papaj2023Andreev,wei2023gapless}, and topological superconductivity~\cite{Zhu2023Topological,Li2023Majorana,ghorashi2023altermagnetic}. These intriguing features have garnered growing interest from both theoretical and experimental perspectives~\cite{Lee2024Broken,Zhou2024Crystal,fedchenko2024observation,han2024electrical,Feng2024Incommensurate,Osumi2024Observation,chen2024emerging,Bhowal2024Ferroically,cheong2024altermagnetism,roig2024minimal,yuan2024non,He2023Nonrelativistic,zhu2024Nature,krempasky2024Nature,reimers2024NatComm,JXHu25PRL,zyuzin2024magnetoelectric,XCZhu25PRL,Duan2025prl,MQGu25PRL,YYChen25PRL,ZMWang25PRL}, paving the way for the development of next-generation superconducting devices.

In line with this objective, non-relativistic spin splitting is not the privilege of collinear altermagnets but is also prevalent in non-collinear antiferromagnets~\cite{Jungwirth2016NNanoTech,vsmejkal2022anomalous}. Unlike collinear counterparts, non-collinear antiferromagnets feature diverse spin orientations among adjacent atoms or ions.
A notable example is the chiral antiferromagnets (cAFMs), which have been identified in a growing number of materials~\cite{Chen2014Anomalous,nakatsuji2015large,Baltz18RMP,manchon2019RMP,jiang2023enumeration,ren2023enumeration,xiao2023spin}, 
including kagome metals such as Mn$_3$Sn~\cite{nakatsuji2015large}, 
Mn$_3$Pt~\cite{liu2018electrical},
Mn$_3$Ge~\cite{Kiyohara2016Giant,nayak2016large}, Mn$_3$Ga~\cite{liu2017SciRep,Song2024AFM} 
and Mn$_3$Ir~\cite{zhang2016giant}. Recent experimental efforts have involved Josephson junctions using Mn$_3$Ge thin films~\cite{jeon2021long,jeon2023chiral}. In contrast to collinear altermagnets, spin is not conserved in cAFMs, leading to non-trivial spin textures that could preserve parity~\cite{Jungwirth2016NNanoTech,vsmejkal2022anomalous}. Nevertheless, investigations into the fundamental interplay between even-parity spin texture and superconductivity in junctions are surprisingly scarce. This gap offers an opportunity to explore new mechanisms for generating spin-triplet pairing, beyond known scenarios, such as (i) magnetic noncollinearity at ferromagnet/superconductor interfaces~\cite{Bergeret2001LongRange,Volkov03PRL,Bergeret05RMP,Buzdin05RMP,Linder09PRL,Cottet11PRL,Eschrig2015review}; (ii) odd-parity spin textures induced by SOC~\cite{Demler97prb,Gorkov01PRL,Frigeri04PRL,Bergeret13PRL,Bergeret14PRB,Crepin15prl,smidman2017superconductivity,Cayao17PRB,Cayao18prb,Fleckenstein18prb}; and (iii) collinear antiferromagnet/superconductor junctions with spin-active interfaces or spin canting~\cite{Bobkova05PRL,Kamra18PRL,Rabinovich19PRR, JakobsenPRB20,Bobkov22prb,Bobkov23PRB,Fyhn23PRL}.

In this work, we study a unique even-parity spin texture in cAFMs and explore its role in proximity-induced pairing correlations in cAFM coupled to conventional $s$-wave superconductors. We discover a mixed singlet-triplet pairing state, wherein both components exhibit finite center-of-mass momentum, akin to the  Fulde-Ferrell-Larkin-Ovchinnikov (FFLO) state, but with a trade-off behavior between them. While the singlet pairing exhibits even parity in both frequency and antiferromagnetic order, the triplet pairings are odd in these aspects. Moreover, the relative phase between the singlet and triplet amplitudes can be controlled by adjusting the junction orientation. We illustrate this SOC-free mechanism for generating triplet pairing through general symmetry analysis and, as a concrete example, by examining the cAFM on a kagome lattice. We construct low-energy effective models for the system and analytically show that the finite-momentum Cooper pairs are associated with inter-band pairing. For experimental implications for candidate materials such as Mn$_3$Ge and Mn$_3$Ga, we find that out-of-plane spin canting causes one of the equal-spin triplet amplitudes to predominate, thus facilitating spin-polarized supercurrents. 

\begin{figure}[t]
\includegraphics[width=0.48\textwidth]{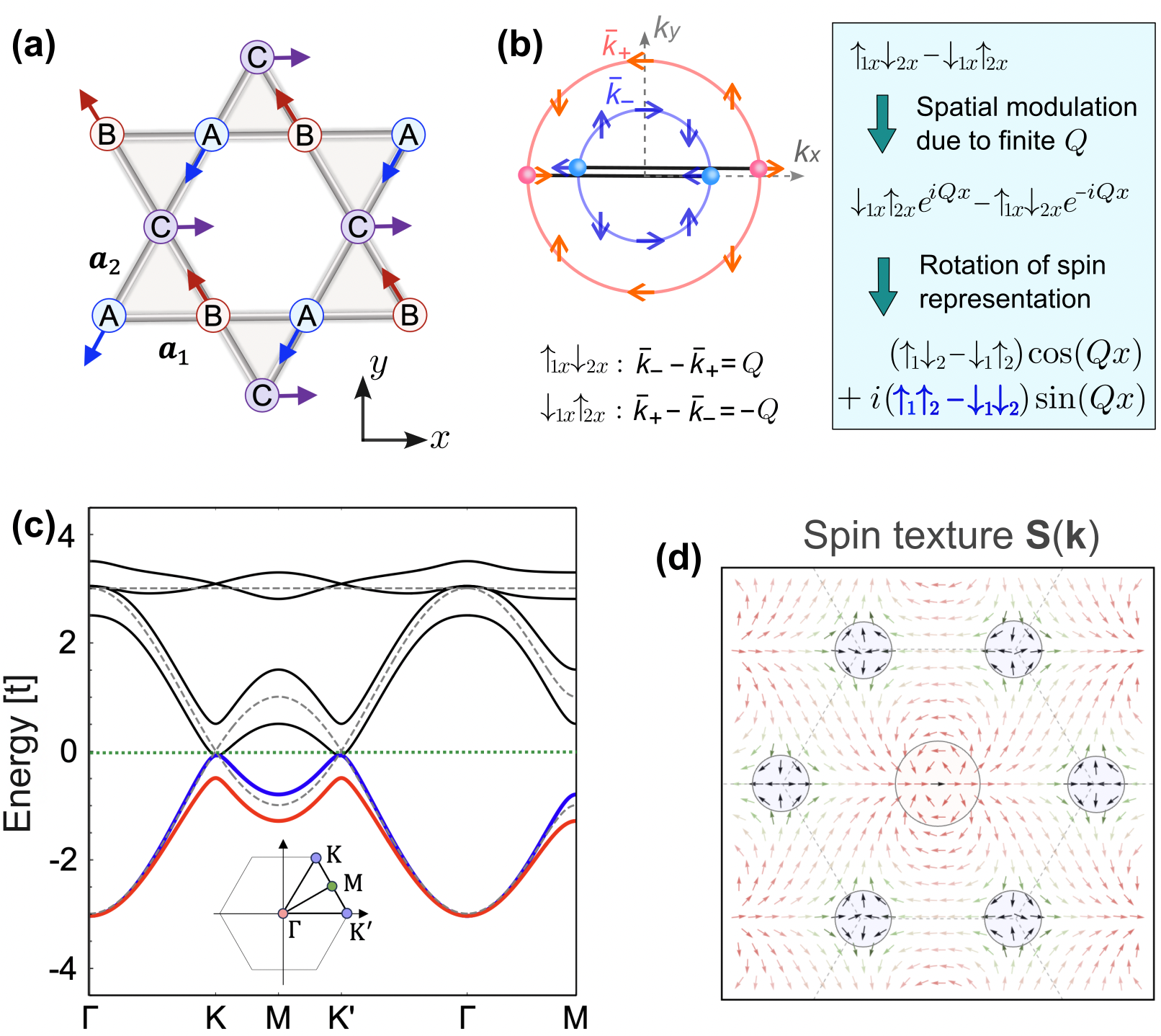}

\caption{(a) Crystal structure of the kagome antiferromagnet with a $120^\circ$ cAFM order. The arrows on the three sublattices $A,B,C$ indicate the local magnetic moments. ${\bm a}_1$ and ${\bm a}_2$ are nearest-neighbor vectors. 
(b) Creation mechanism of mixed singlet and equal-spin triplet pairing. For illustration, we consider the spins along $x$-direction and thus have two Cooper pairs with opposite center-of-mass momenta $\pm Q$. 
(c) Band structure along high symmetry lines in the absence (dashed) and presence (solid) of cAFM order. The inset shows the first Brillouin zone with the high symmetry points indicated. 
(d) Even-parity spin texture of the lowest band in the Brillouin zone.  
Parameters are $t=1$ and $J=0.3t$.
}
\label{fig:band-structure}
\end{figure}

{\color{blue}\emph{Mixed singlet-triplet pairing from even-parity spin texture.}--}
Non-relativistic spin-split bands can be classified by spin-space group, which involves rotation operators acting independently on space and spin coordinates~\cite{ifmmode2022Beyond,ifmmode2022Emerging,jiang2023enumeration,ren2023enumeration,xiao2023spin,Liu2022spin,schiff2023spin}. For concreteness, we consider the non-collinear cAFM phase on the kagome lattice [Fig.~\ref{fig:band-structure}(a)]. This phase can be characterized by $\{ U_{z}(\tfrac{2\pi}{3}) || C_z(\tfrac{\pi}{3})\}$, which combines a six-fold spatial rotation $C_z(\tfrac{\pi}{3})$ and a three-fold spin rotation $U_z(\tfrac{2\pi}{3})$ about $z$-axis. Additional symmetries include $\{U_x(\pi)||C_x(\pi)\}$ (reflection), $\{E || \mathcal{I}\}$ (inversion) and $\{U_z(\pi)\mathcal{T} || \mathcal{I}\}$ (mirror), where $\mathcal{I}$ and $\mathcal{T}$ are inversion and time-reversal symmetries, respectively. These operators $g$ expand the little group at the $\Gamma$ point in momentum space, determining the spin texture of each band by satisfying $g {\bf S}({\bf k}) g^{-1} =  {\bf S}(g {\bf k})$, where ${\bf S}({\bf k})$ denotes the spin polarization of a state with momentum ${\bf {\bf k}}=(k_{x},k_{y})$. The in-plane reflection
enforces $S_z({\bf k})=0$, while both in-plane components are even in ${\bf k}$ due to inversion symmetry, as denoted by ${\bf S}({\bf k}) = {\bf S}(-{\bf k})$. Thus, we refer to this as the even-parity spin texture. 
More details can be found in the Supplemental Material (SM)~\cite{SM-AFM2024}.
It fundamentally differs from the odd-parity spin texture arising from SOC, which obeys ${\bf S}({\bf k})=- {\bf S}(-{\bf k})$.

The even-parity spin texture drives the emergence of mixed singlet-triplet pairing in the cAFM in proximity to a conventional superconductor. To elucidate the basic mechanism, we present an intuitive picture in Fig.~\ref{fig:band-structure}(b). Given proximity-induced singlet Cooper pairs, only inter-band pairing is allowed due to the even-parity spin texture~\cite{SM-AFM2024}. For illustration, we consider a junction oriented along $x$-direction, with dominant electronic states spin-polarized in $x$-direction. In this case, electrons with spin $\uparrow_{1x}$ at $-\bar{k}_+$ ($\downarrow_{1x}$ at $-\bar{k}_-$) must pair with those with $\downarrow_{2x}$ at $\bar{k}_-$ ($\uparrow_{2x}$ at $\bar{k}_+$). The pairing may occur non-locally in time, as the subscripts $1$ and $2$ imply. Thus, the Cooper pairs carry finite center-of-mass momentum given by $\pm Q=\pm (\bar{k}_- -\bar{k}_+)$. The pairing amplitude acquires spatially varying phases $e^{\pm iQx}$, resulting in an FFLO-like state~\cite{Eschrig2015review}, i.e., $\uparrow_{1x}\downarrow_{2x}e^{iQx}-\downarrow_{1x}\uparrow_{2x}e^{-iQx}$. To reveal equal-spin triplet component polarized in $z$-direction, a spin rotation can be performed as $\uparrow_{\nu z}=(\uparrow_{\nu x}+\downarrow_{\nu x})/\sqrt{2}$ and $\downarrow_{\nu z}=(\uparrow_{\nu x}-\downarrow_{\nu x})/\sqrt{2}$ with $\nu\in\{1,2\}$ that are eigentates of $s_z$. This yields a mixed singlet-triplet pairing
$(\downarrow_{1z}\uparrow_{2z}-\uparrow_{1z}\downarrow_{2z})\cos(Qx)+i(\uparrow_{1z}\uparrow_{2z}-\downarrow_{1z}\downarrow_{2z})\sin(Qx)$. 
According to the Pauli principle, these $s$-wave triplet components are only permitted for the non-static case, implying that they are odd in frequency~\cite{Bergeret05RMP,Tanaka07PRL,Linder19RMP}.

Moreover, this argument holds for junctions oriented in any other direction, where dominant electronic states are spin-polarized in different directions. The mixed pairing reads generally $(\downarrow_{1z}\uparrow_{2z}-\uparrow_{1z}\downarrow_{2z})\cos(Qx)+i(e^{i2\phi}\uparrow_{1z}\uparrow_{2z}-e^{-i2\phi} \downarrow_{1z}\downarrow_{2z})\sin(Qx)$, where $\phi$ is the junction orientation relative to $x$-axis and factor $2$ reflects the symmetry $\{ U_z(\tfrac{2\pi}{3}) || C_z(\tfrac{\pi}{3}) \}$.  Thus, when $\phi$ varies, the phases of the triplet components change, while the phase of the singlet component remains the same as that in the superconductor. This analysis shows that opposite-spin triplet pairing correlation vanishes, aligning with the symmetry analysis provided in the SM~\cite{SM-AFM2024} and validated by our numerical calculations [Fig.~\ref{fig:Mz-canting}(c)]

Hence, the even-parity spin texture uniquely induces the mixed singlet-triplet pairing at finite frequencies in the junction. It showcases several intriguing features: (i) equal magnitudes for two equal-spin triplet components, (ii) a trade-off in the spatial oscillations between singlet and triplet components,
(iii) a tunable $\pi$-periodic phase difference between singlet and triplet components controlled by junction orientation, and
(iv) no need for net magnetization, SOC, multi-band superconductivity, spin-active interfaces, or large-scale magnetic inhomogeneities (e.g., domain walls). Below, we verify these features numerically and analytically, establishing this mixed pairing as a novel state distinct from previous proposals~\cite{Bergeret2001LongRange,Volkov03PRL,Bergeret05RMP,Buzdin05RMP,Linder09PRL,Cottet11PRL,Eschrig2015review,Demler97prb,Gorkov01PRL,Frigeri04PRL,Bergeret13PRL,Bergeret14PRB,Crepin15prl,smidman2017superconductivity,Cayao17PRB,Cayao18prb,Fleckenstein18prb,Bobkova05PRL,Andersen06PRL,Kamra18PRL,Rabinovich19PRR, JakobsenPRB20,Bobkov22prb,Fyhn23PRL,Tanaka07PRL,Tanaka07PRB,Black-Schaffer13PRB,Linder19RMP}.

{\color{blue}\emph{Odd-frequency equal-spin triplet pairing.}--}To validate the general analysis above, we consider a tight-binding model on the kagome lattice~\cite{Chen2014Anomalous,Liu2022spin}, 
\begin{align}\label{eq-kagome-ham}
\mathcal{H}({\bf k}) = 
\begin{pmatrix} t & f_{1,{\bf k}} & f_{2,{\bf k}} \\
f_{1,{\bf k}} & t & f_{3,{\bf k}} \\
f_{2,{\bf k}} & f_{3,{\bf k}} & t
\end{pmatrix} s_0 +
\begin{pmatrix}
   {\bm d}_{A}   & 0 & 0 \\
   0  & {\bm d}_{B}  & 0 \\ 
   0  & 0 & {\bm d}_{C} 
\end{pmatrix} \cdot {\bf{s}},    
\end{align}
where $f_{i,{\bf k}}=-2t\cos({\bf k}\cdot{\bf a}_i)$, $t$ is the strength of nearest-neighbor hopping between $s$-electrons and is taken as energy unit. The three nearest-neighbor vectors are defined as ${\bf a}_{1}=(1,0)$, ${\bf a}_{2}=(1,\sqrt{3})/2$, and ${\bf a}_{3}={\bf a}_{2}-{\bf a}_{1}$. For concreteness, we consider a $120^{\circ}$ cAFM configuration with local magnetic moments ${\bm d}_{j}=J(\cos\theta_j,\sin\theta_j,0)$, where $J$ measures the strength and $\theta_{A/B/C}=\{4\pi/3, 2\pi/3, 0 \}$. The band structure for $J=0$ along high-symmetry lines is depicted in Fig.~\ref{fig:band-structure}(c) [dashed curves]. It possesses Schr\"odinger-like and Dirac-like electrons at the band edge (around the $\Gamma$ point) or center (around the $K/K'$ point), respectively. The presence of $J\neq0$ breaks the spin degeneracy of the band structure [solid lines in Fig.~\ref{fig:band-structure}(c)], and leads to even-parity spin textures for both types of electrons [Fig.~\ref{fig:band-structure}(d)].

We compute the pairing correlations in an $s$-wave superconductor/cAFM junction oriented along $x$-direction from the anomalous Matsubara Green function, using a recursive technique~\cite{Sancho1984JPFMP,Sancho1985JPFMP}. The superconducting lead is modeled on the same kagome lattice without magnetism but a uniform $s$-wave pairing potential, described by  $H_{\Delta}
=\sum_{i,\alpha \in \{A,B,C\}}
\bigl(
\Delta c_{i,\alpha,\uparrow}^\dagger \, c_{i,\alpha,\downarrow}^\dagger +h.c.
\bigr)$, where $i$ runs over all unit cells and $\alpha$ labels the three sublattices. For simplicity, we impose translation symmetry in $y$-direction. The anomalous Green function is decomposed into~\cite{Breunig19prb} 
\begin{equation}
 {\bm G}_{eh} 
 = -is_y \sum_{j={0,x,y,z}} {\bm f}_j(x,x',k_y,i\omega) s_j, 
\end{equation} 
where ${\bm f}_0$ corresponds to the singlet, and ${\bm f}_z$ as well as ${\bm f}_{\uparrow\uparrow/\downarrow\downarrow}= \mp {\bm f}_x - i {\bm f}_y$ are triplet amplitudes. $x$ and $x'$ denote layer positions of unit cells of the kagome lattice along $x$-axis, and $\omega$ the Matsubara frequency. We focus on local pairings with $x=x'$. Summing over the three sublattices within each unit cell, we obtain the pairing amplitudes as $\mathcal{F}_j(x,k_y,\omega) = \text{Tr} [{\bm f}_j(x,x,k_y,i\omega)]$, where $j\in\{0,z,\uparrow\uparrow,\downarrow\downarrow\}$. We use the parameters $J=0.5t$ and $\Delta=0.1t$ in the calculations.

\begin{figure}[t]
\includegraphics[width=0.48\textwidth]{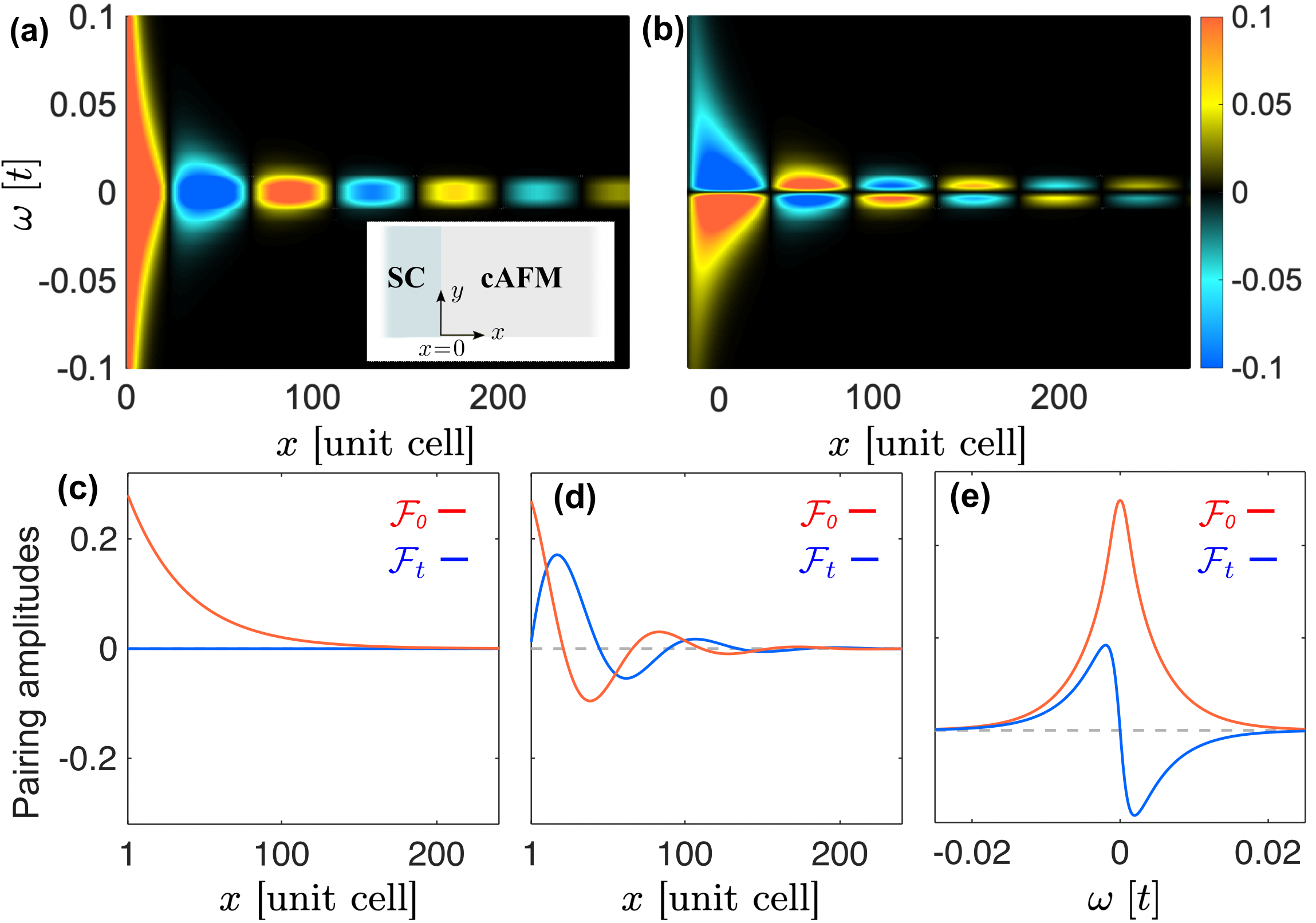}

\caption{(a) Singlet amplitude $\mathcal{F}_0$ (in units of $t^{-1}$) as a function of frequency $\omega$ and position $x$ (in units of unit-cell layers) in the cAFM. Inset sketches the SC-cAFM junction oriented in $x$-direction.
(b) Same as (a) but for equal-spin triplet pairing amplitudes, $\mathcal{F}_t=i\mathcal{F}_{\uparrow\uparrow}=-i\mathcal{F}_{\downarrow\downarrow}$.  
(c) $\mathcal{F}_0$ (red) and $\mathcal{F}_t$ (blue) as functions of $x$ in the cAFM  for $J=0$ and $\omega=0.1\Delta$.
(d) Same as (c) but for a finite cAFM order ($J=0.5t$). 
(e) $\mathcal{F}_0$ (red) and $\mathcal{F}_t$ (blue) as functions of $\omega$ for $x=80$.
}

\label{fig:triplet-FFLO}
\end{figure}

\begin{figure}[t]
\includegraphics[width=0.48\textwidth]{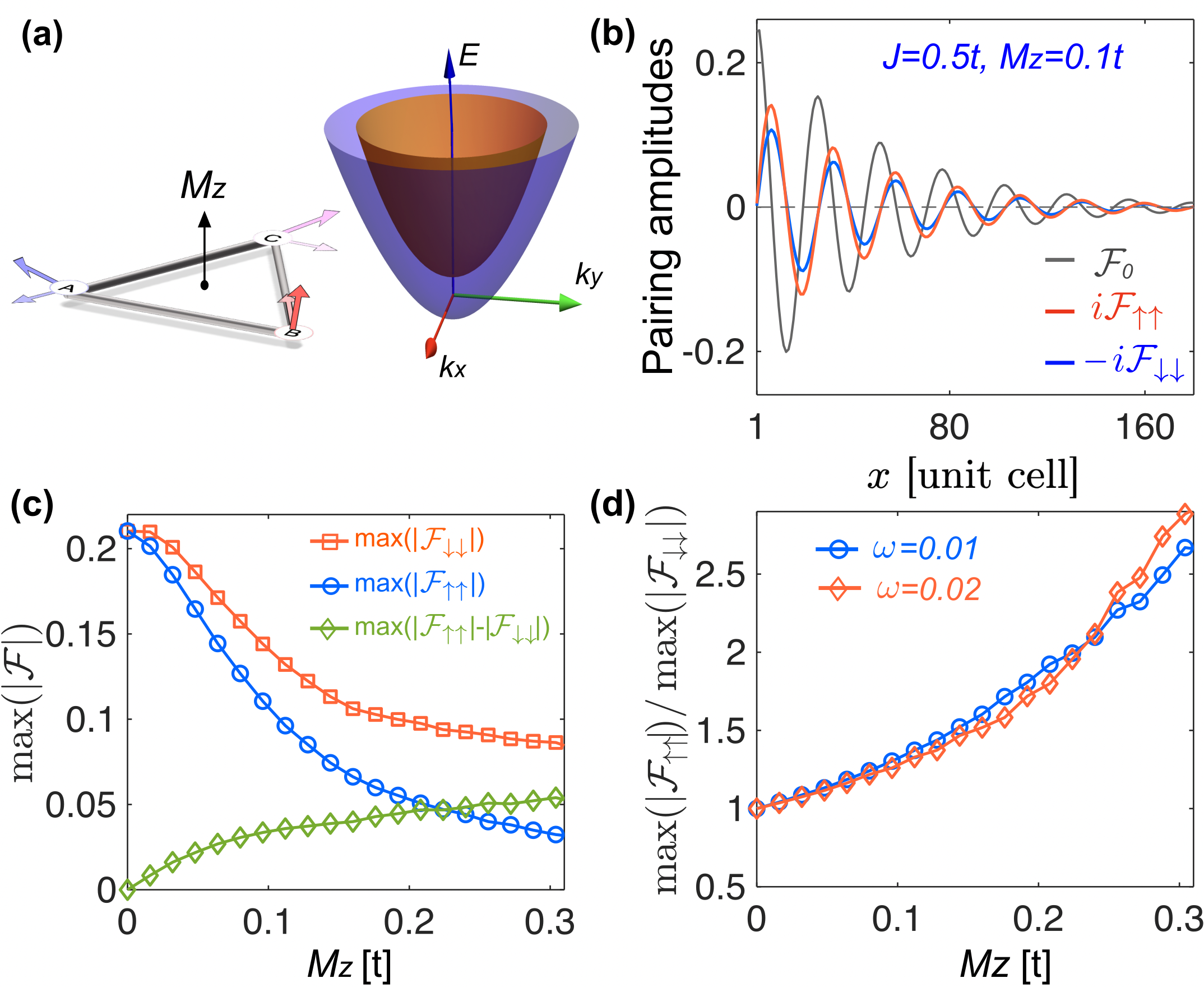}

\caption{
(a) Schematic of out-of-plane canting on the cAFM and the band structure near the band edge. 
(b) Spin-singlet (back) and equal-spin-triplet amplitudes (red and blue), $\mathcal{F}_0$, $\mathcal{F}_{\uparrow\uparrow}$ and $\mathcal{F}_{\downarrow\downarrow}$ (in units of $t^{-1}$), as functions of position $x$ in the cAFM under out-of-plane spin canting (i.e., $J=0.5t$ and $M_z=0.1t$).  
(c) $\max(|\mathcal{F}_{\uparrow\uparrow}|)$, $\max(|\mathcal{F}_{\downarrow\downarrow}|)$ and their difference as functions of canting strength $M_z$. (d) $\max(|\mathcal{F}_{\uparrow\uparrow}|)/\max(|\mathcal{F}_{\downarrow\downarrow}|)$ as a function of $M_z$. Other parameters: $\mu=\mu_S=-2t$, $\omega=0.01t$ and $\Delta=0.05t$. }

\label{fig:Mz-canting}
\end{figure}

Figure~\ref{fig:triplet-FFLO} presents the results for $k_y=0$ and a chemical potential near the band edge (i..e, $\mu=-2.5t$). We observe a purely real singlet amplitude $\mathcal{F}_0$ induced in the cAFM, even at finite frequency $\omega$. It maximizes in the static limit ($\omega=0$) and decays monotonically to zero as $|\omega|$ grows. For $|\omega|<\Delta$, pronounced oscillations around zero appear near the interface, similar to the FFLO state [Fig.~\ref{fig:triplet-FFLO}(a)]. More strikingly, for $\omega\neq0$, while $\mathcal{F}_z=0$, two purely imaginary equal-spin triplet components emerge and have the same amplitudes, $\mathcal{F}_{\downarrow\downarrow}=-\mathcal{F}_{\uparrow\uparrow}=i\mathcal{F}_t$ [Fig.~\ref{fig:triplet-FFLO}(b)]. $\mathcal{F}_t$ is odd in $\omega$, in contrast to $\mathcal{F}_0$. It is of the same order of magnitude and oscillates in $x$ as $\mathcal{F}_0$. In a nonmagnetic metal ($J=0$), only the singlet pairing exists ($\mathcal{F}_0\neq0$ but $\mathcal{F}_z=\mathcal{F}_t=0$) [Fig.~\ref{fig:triplet-FFLO}(c)]. Once $J\neq0$ is present, FFLO-like pairings are developed in both singlet and equal-spin triplet channels at finite $\omega$, as evidenced in Fig.~\ref{fig:triplet-FFLO}(d). This comparison demonstrates that the mixed singlet-triplet pairing is driven by the even-parity spin texture stemming from the cAFM order. Additionally, the oscillations of $\mathcal{F}_0$ and $\mathcal{F}_t$ appear to have a $\pi/2$ difference in phase, resembling a trade-off between them.

Moreover, as $\omega$ increases, $\mathcal{F}_t$ undergoes a rapid increase until reaching a maximum value, followed by a monotonic decay to zero [Figs.~\ref{fig:triplet-FFLO}(b) and \ref{fig:triplet-FFLO}(e)]. The amplitude of $\mathcal{F}_t$ increases as $J$ increases, and flips sign when $J$ flips sign. The results of summing over all $k_y$ are provided in the SM~\cite{SM-AFM2024}, showing qualitatively the same features. We also examine the proximity effect for other energy regimes and demonstrate that these intriguing observations persist~\cite{SM-AFM2024}. Rotating the junction orientation varies the phases of $\mathcal{F}_{\uparrow\uparrow(\downarrow\downarrow)}$. All these results agree with our general argument.

{\color{blue}
\emph{Finite-momentum Cooper pairing.}--}
Next, we prove the emergence of finite-momentum pairing analytically. Since the singlet and triplet pairings share the same finite momentum at any frequency, it is sufficient to consider only the singlet pairing in the static limit. To this end, we derive two representative low-energy effective Hamiltonians near the band edge ($E_{\text {edge}}=-3t$) and center ($E_{\text {cen}}=0$), respectively (see SM~\cite{SM-AFM2024} for derivation). The first one showcases Schr\"odinger-like electrons around the $\Gamma$ point,
\begin{equation}
\mathcal{H}_{\text{\ensuremath{\Gamma}}}({\bf k})=t({k_x^2+k_y^2}) s_{0}+m_\Gamma[(k_{x}^{2}-k_{y}^{2})s_{x} - 2k_{x}k_{y}s_{y}],
\label{eq:gamma}
\end{equation}
where $m_\Gamma=J/12$~\cite{SM-AFM2024,lee2024prl,LHHu25SCPMA}. The second model features Dirac-like electrons around the $K/K'$ points,
\begin{equation}
\mathcal{H}_{K(K')}({\bf k})
  =\chi v(k_{x}\sigma_{1}+k_{y}\sigma_{3})s_{0}+m_K(\sigma_{1}s_{x}-\sigma_{3}s_{y}),
 \label{eq:Kpoint}
\end{equation}
with $v=\sqrt{3}t$, $m_K=J/2$, $\chi=+1$ ($-1$) for the valley around the $K$ ($K'$) point, and $\sigma_i$ are Pauli matrices acting on pseudospin~\cite{SBZhang2025RRP}. Using the approach of Cooper-pair propagator, we calculate observables such as local pairing correlations and Josephson supercurrent~\cite{SM-AFM2024}. Explicitly, for large widths and Fermi surfaces, the induced pairing correlations $\langle|\Psi({\bf r})|\rangle$ for the Sch\"odinger and Dirac electrons can be found as 
\begin{align}\label{eq-order-para-both}
\langle|\Psi({\bf r})|\rangle_{\Gamma(K)} 
& \propto \dfrac{1}{x^{3/2}}\cos\Big( Q_{\Gamma(K)} x-\frac{\pi}{4} \Big), 
\end{align}
where we have $Q_\Gamma \approx \sqrt{\mu_\Gamma} m_\Gamma$ and $Q_K \approx 2m_K/v$ for small $J$. The spatial oscillation of $\langle|\Psi({\bf r})|\rangle_{\Gamma(K)}$ agrees well with our numerical findings. The periodicity of both oscillations is determined by the Cooper-pair momentum, i.e., $P_{\Gamma(K)} = 2\pi/Q_{\Gamma/K}$. Notably, $P_\Gamma$ increases as $\mu_\Gamma$ decreases, while $P_K$ remains independent of $\mu_K$, which may be experimentally verified. The finite-momentum pairing fundamentally differs from that of N\'eel-type triplet pairing in collinear antiferromagnet junctions, which stems from Umklapp scattering and cancels over sublattices~\cite{Bobkov23PRB}. Note that here we consider the cases with large $\mu_\Gamma$ or $\mu_K$ so that two Fermi surfaces form at $\Gamma$ or $K$. For small $|\mu_K|(<|J|)$, only one Fermi surface exists at each valley and triplet pairing becomes dominant~\cite{SBZhang2025RRP,JXHou25PRB}. In the SM~\cite{SM-AFM2024}, we further calculate supercurrents in Josephson junctions and show that the finite-momentum pairing generates 0-$\pi$ transitions, which can be achieved by adjusting junction length or gating/doping on the cAFM.

{\color{blue}\emph{Spin canting and experimental implications.}--}
As discussed above, the equal-spin triplet pairings $\mathcal{F}_{\uparrow\uparrow/\downarrow\downarrow}$ exhibit equal magnitude. However, real materials may have out-of-plane spin canting, as signified by sizable anomalous Hall effects in experiments~\cite{Kiyohara2016Giant}. This spin canting induces a weak net magnetization and makes one triplet component predominant. To illustrate this, we capture the spin canting $M_z$ by incorporating an additional Zeeman term into the model. Thus, the inclusion of $M_zs_z$ (or $M_zs_z\sigma_0$) in the effective Hamiltonians in Eqs.~\eqref{eq:gamma} and \eqref{eq:Kpoint} gives rise to the band splitting and the opening of spin gaps at both $\Gamma$ and $K/K'$ points [Fig.~\ref{fig:Mz-canting}(a)].

In Fig.~\ref{fig:Mz-canting}, we calculate numerically the pairing correlations with the kagome lattice model to explore the effect of $M_z$. We find that the triplet amplitudes $\mathcal{F}_{\uparrow\uparrow/\downarrow\downarrow}$ remain purely imaginary, while the singlet one $\mathcal{F}_{0}$ is real-valued. Their trade-off behavior persists [Fig.~\ref{fig:Mz-canting}(b)]. 
A finite opposite-spin triplet component $\mathcal{F}_z$ (corresponding to $\uparrow\downarrow+\uparrow\downarrow$) emerges as well, which is purely imaginary and oscillates in phase with $\mathcal{F}_{\uparrow\uparrow/\downarrow\downarrow}$. 
Interestingly, a finite $M_z$ indeed creates an imbalance between $\mathcal{F}_{\uparrow\uparrow}$ and $\mathcal{F}_{\downarrow\downarrow}$.  Both ${\cal F}_{\uparrow\uparrow/\downarrow\downarrow}$ are generally suppressed by increasing $M_z$. However, they are suppressed differently, resulting in a substantial discrepancy between them. For positive $M_z$, we have always $|\mathcal{F}_{\uparrow\uparrow}|>|\mathcal{F}_{\downarrow\downarrow}|$, while for negative $M_z$, $|\mathcal{F}_{\uparrow\uparrow}|<|\mathcal{F}_{\downarrow\downarrow}|$ (detailed in the SM~\cite{SM-AFM2024}).

In Fig.~\ref{fig:Mz-canting}(c), we calculate the maximum values of $|\mathcal{F}_{\uparrow\uparrow/\downarrow\downarrow}|$ and their difference as a function of $M_z$. For a small canting tendency, $\mathcal{F}_{\uparrow\uparrow}$ and $\mathcal{F}_{\downarrow\downarrow}$ decrease monotonically at rather different rates as $M_z$ increases. Figure~\ref{fig:Mz-canting}(d) shows the ratio $\max(|\mathcal{F}_{\uparrow\uparrow}|)/\max(|\mathcal{F}_{\downarrow\downarrow}|)$ as a function of $M_z$~\footnote{The ratio of between $\mathcal{F}_{\uparrow\uparrow}/\mathcal{F}_{\downarrow\downarrow}$ is approximately independent of position $x$.}. This ratio increases monotonically as $M_z$ grows for any frequencies. Thus, a stronger $M_z$ yields more excess Cooper pairs with specific spin polarization, implying a net spin supercurrent in the superconducting junctions. These general features occur for both Schr\"odinger and Dirac electrons. 
Note that instead of spin-canting, a significant $M_z$ could also originate from orbital magnetization under an external magnetic field~\cite{chen2020prb}, leading to the same conclusion.

Finally, we discuss cAFM candidate materials Mn$_3$Ga and Mn$_3$Ge~\cite{Kiyohara2016Giant,nayak2016large,liu2017SciRep,Song2024AFM} 
as possible platforms for testing our predictions. The band structures of these materials have been investigated using \emph{ab initio} calculations~\cite{zhang2017Strong}, showing pronounced spin-split electronic states with quadratic dispersion around the $\Gamma$ point. 
We estimate the shifts in $\mu$ needed to bring the quadratic electrons to the Fermi level, which are around $-0.1$ eV and $-0.3$ eV for Mn$_3$Ga and Mn$_3$Ge, respectively~\footnote{We neglect changes in the chemical potential $\mu$ when the materials are made into thin films.}. For Mn$_3$Ga with N\'eel temperature 470 K, a significant spin splitting $m_\Gamma/t=0.35$ can be estimated for the effective model in Eq.~\eqref{eq:gamma}. Using $\mu_\Gamma \equiv \mu-E_{\text {edge}}= 0.1$-$0.3 t$ in Eq.~\eqref{eq:gamma} and the lattice constant of $a=0.53$ nm, we estimate the oscillation periodicity of order parameters and Josephson currents as $\sim 2\pi a t/ (\sqrt{\mu_\Gamma/t} \, m_\Gamma) \approx 17$-30 nm. This suggests that $0$-$\pi$ transitions could be observed in Josephson junctions with lengths exceeding 30 nm, a feasible length scale within current experimental settings. For Mn$_3$Ge, the N\'eel temperature is around 380 K and the cAFM phase remains robust at low temperatures~\cite{jeon2021long,jeon2023chiral}. The spin splitting is weaker ($m_\Gamma/t\sim 0.1$), compared to Mn$_3$Ga. Thus, for the same $\mu_\Gamma$, the oscillation periodicity would be larger.

{\color{blue}\emph{Conclusions and discussions.}--}
In summary, we have proposed a general theory for creating mixed singlet-triplet pairing states through even-parity spin texture in cAFMs that preserve inversion symmetry. For kagome systems, our general symmetry analysis, numerical calculations, and analytical results have consistently demonstrated that the even-parity spin texture gives rise to a mixed state of even-frequency singlet and odd-frequency equal-spin triplet Cooper pairs with finite momentum in the cAFM when interfaced with conventional superconductors. These finite-momentum pairings can be detected via damped oscillations in order parameters and controllable 0-$\pi$ transitions in Josephson junctions. Additionally, we have shown that the out-of-plane spin canting leads to an imbalance between Cooper pairs with opposite spins. 
This work demonstrates the singlet-triplet mixed state is not a privilege of spin-orbit coupled system but more broadly associated with the spin texture. In realistic systems, inhomogeneity might exist in the interfaces, leading to spatial variations in the coupling between the cAFM and the superconductor. However, as long as these variations are much smaller than the average coupling strength, our results remain qualitatively unaffected.

While our discussion primarily focuses on kagome-type systems, our main findings also apply to other lattice structures hosting the cAFM phase~\cite{Martin08PRL}, such as in the Mn$_3A$N family ($A$ = Ni, Ga, Sn) with triangular lattices and new candidate materials enumerated in recent studies~\cite{jiang2023enumeration,ren2023enumeration,xiao2023spin}. 

We thank Qian~Niu and Zhenyu~Zhang for valuable discussions.
S.B.Z. was supported by the start-up fund at HFNL, the Innovation Program for Quantum Science and Technology (Grant No. 2021ZD0302801), and the National Natural Science Foundation of China (Grant No. 12488101). L.H.H. was supported by the start-up fund of Zhejiang University, the Fundamental Research Funds for the Central Universities (Grant No. 226-2024-00068), the Jane and Aatos Erkko Foundation and the Keele Foundation as part of the SuperC collaboration.

%

\appendix

\onecolumngrid
 \setcounter{figure}{0}
\renewcommand{\thefigure}{A\arabic{figure}}

\tableofcontents

\section{${\bf k}\cdot {\bf p}$ effective models\label{sm-sec-A}}
\subsection{Construction of the effective models}

In this subsection, we derive the ${\bf k}\cdot {\bf p}$ models for the Schr\"odinger and Dirac electrons, respectively. To this end, we write the Hamiltonian in Eq. (1) in the main text into two parts: $\mathcal{H}({\bf k})=\mathcal{H}_0({\bf k})+H_{\text{cAFM}}$, where 
\begin{equation} \label{eq:HJ}
\mathcal{H}_0({\bf k}) = 
\begin{pmatrix} t & f_{1,{\bf k}} & f_{2,{\bf k}} \\
f_{1,{\bf k}} & t & f_{3,{\bf k}} \\
f_{2,{\bf k}} & f_{3,{\bf k}} & t
\end{pmatrix} \otimes s_0,\;\;\;\;
H_{\text{cAFM}}= 
\begin{pmatrix}
   {\bm d}_{A}\cdot {\bf{s}}   & 0 & 0 \\
   0  & {\bm d}_{B}\cdot {\bf{s}}  & 0 \\ 
   0  & 0 & {\bm d}_{C}\cdot {\bf{s}} 
\end{pmatrix},  
\end{equation}
where $f_{i,{\bf k}}=-2t\cos({\bf k}\cdot{\bf a}_i)$, ${\bf a}_{1}=(1,0)$, ${\bf a}_{2}=(1,\sqrt{3})/2$, and ${\bf a}_{3}={\bf a}_{2}-{\bf a}_{1}$. ${\bm d}_{A,B,C}$ are given in the main text. The Hamiltonian is written in the basis $\{c_{A,\uparrow},c_{A,\downarrow},c_{B,\uparrow},c_{B,\downarrow},c_{C,\uparrow},c_{C,\downarrow}\}$.

We first look at the Schr\"odinger electrons around the $\Gamma$ (${\bf k}=0$) point near the band edge ($E_{\text{edge}}=-3t$). The effective model is constructed in a basis formed by two electronic states around the $\Gamma$ point with opposite spins, evaluated in the absence of cAFM order (i.e., setting $H_{\text{cAFM}} = 0$). These two electron states are obtained by diagonalizing $\mathcal{H}_0({\bf k})$ near the $\Gamma$ point. Expanding to the quadratic order in $\bf k$ around the $\Gamma$ point, their wavefunctions can be written as 
\begin{subequations}
\begin{align}
\left|\psi_{\Gamma,\uparrow}({\bf k})\right>= & \frac{1}{N_0} 
\begin{pmatrix} 1 + \dfrac{k_x^2-k_y^2}{8} - \dfrac{k_xk_y}{4\sqrt{3}}, &0, & 1 - \dfrac{k_x^2-k_y^2}{8} + \dfrac{k_xk_y}{4\sqrt{3}}, & 0, & 1 , & 0 
\end{pmatrix}^T,  \\
\left|\psi_{\Gamma,\downarrow}({\bf k})\right> =  & \frac{1}{N_0} 
\begin{pmatrix} 0, & 1 + \dfrac{k_x^2-k_y^2}{8} - \dfrac{k_xk_y}{4\sqrt{3}}, & 0, &1 - \dfrac{k_x^2-k_y^2}{8} + \dfrac{k_xk_y}{4\sqrt{3}}, & 0, & 1  
\end{pmatrix}^T,
\end{align}
\end{subequations}
where $N_0$ denotes the normalization factor and the superscript $T$ means tranpose. Projecting the full kagome lattice Hamiltonian (including the cAFM order) into the space spanned by the bare states,  
\begin{equation}
    H_{s,s'} = \langle \psi_{\Gamma,s}| \mathcal{H}({\bf k}) | \psi_{\Gamma,s'}\rangle, \;\;\; s,s'\in\{\uparrow,\downarrow\},
\end{equation}
we obtain the ${\bf k}\cdot {\bf p}$ Hamiltonian as
\begin{equation}
\mathcal{H}_{\text{\ensuremath{\Gamma}}}({\bf k})=t(k_{x}^{2}+k_{y}^{2})s_{0}+m_\Gamma[(k_{x}^{2}-k_{y}^{2})s_{x} - 2k_{x}k_{y}s_{y}].
\label{eq:G-model}
\end{equation}
Here, we keep only up to the quadratic order in momentum again. 
The first term corresponds to the usual kinetic energy. The second term originates from the cAFM order and exhibits a unique spin-momentum locking. The two spin-split bands are $\varepsilon_\pm({\bf k}) = (t \pm m_\Gamma) k^2$, and they carry nontrivial spin textures given by ${\bf S}({\bf k})=\pm(\hbar/2)(-\cos2\theta_{\bf{k}},\sin2\theta_{\bf{k}})$. $\theta_{\bf{k}}=\arctan(k_y/k_x)$ is the angle characterizing the direction of momentum. 

To quantify the topology of ${\bf S}({\bf k})$, we calculate the winding number 
\begin{align}
	\mathcal{N}_w = \frac{1}{2\pi} \int_{0}^{2\pi} d\theta \; \hat{z}\cdot\Big[{\bf \hat S}(k,\theta)\times \frac{d {\bf \hat S}(k,\theta)}{d\theta}\Big],
\end{align}
Note that although the two bands have opposite chiral spin textures, they carry the same spin winding number $\mathcal{N}_{w}= 2$. The two bands touch quadratically at the $\Gamma$ point, as protected by the spin-space group. This contrasts with classical ferromagnets in which all bands are fully separated.

For the Dirac electrons in the band center regime, the electrons are mainly around the $K$ and $K'$ points. We take one valley around $K$ for illustration. The result for the other valley can be obtained by exploiting inversion symmetry. There are two degenerate states at the $K$ point in the absence of cAMF order. In an orthogonal way, we can write the four bare basis states (considering spin degrees of freedom) as
\begin{subequations}
\begin{align}
|\psi_{\alpha,\uparrow}\rangle & = \begin{pmatrix}
    \dfrac{\sqrt{3}+3}{6},& 0, & \dfrac{\sqrt{3}-3}{6}, & 0, &\dfrac{1}{\sqrt{3}}, & 0
\end{pmatrix}^T,  \\
|\psi_{\alpha,\downarrow}\rangle & = \begin{pmatrix} 0, &
    \dfrac{\sqrt{3}+3}{6}, & 0, & \dfrac{\sqrt{3}-3}{6}, &0, &\dfrac{1}{\sqrt{3}}
\end{pmatrix}^T,  \\
|\psi_{\beta,\uparrow}\rangle & = \begin{pmatrix} \dfrac{\sqrt{3}-3}{6}, &0, & \dfrac{\sqrt{3}+3}{6},&0, &\dfrac{1}{\sqrt{3}}, & 0
\end{pmatrix}^T, \\
|\psi_{\beta,\downarrow}\rangle & = \begin{pmatrix} 0, & \dfrac{\sqrt{3}-3}{6}, & 0, & \dfrac{\sqrt{3}+3}{6}, & 0, & \dfrac{1}{\sqrt{3}}
\end{pmatrix}^T.
\end{align}
\end{subequations}
Note that we keep only up to the linear order in momentum $\bf k$ (measured from the $K$ point). Similar to the case of the band edge, we perform the projection and obtain the ${\bf k}\cdot {\bf p}$ Hamiltonian as 
\begin{subequations}
\begin{align}
H_{\text{K}}({\bf k}) & =\begin{pmatrix} vk_{y} & v k_{x} & im_K & m_K\\
v k_{x} & -v k_{y} & m_K & -im_K\\
-im_K & m_K & v k_{y} & v k_{x}\\
m_K & im_K & v k_{x} & -v k_{y}
\end{pmatrix} \\
&=v(k_{x}\sigma_{1}+k_{y}\sigma_{3})s_{0}+m_K(\sigma_{1}s_{x}-\sigma_{3}s_{y}),
\label{eq:KKpoint}
\end{align}
\end{subequations}
where the effective velocity is $v=\sqrt{3}t$, and $m_K=J/2$. The model is written in the basis
$(\psi_{\alpha,\uparrow},\psi_{\beta,\uparrow},\psi_{\alpha,\downarrow},\psi_{\beta,\downarrow})$.
$(\sigma_{1},\sigma_{2},\sigma_{3})$ are the Pauli matrices in the
$\{\alpha,\beta\}$ space. Using inversion symmetry, the model for the other valley at $K'$ can be found as 
\begin{align}
H_{\text{K}'}({\bf k}) & = - v(k_{x}\sigma_{1}+k_{y}\sigma_{3})s_{0}+m_K(\sigma_{1}s_{x}-\sigma_{3}s_{y}).
\label{eq:K'point}
\end{align}
It is the same as Eq.~\eqref{eq:KKpoint} but with an opposite velocity. 
The opposite signs of the Dirac fermion's velocity at the two valleys are enforced by inversion symmetry, i.e., $\mathcal{I}\mathcal{H}_{K}({\bf k}) \mathcal{I}^{-1} = \mathcal{H}_{K'}(-{\bf k})$ with $\mathcal{I}=\sigma_0s_0$. The energy bands are $E_{a,b}({\bf k}) = a m_K + b \sqrt{v^2k^2+m_K^2}$ with indices band $a,b=\pm$. Note that there is also a two-fold degeneracy (i.e.,~$E_{+,-}(0)=E_{-,+}(0)$) at the $K/K'$ points, protected by the spin-space group. Up to the lowest order of $J$, the spin-splitting term ($m_K$) turns out to be momentum-independent. 
Nevertheless, we find a nontrivial spin texture of the $E_{a,b}$ band at valley $\chi$ as ${\bf S}({\bf k}) \propto {\chi ab (\hbar/2) (\cos \theta_{\bf k}, - \sin\theta_{\bf k})}$. All the bands exhibit the spin winding number $\mathcal{N}_{w} = -1$, again consistent with the numerical results presented in the main text.

We briefly comment on the above two ${\bf k} \cdot {\bf p}$ models for cAFMs. First, both models explicitly break time-reversal symmetry but retain zero net magnetization. For the Schr\"odinger electrons, the AFM order behaves as a $\bf k$-dependent Zeeman field, $\mu_B{\bf B}_{\text{eff}}({\bf k})=m_\Gamma (k_x^2-k_y^2,-2k_xk_y)$, and couples directly to the electron spin. A similar term can be realized in AFM systems with hexagonal point groups~\cite{Fernandes2024Topological}. For the Dirac electrons, each valley has a ${\bf k}$-independent spin-splitting term that subtly couples to both the spin and sublattice (or called pseudospin). The interplay of this term with the pseudospin-momentum locking [i.e., the first term in Eqs.~\eqref{eq:KKpoint} and \eqref{eq:K'point}] then leads to vanishing net spin polarization, even within each valley. The two valleys also possess opposite spin textures. Furthermore, from both models, we see that all electron spins are polarized in the kagome plane and the direction of polarization is determined by the direction $\theta_{\bf k}$ of momentum.  Last but not least, by comparing the two models, we find that $m_K=6m_\Gamma$. This explains the observation that for given AFM strength $J$ [see Fig.~1 in the main text], the kagome cAFM exhibits a more pronounced band spin splitting for the Dirac electrons around the band center compared to that for the Sch\"odinger-like electrons near the band edge. This pronounced splitting may be responsible for the larger spin Hall conductivity observed when the chemical potential approaches the Dirac electrons~\cite{BHYan17PRL,zhang2018spin}.

\subsection{Pairing in the chiral antiferromagnet}
In this subsection, we show that the pairing correlation in the cAFM is predominantly inter-band. To illustrate this, we take the energy regime featuring Sch\"odinger electrons and use the $\bf{k}\cdot\bf{p}$ Hamiltonian. We begin with conventional Bardeen-Cooper-Schrieffer (BCS) superconductivity, which is described by an on-site $s$-wave pairing potential. By projecting the local pairing potential onto the low-energy Hamiltonian, the resulting Bogoliubov-de Gennes (BdG) Hamiltonian in spin and Nambu spaces is
\begin{align}
{\cal H}_{\text{BdG}} = t(k_x^2+k_y^2)s_0\gamma_z + J(k_x^2-k_y^2)s_x\gamma_z + 2Jk_xk_y s_y\gamma_0 -\mu s_0\gamma_z  + \Delta s_y\gamma_y,   
\end{align}
where $\bm s$ and $\bm \gamma$ are Pauli matrices in spin and Nambu spaces, respectively, $\mu$ is the chemical potential and $\Delta$ is the pairing potential. By performing a unitary transformation, the Hamiltonian can be rewritten in the band basis as 
\begin{align}
{\cal H}_{\text{BdG}}' = \left(\begin{matrix}
    (t+J)k^2-\mu & 0 & 0  & -e^{2i\theta_{\bm k}}\Delta \\
    0 & (t-J)k^2 -\mu & e^{2i\theta_{\bm k}}\Delta &  0 \\
    0 & e^{-2i\theta_{\bm k}}\Delta & -(t+J)k^2+\mu & 0 \\
    -e^{-2i\theta_{\bm k}}\Delta & 0 &0 & -(t-J)k^2+\mu
\end{matrix}  \right).
\end{align} 
Here, $\theta_{\bm k} =\arctan(k_y/k_x)$ is the angle defined in momentum space. 
This result shows that the pairing occurs between different bands (Fermi surfaces).

\section{Pairing correlations and Josephson currents for Schr\"odinger electrons}
\label{sm-sec-B}

\subsection{Cooper-pair propagator}

We derive the Cooper-pair propagator in the cAFM.  
We start with the ${\bf k}\cdot {\bf p}$ Hamiltonian, Eq.~\eqref{eq:G-model}, for the Schr\"odinger electrons. The model has two bands given by 
\begin{align}
\varepsilon_{{\bf k},\eta}  =(t+\eta m_\Gamma)k^{2},
\end{align}
where $k=(k^2_x+k_y^2)^{1/2}$.
Defining the projection matrices as 
\begin{align}
\begin{split}
P_{\eta} 
  &=\dfrac{s_{0}+\eta[-\cos(2\phi)s_{x}+\sin(2\phi)s_{y}]}{2},
\end{split}
\end{align}
where $k_{x}=k\cos\phi$ and $k_{y}=k\sin\phi$, the non-interacting
retarded Green function can be written as
\begin{equation}
G_{0}^{\text{ret}}({\bf k},\epsilon)=\sum_{\eta}\dfrac{P_{\eta}}{\epsilon+\mu_\Gamma-\varepsilon_{{\bf k},\eta}+i\delta},
\end{equation}
where $\mu_\Gamma$ is the chemical potential measured from the band edge. 
The spectral matrix function in momentum space reads
\begin{equation}
A_{0}({\bf k},\epsilon) \equiv 2\pi\sum_{\eta}\delta(\epsilon+\mu_\Gamma-\varepsilon_{{\bf k},\eta})P_{\eta}.\label{eq:spectral-function}
\end{equation}
Fourier transforming Eq.\ (\ref{eq:spectral-function}),
the spectral matrix function in real space is given by
\begin{align}\label{eq:SpectralFunction}
\begin{split}
 g_{0}({\bf r},\epsilon) 
 & = \int\dfrac{d^{2}{\bf k}}{(2\pi)^{2}}e^{i{\bf k}\cdot{\bf r}} A_{0}({\bf k},\epsilon)  \\
 & =\dfrac{1}{2\pi}\sum_{\eta}\int_{0}^{\infty}dkk\int_{0}^{2\pi}d\phi e^{ikr\cos(\phi-\theta)}\dfrac{\delta(k-k{}_{\eta})}{|\partial\varepsilon_{{\bf k},\eta}/\partial_{k}|_{k=k_{\eta}}}P_{\eta} \\
 & =\dfrac{1}{4\pi}\sum_{\eta}\dfrac{1}{t+\eta m_\Gamma }\int_{0}^{2\pi}d\phi'e^{ik_{\eta}r\cos\phi'}P_{\eta}',
\end{split}
\end{align}
where $\theta$ is the direction of the propagation ${\bf r}$, we have changed the integral to polar coordinates and redefined the angle variable $\phi'\equiv\phi-\theta$. The Fermi wave numbers are given by 
\begin{equation}
k_{\eta}=\sqrt{\dfrac{\epsilon+\mu_\Gamma}{t+\eta m_\Gamma}}.
\end{equation}
Explicitly, we find
\begin{align}
g_{0}({\bf r},\epsilon) & =\dfrac{1}{4}\sum_{\eta}\dfrac{1}{t+\eta m_\Gamma}\{J_{0}(k_{\eta}r)s_{0}+\eta J_{2}(k_{\eta}r)[\cos(2\theta)s_{x}-\sin(2\theta)s_{y}]\},
\end{align}
where $s_{\pm}=(s_{x}\pm is_{y})/2$ and $J_n$ with $n$ being integers are Bessel functions 
\begin{align}
J_{n}(z) & =(-1)^{n}J_{-n}(z) =\dfrac{1}{2\pi i^{n}}\int_{0}^{2\pi}d\phi e^{iz\cos\phi+in\phi}.
\end{align}

Next, we calculate the $\mathcal{T}$ function
\begin{align}
\begin{split}
\mathcal{T}({\bf r},\epsilon,\epsilon') & =\text{Tr}[g_{0}({\bf r},\epsilon)s_{y}g_{0}^{T}({\bf r},\epsilon')s_{y}] \\
&=\dfrac{1}{8}\sum_{\eta,\eta'}\dfrac{1}{t+\eta m_\Gamma}\dfrac{1}{t+\eta'm_\Gamma}[J_{0}(k_{\eta}r)J_{0}(k_{\eta'}'r)-\eta\eta'J_{2}(k_{\eta}r)J_{2}(k_{\eta'}'r)].
\end{split}
\end{align}
We consider large Fermi surfaces such that $\mu\gg|\epsilon|$
and $k_{\eta}r\gg1$. In this case, we approximate
\begin{subequations}
\begin{align}
J_{n}(k_{\eta}r) & \approx \sqrt{\dfrac{2}{\pi {\bar{k}_{\eta}}r}}\cos\Big(k_{\eta}r-\dfrac{n\pi}{2}-\dfrac{\pi}{4}\Big), \\
k_{\eta}  &\approx \sqrt{\dfrac{\mu_\Gamma}{t+\eta m_\Gamma}}\Big(1+\dfrac{\epsilon}{2\mu_\Gamma}\Big)=\bar{k}_{\eta}\Big(1+\dfrac{\epsilon}{2\mu_\Gamma}\Big),
\end{align}
\end{subequations}
and find that only the ``interband'' terms with $\eta'=-\eta$ are important. Here, we define $\bar{k}_{\eta}\equiv \sqrt{{\mu_\Gamma}/({t+\eta m_\Gamma})}$. This yields 
\begin{align}
\mathcal{T}({\bf r},\epsilon,\epsilon') & =\dfrac{1}{2\pi r(t^2-m_\Gamma^{2})}\sum_{\eta}\sqrt{\dfrac{1}{k_{\eta}k_{-\eta}'}}\cos\Big(k_{\eta}r-\dfrac{\pi}{4}\Big)\cos\Big(k_{-\eta}'r-\dfrac{\pi}{4}\Big).
\end{align}

The Cooper-pair propagator can be written as an integral of the $\mathcal{T}$ function and found as
\begin{align}
\begin{split}
D_\Gamma({\bf r}) & =\dfrac{1}{2}\int_{0}^{\infty}\dfrac{d\epsilon}{2\pi}\dfrac{d\epsilon'}{2\pi}\dfrac{\mathcal{T}({\bf r},\epsilon,\epsilon')+\mathcal{T}({\bf r},-\epsilon,-\epsilon')}{\epsilon+\epsilon'} \\
&={\dfrac{A_\Gamma}{8\pi^2 r^2}}\cos[(\bar{k}_{+}-\bar{k}_{-})r],
\end{split}
\end{align}
where $A_\Gamma={\mu_\Gamma^{1/2}}/[{(t^2-m_\Gamma^{2})^{3/4}}(\bar{k}_{+}+\bar{k}_{-})]$, we have considered $a=J_{+}/J_{_{-}}>0$ (i.e., $|J|<2$) and
used 
\begin{subequations}
\begin{align}
\int_{0}^{\infty}dxdy\dfrac{\cos(ax-y)}{x+y} & =\dfrac{\pi}{1+a}, \\ 
\int_{0}^{\infty}dxdy\dfrac{\cos(ax+y)}{x+y} &=0.
\end{align}
\end{subequations}

\subsection{Pairing correlations}

We consider a planar junction made by the cAFM and an $s$-wave superconductor. The coupling between the two layers is assumed as a constant $\lambda$ along the interface at $x=0$. The width of the junction is $W$. For large width $W$ and Fermi surfaces, the proximity-induced local pairing correlation becomes independent of the $y$ coordinate, i.e., $\langle|\Psi(x,y)|\rangle =\langle|\Psi(x,0)|\rangle$. Thus, we find the pairing correlation as
\begin{align}
\begin{split}
\langle|\Psi(x,y)|\rangle_\Gamma & =\lambda\int_{-W/2}^{W/2} { dy'D_\Gamma (x,0;0,y') }\\
 &{ =\dfrac{\lambda A_\Gamma}{8\pi^2} \int_{-W/2}^{W/2}dy'\dfrac{\cos[(\bar{k}_{+}-\bar{k}_{-})\sqrt{x^2+(h')^2}]}{x^2+(y')^2}} \\
 & =\dfrac{\lambda A_\Gamma}{8\pi^2 x} \int_{\alpha}^{\pi-\alpha}d\theta\cos[(\bar{k}_{+}-\bar{k}_{-})x\csc\theta] \\
 & \approx \dfrac{\lambda A_\Gamma}{8(\pi x)^{3/2}} \dfrac{1}{|\bar{k}_{+}-\bar{k}_{-}|^{1/2}}\cos\Big[(\bar{k}_{+}-\bar{k}_{-})x-\dfrac{\pi}{4}\Big],
\end{split}
\end{align}
where $\alpha=\arctan(2x/W)$ and $(\bar{k}_{+}-\bar{k}_{-})x\gg 1$ is assumed. {Note that in this derivation, we consider the NS junction where Cooper pairs propagate from an initial position ${\bf r}_1=(0,y')$ at the interface to a final point ${\bf r}_1=(x,y)$ in the cAFM. Accordingly, we restore the full position dependence and write the propagator as ${\cal D}({\bf r}_2;{\bf r}_1)$.}

\subsection{Critical Josephson current}

We consider the junction oriented along $x$-direction. We assume the couplings at the interfaces are constant $\lambda_{1(2)}$ and that the width $W$ of the junction is much longer than the length $L$. Under these considerations, 
the critical Josephson current can be found as
\begin{equation}
I_{c}^\Gamma= \dfrac{4e\lambda_{1}\lambda_{2}}{\hbar}  \int_{-W/2}^{W/2}dy'dy_{1}'D_\Gamma (L,y';0,y_{1}').
\label{eq:I_c}
\end{equation}
We change the variables $(y_{1}',y')\rightarrow(s,\zeta)=(y_{1}'+y',y'-y_{1}')$,
and rewrite the integral as 
\begin{equation}
\int_{-W/2}^{W/2}dy'dy_{1}'=\dfrac{1}{2}\int_{-W}^{W}d\zeta\int_{|\zeta|-W}^{W-|\zeta|}ds.
\end{equation}
Thus, the critical current is given by
\begin{align}
\begin{split}
I_{c}^\Gamma  & = \dfrac{2e \lambda_{1}\lambda_{2}}{\hbar} {\dfrac{A_\Gamma}{8\pi^2}} 
\int_{-W}^{W}d\zeta\int_{|\zeta|-W}^{W-|\zeta|}ds\dfrac{\cos[(\bar{k}_{+}-\bar{k}_{-})\sqrt{t^{2}+L^{2}}]}{t^{2}+L^{2}} \\
 & = \dfrac{e \lambda_{1}\lambda_{2} A_\Gamma}{2\pi^2 \hbar} \int_{-W}^{W}d\zeta (W-|\zeta|)\dfrac{\cos[(\bar{k}_{+}-\bar{k}_{-})\sqrt{\zeta^{2}+L^{2}}]}{\zeta^{2}+L^{2}} \\
 & =\dfrac{e \lambda_{1}\lambda_{2} A_\Gamma}{2\pi^2 \hbar}
 \int_{\pi-\alpha'}^{\alpha'}(-L\csc^{2}\theta')d\theta'(W-L|\cot\theta'|)\dfrac{\cos[(\bar{k}_{+}-\bar{k}_{-})L\csc\theta']}{L^{2}\csc^{2}\theta'} \\
 & =\dfrac{e \lambda_{1}\lambda_{2}W }{2\hbar (\pi L)^{3/2}} \dfrac{A_\Gamma}{|\bar{k}_{+}-\bar{k}_{-}|^{1/2}}\cos\Big[(\bar{k}_{+}-\bar{k}_{-})L-\dfrac{\pi}{4}\Big],
\end{split}
\end{align}
where $\zeta=x'-x_{1}'=L\cot\theta'$, $\alpha'=\text{arctan}(L/W$)), and in the third line, we have converted the integral over $\zeta$ to an integral over angle $\theta'$ (i.e., $\zeta=L\cot\theta'$ and $d\zeta=-L\csc^{2}\theta'd\theta'$). From this result, we also see that the dominant Cooper-pair propagation direction is along the junction direction.

\section{Pairing correlations and Josephson currents for Dirac electrons}
\label{sm-sec-C}

This section considers the case of Dirac electrons near the band center regime. We start with the ${\bf k}\cdot {\bf p}$ effective model, Eqs.~\eqref{eq:KKpoint} and \eqref{eq:K'point}, at the $K/K'$ point. The Green function for the $K$ point can be written as
\begin{align}
G_{0K}^{\text{ret}}({\bf k},\epsilon) & =\dfrac{1}{\epsilon+\mu_K-H_{\text{K}}({\bf k})+i\delta} 
= \sum_{a,b=\pm}\dfrac{P_{a,b}({\bf k})}{\epsilon-bE_{a}({\bf k})+i\delta},\label{eq:Gfunction}
\end{align}
where $\mu_K$ is the chemical potential measured from $E_{\text {cen}}=0$ and
\begin{subequations}
\begin{align}
E_{0}({\bf k}) & =\sqrt{v^{2}k^{2}+2m_K^{2}}, \\
E_{\pm}({\bf k}) & 
=\pm m_K + \sqrt{v^{2}k^{2}+m_K^{2}}, \\
H_{1}({\bf k}) & = v \sigma_{0}(k_{x}s_{x}-k_{y}s_{y})-m_K\sigma_{2}s_{z}, \\
P_{a,b}({\bf k}) & =\dfrac{1}{4}\left[1+a\dfrac{H_{1}({\bf k})}{E_{0}({\bf k})}\right]\left[1+b\dfrac{H_{\text{K}}({\bf k})}{E_{a}({\bf k})}\right].
\end{align}
\end{subequations}
The result for the $K'$ point is the same but with flipping the sign of the velocity.

\subsection{Cooper-pair propagator}

Without loss of generality, we consider $\mu_K>2|m_K|$ such that the Fermi energy crosses the two positive bands. The resulting $g$ matrix function
is given by (we drop $b=+$ below)
\begin{align} \label{eq:gK0-function}
\begin{split}
g_{K0}({\bf r},\epsilon) & =2\pi\sum_{a}\int\dfrac{d^{2}{\bf k}}{(2\pi)^{2}}e^{i{\bf k}\cdot{\bf r}}\delta(\epsilon+\mu_K-E_{{\bf k},a})P_{a}({\bf k}) \\
 & =\dfrac{1}{2\pi}\sum_{a}\int_{0}^{\infty}dkk\int_{0}^{2\pi}d\phi e^{ikr\cos(\phi-\theta)}\dfrac{\delta(k-k{}_{a})}{|\partial E_{{\bf k},a}/\partial_{k}|}P_{a}({\bf k}) \\
 & =\dfrac{1}{2\pi}\sum_{a}\dfrac{\mu_K-am_K}{v^{2}}\int_{0}^{2\pi}d\phi e^{ikr\cos(\phi-\theta)}P_{a}({\bf k})\Big|_{k=k_{a}'},
\end{split}
\end{align}
where
\begin{subequations}
\begin{align}
k_{a}' & =\dfrac{\sqrt{(\mu_K+\epsilon-am_K)^{2}-m_K^{2}}}{v}, \\
\dfrac{\partial E_{a}}{\partial_{k}}\Big|_{k=k_{a}}  &=\dfrac{v^{2}k_{a}'}{\sqrt{m_K^{2}+v^{2}k_a^{\prime2}}} =\dfrac{v^2 k_{a}'}{\mu_K-am_K}.
\end{align}
\end{subequations}
At $k=k_{a}'$, we have $E_{0}({\bf k})
=\sqrt{(\mu_K-am_K)^{2}+m_K^{2}}$, $E_{\pm}({\bf k}) =\mu$, and
\begin{align}
\begin{split}
P_{a}({\bf k})  =\dfrac{1}{4(\mu_K-am_K)}\Big[
&(\mu_K-am_K)s_{0}\sigma_{0}+m_K(as_{3}\sigma_{3}+s_{2}\sigma_{3}-s_{1}\sigma_{1})  \\
+& v(ak_{x}s_{1}\sigma_{0}-ak_{y}s_{2}\sigma_{0}+k_{x}s_{0}\sigma_{1}+k_{y}s_{0}\sigma_{3}) \\
+& av^{2}(k_{x}^{2}s_{1}\sigma_{1}-k_{y}^{2}s_{2}\sigma_{3}+k_{x}k_{y}s_{1}\sigma_{3}+k_{x}k_{y}s_{2}\sigma_{1})\Big].
\end{split}
\end{align}
Thus, plugging the above equation into Eq.~\eqref{eq:gK0-function}, we obtain
\begin{align}
g_{K0}({\bf r},\epsilon)= & \dfrac{1}{8\pi}\sum_{a}\dfrac{1}{v^{2}}\int_{0}^{2\pi}d\phi e^{ik_{a}'r\cos(\phi-\theta)}\Big[(\mu_K-am_K)s_{0}\sigma_{0}- m_K(as_{3}\sigma_{3}+s_{2}\sigma_{3}-s_{1}\sigma_{1})\nonumber \\
& \;\;\; +v(ak_{x}s_{1}\sigma_{0}-ak_{y}s_{2}\sigma_{0}+k_{x}s_{0}\sigma_{1}+k_{y}s_{0}\sigma_{3}) 
 +av^{2}(k_{x}^{2}s_{1}\sigma_{1}-k_{y}^{2}s_{2}\sigma_{3}+k_{x}k_{y}s_{1}\sigma_{3}-k_{x}k_{y}s_{2}\sigma_{1}) \Big] \nonumber \\
= & \dfrac{1}{8v^{2}}\sum_{a}[2J_{0}(k_{a}'r)[(\mu_K-am_K)s_{0}\sigma_{0}-m_K(as_{3}\sigma_{3}+s_{2}\sigma_{3}-s_{1}\sigma_{1})] 
+2vk_{a}'J_{1}(k_{a}'r)(-a\sin\theta s_{1}\sigma_{0} \nonumber \\
 & \;\;\;\;\;\;\; \;\;\;\; -a\cos\theta s_{2}\sigma_{0}-\sin\theta s_{0}\sigma_{1}+\cos\theta s_{0}\sigma_{3}) + av^{2}k_{a}^{\prime 2}\{[-J_{2}(k_{a}'r)\cos(2\theta)+J_{0}(k_{a}'r)]s_{1}\sigma_{1} \notag \\
 & \;\;\;\;\;\;\;\;\;\;\;  -[J_{2}(k_{a}'r)\cos(2\theta)+J_{0}(k_{a}'r)]s_{2}\sigma_{3} -J_{2}(k_{a}'r)\sin(2\theta)(s_{1}\sigma_{3}-s_{2}\sigma_{1})\}]. 
\end{align}

Similarly, we find (by changing all the signs of the terms associated
with $k_{a}$ since the two valleys are related by ${\bf k}\rightarrow-{\bf k}$)
\begin{align}
g_{K'0}({\bf r},\epsilon)= & \dfrac{1}{8v^{2}}\sum_{a}[2J_{0}(k_{a}'r)[(\mu_K-am_K )s_{0}\sigma_{0}-m_K (as_{3}\sigma_{3}+s_{2}\sigma_{3}-s_{1}\sigma_{1})]-2vk_{a}'J_{1}(k_{a}'r)(-a\sin\theta s_{1}\sigma_{0} \nonumber \\
 & \;\;\;\;\;\;\;\;\;\; -a\cos\theta s_{2}\sigma_{0}-\sin\theta s_{0}\sigma_{1}+\cos\theta s_{0}\sigma_{3}) +av^{2}k_{a}^{\prime 2}\{[-J_{2}(k_{a}'r)\cos(2\theta)+J_{0}(k_{a}'r)]s_{1}\sigma_{1} \nonumber \\
 & \;\;\;\;\;\;\;\;\;\; -[J_{2}(k_{a}'r)\cos(2\theta)+J_{0}(k_{a}'r)]s_{2}\sigma_{3} -J_{2}(k_{a}'r)\sin(2\theta)(s_{1}\sigma_{3}-\sigma_{1})\}] .
\end{align}

Next, we use the results of $g_{K0}({\bf r},\epsilon)$ and $g_{K'0}({\bf r},\epsilon)$ to calculate the $\mathcal{T}$ matrix function, yielding
\begin{align}
\mathcal{T}({\bf r},\epsilon,\epsilon') & =\text{Tr}[g_{K0}({\bf r},\epsilon)s_{y}g_{K'0}^{T}({\bf r},\epsilon')s_{y}] + \text{Tr}[g_{K'0}({\bf r},\epsilon)s_{y}g_{K0}^{T}({\bf r},\epsilon')s_{y}] \nonumber \\
& = \dfrac{1}{16v^{4}}\sum_{a,a'}\dfrac{1}{(\mu_K-am_K)(\mu_K-a'm_K)}\{[4aa'm_K^{2}-2m_K\mu_K(a+a')+\mu_K^{2}(2-aa')]J_{0}(k_{a}'r)J_{0}(k_{a'}'r) \label{eq:T-function} \notag \\
 & \;\;\;\;\; \;\;\;\;\; \;\;\;\;\; \;\;\;\;\; \;\;\;\;\; \;\;\;\;\; \;\;\;\;\; \;\;\;\;\; \;\;\;\;\; \;\;\;\;\; \;\;\;\;\;\;  +2(aa'-1)\mu_K\sqrt{(\mu_K-2am_K)(\mu_K-2a'm_K)}J_{1}(k_{a}'r)J_{1}(k_{a'}'r) \notag \\
 & \;\;\;\;\;\; \;\;\;\;\; \;\;\;\;\; \;\;\;\;\; \;\;\;\;\; \;\;\;\;\; \;\;\;\;\; \;\;\;\;\; \;\;\;\;\; \;\;\;\;\; \;\;\;\;\;
 -aa'(\mu_K-2am_K)(\mu_K-2a'm_K)J_{2}(k_{a}'r)J_{2}(k_{a'}'r)\}.
\end{align}

Again we consider large Fermi surfaces such that $\mu_K\gg|\epsilon|$ and $k_{\eta}r\gg1$. In this case, we obtain 
\begin{align}
J_{n}(k_{a}'r) & \approx\sqrt{\dfrac{2}{\pi k_{a}'r}}\cos\Big(k_{a}'r-\dfrac{n\pi}{2}-\dfrac{\pi}{4}\Big), \; \text{and}\;\;
k_{a}' 
\approx\bar{k}_{a}'+\xi_{a}\epsilon,
\end{align}
where
\begin{align}
\bar{k}_{a}' &= \dfrac{\sqrt{\mu_K^{2}-2am_K\mu_K}}{v}, \;\;\;\;\;
\xi_{a}  
=\dfrac{\mu_K-am_K}{v\sqrt{\mu^{2}_K-2am_K\mu_K}}.
\end{align}
In the summation over $a$ and $a'$, it turns out that only the terms with $a'=-a$ are important. Thus, Eq.~\eqref{eq:T-function} simplifies to
\begin{align}
\mathcal{T}({\bf r},\epsilon,\epsilon')
= & \dfrac{1}{2\pi v^{4}r}\sum_{a}\dfrac{1}{\mu_K^{2}-m_K^{2}}\dfrac{1}{\sqrt{\bar{k}_{a}'\bar{k}_{-a}'}}\Big[(\mu_K^{2}-2m_K^{2})\cos\Big(k_{a}'r-\dfrac{\pi}{4}\Big)\cos\Big(k_{-a}'r-\dfrac{\pi}{4}\Big) \notag \\
 & \;\;\;\;\;\;\;\;\;\;\;\;\;\;\;\;\;\;\; \;\;\;\;\;\;\;\;\;\;\;\;\;\; \;\;\;\;\;\;\;\;\;\;\;\;\;
 -\mu\sqrt{\mu_K^{2}-4m_K^{2}}\sin\Big(k_{a}'r-\dfrac{\pi}{4}\Big)\sin\Big(k_{-a}'r-\dfrac{\pi}{4}\Big)\Big].
\end{align}

Finally, we find the Cooper-pair propagator as
\begin{align}
\begin{split}
D_K ({\bf r})= & \dfrac{1}{2}\int_{0}^{\infty}\dfrac{d\epsilon}{2\pi}\dfrac{d\epsilon'}{2\pi}\dfrac{\mathcal{T}({\bf r},\epsilon,\epsilon')+\mathcal{T}({\bf r},-\epsilon,-\epsilon')}{\epsilon+\epsilon'} 
= \dfrac{A_K}{8\pi^{2}r^{2}}\cos[(\bar{k}_{+}'-\bar{k}_{-}')r],
\end{split}
\end{align}
where $A_K = [\mu_K-(\mu_K^{2}-4m_K^{2})^{1/2} ]^{2} / [v^3\mu_K^{1/2}(\xi_{+}+\xi_{-})(\mu_K^{2}-4m_K^{2})^{1/4}(\mu_K^{2}-m_K^{2})]$. 
The propagator is isotropic for the propagation direction and oscillates as a function of the propagation distance $r$.

\subsection{Pairing correlations and Josephson currents}

Following a similar procedure as the case of Schr\"odinger electrons, it is straightforward to find the proximity-induced pairing correlation as
\begin{align}
\begin{split}
\langle|\Psi(x,y)|\rangle_K & =\lambda\int_{-W/2}^{W/2}dy'D_K (x,0;0,y')  
=\dfrac{\lambda }{8(\pi x)^{3/2}} \dfrac{A_K}{|\bar{k}_{+}'-\bar{k}_{-}'|^{1/2}}\cos\Big[(\bar{k}_{+}'-\bar{k}_{-}')x-\dfrac{\pi}{4}\Big],
\end{split}
\end{align}
and the critical Josephson current as
\begin{align} \label{eq:I_c_K}
\begin{split}
I_{c}^K &= \dfrac{4e\lambda_{1}\lambda_{2}}{\hbar}  \int_{-W/2}^{W/2}dy'dy_{1}'D_K (L,y';0,y_{1}')  
=\dfrac{e \lambda_{1}\lambda_{2}W }{2\hbar (\pi L)^{3/2}} \dfrac{A_K}{|\bar{k}_{+}'-\bar{k}_{-}'|^{1/2}}\cos\Big[(\bar{k}_{+}'-\bar{k}_{-}')L-\dfrac{\pi}{4}\Big] .
\end{split}
\end{align}
They take similar forms as those for the Schr\"odinger electrons.

\section{Comparison between results for Schr\"odinger and Dirac electrons}
\label{compare two electrons model}

\begin{figure}[t]
\includegraphics[width=0.7\textwidth]{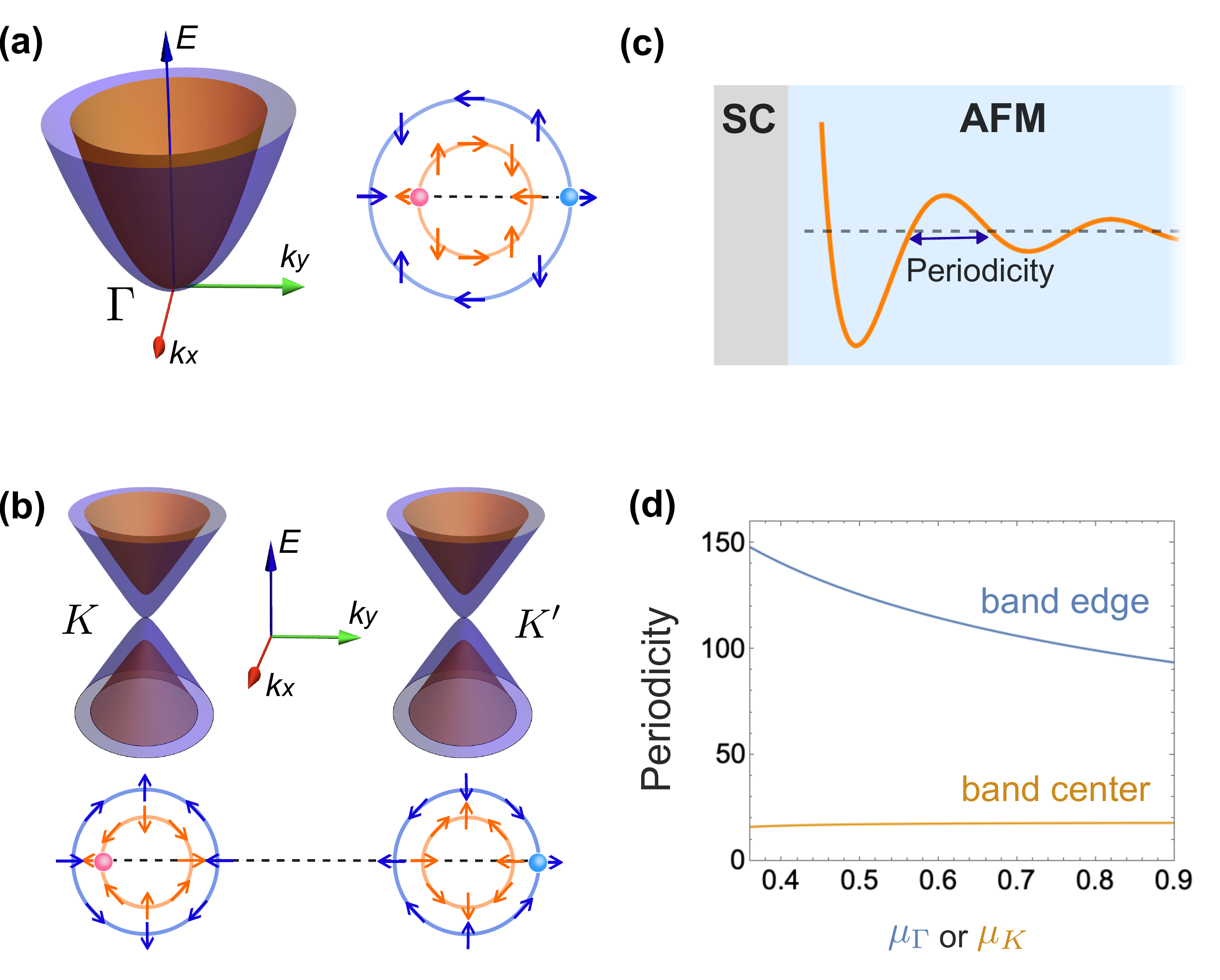}
	
\caption{(a) Band structure near the band edge at the $\Gamma$  point. The Cooper pairing occurs predominantly between the two Fermi surfaces, leading to finite-momentum Cooper pairs. The spin textures of the Fermi surfaces are indicated by the arrows. 
(b) Band structure near the band center at the $K/K'$ point. The Cooper pairing occurs predominately between different valleys. 
(c) Schematic for the SC/cAFM setup. The proximity-induced pairing correlations exhibit damped oscillations near the interface. 
(d) Periodicity of oscillations as a function of the chemical potential $|\mu_\Gamma|$ ($|\mu_K|$) measured from the band edge (center).   Parameters are $t=1$ and $J=0.3t$.}
	
\label{fig:FFLO}
\end{figure}

As shown in the previous sections, the Cooper pair propagators for Sch\"odinger and Dirac electrons are given by
\begin{subequations}
\begin{align}
	D_{\Gamma}({\bf r}) 
	& =\dfrac{A_\Gamma}{8\pi^2r^2} \cos[(\bar{k}_{-}-\bar{k}_{+}) r], \\
	D_K({\bf r}) &= 
	 \dfrac{A_K}{8\pi^{2} r^{2}} \cos[(\bar{k}_{+}'-\bar{k}_{-}')r].
\end{align}
\end{subequations}
Here, $\bar{k}_\pm = [\mu_\Gamma/(t \pm m_\Gamma)]^{1/2} $ correspond to the Fermi momentum modulus of the inner and outer Fermi surfaces for the Sch\"odinger electrons [see Fig.~\ref{fig:FFLO}(a)]. 
As for the Dirac electrons, we define $\bar{k}_{\pm}'={\sqrt{\mu_K^{2}\mp2m_K\mu_K}}/v$ and $\xi_{\pm}=(\mu_K\mp m_K)/({\mu_K^{2}\mp2 m_K \mu_K})^{1/2}$.
We have considered the case with $\mu_K>|2m_K|$, where two Fermi surfaces exist at each valley [see Fig.~\ref{fig:FFLO}(b)].
For both cases, the Cooper pair carries a finite momentum (i.e.,~electrons from inner Fermi surface must pair with electrons from outer Fermi surface). 
It leads to
\begin{subequations}
\begin{align}
Q_{\Gamma} &=\bar{k}_{-}-\bar{k}_{+} , \\
Q_K&=\bar{k}_{+}'-\bar{k}_{-}'.
\end{align}
\end{subequations}

We note that in both cases, the Cooper-pair propagator is isotropic in the propagation direction, owing to the approximately isotropic Fermi surfaces in the band edge and center regimes. When moving away from these regimes, higher momentum corrections become important and the propagator becomes anisotropic. However, the Cooper-pair momentum is retained in any direction, in contrast to that in altermagnets which vanishes in specific directions~\cite{SBZ2023arXiv}. It is also noteworthy that for weak cAFM strengths ($J<0.3t$), the Cooper-pair momentum exhibits different dependences on the chemical potential in the two regimes. Specifically, for the band edge it increases when the chemical potential grows, i.e., $Q_\Gamma \approx \sqrt{\mu_\Gamma}m_\Gamma$, whereas, for the band center, it becomes insensitive to the chemical potential, i.e., $Q_K \approx 2m_K/v$. This distinction reflects the difference between Cooper pairs originating from quadratic Sch\"odinger and linear Dirac fermions, which is a general feature of cAFMs.

As derived in the previous sections, the local pairing correlations for Sch\"odinger and Dirac electron are given by
\begin{subequations}\label{eq-order-para-both}
	\begin{align}
		\langle|\Psi({\bf r})|\rangle_{\Gamma} 
		& \propto \dfrac{1}{x^{3/2}}\cos\Big(Q_\Gamma x-\dfrac{\pi}{4}\Big), \\
		\langle|\Psi({\bf r})|\rangle_K 
		& \propto \dfrac{1}{x^{3/2}} \cos\big(Q_K x-\dfrac{\pi}{4}\Big).
	\end{align}
\end{subequations}
The pairing correlations exhibit several salient features, as shown in Eq.~\eqref{eq-order-para-both}. First, due to the finite momentum of Cooper pairs, both $\langle|\Psi({\bf r})|\rangle_{\Gamma}$ and $ \langle|\Psi({\bf r})|\rangle_K$ oscillate around zero in real space, analogous to the FFLO state. The periodicity of oscillations is determined exactly by the Cooper-pair momentum, i.e., $P_\Gamma = 2\pi/Q_\Gamma$ and $P_K = 2\pi/Q_K$. This supports our discussion made in the main text.  Second, for weak AFM strengths, $Q_\Gamma \approx \sqrt{\mu_\Gamma} m_\Gamma$ and $Q_K \approx 2m_K/v$. Thus, although the Sch\"odinger and Dirac electrons in the two complementary energy regimes exhibit similar FFLO-like states, they display different behaviors in terms of dependence on chemical potential due to different types of electrons, as shown in Fig.~\ref{fig:FFLO}(d). This result also indicates that for given position $x$, while for the Dirac electrons around the band center, the pairing correlation is insensitive to the change of chemical potential, for the Sch\"odinger-like electrons near the band edge, it oscillates pronouncedly about zero as a function of chemical potential.  
Third, the pairing amplitudes decay as $x^{-3/2}$ independent of the type of electrons. This universal decay behavior results from the joint effect of the inherent attenuation of the Cooper-pair propagator over the propagation distance and the phase interference between the propagators. 
Finally, we note that owing to the isotropic nature of the Cooper-pair propagator, the pairing correlation is always dominated by the propagator in the direction normal to the SC/cAFM interface. Thus, the superconducting transport properties and the resulting pairing correlation are essentially independent of the junction orientation.

\section{Symmetry analysis on the spin texture and pairing correlations \label{sec:symmetry-analysis}
}
\begin{itemize}
\item[(1)] {\bf Spin-space symmetry constraints on the spin texture.} \\   
The spin vector  at each momentum $\bm k$ is given by $\Vec{s}_n({\bm k}) = \langle \psi_n({\bm k}) | \Vec{s} | \psi_n({\bm k}) \rangle$, where $| \psi_n({\bm k}) \rangle$ is the eigen wavefunction at the $n$-th band, and $\Vec{s}$ is the spin operator. To illustrate the effect of symmetry on the spin texture, we consider the spin-space symmetry $\{U_x(\pi) || C_x(\pi)\}$ as an example. This symmetry operation transforms the spin components as follows: 
\begin{align}
\{U_x(\pi) || C_x(\pi)\}: \; \left(s_x({\bm k}),s_y({\bm k}),s_z({\bm k})\right)  \to  \left(s_x(R_{2x}{\bm k}),-s_y(R_{2x}{\bm k}),-s_z(R_{2x}{\bm k})\right),
\end{align}
where $R_{2x}{\bm k} = (k_x,-k_y,-k_z)$. Thus, for an arbitrary ${\bm k}$ point, we arrive at the following result:
\begin{align}
\Vec{s}_n({\bm k}) + \Vec{s}_n(R_{2x}{\bm k}) = (s_x({\bm k}),0,0),    
\end{align}
which indicates that the total magnetization can generally be non-zero and must be polarized in $x$ axis. Similarly, we are able to get all the symmetry constraints on the spin vector as
\begin{align}
\begin{split}
\{U_z(\frac{2\pi}{3}) || C_z(\frac{\pi}{3})\} &:  (s_x(\bm{k}),s_y(\bm{k}),s_z(\bm{k}))  \to  (s_x(R_{6z}\bm{k}),s_y(R_{6z}\bm{k}),s_z(R_{6z}\bm{k})) {\cal D}[R_{3z}] \\
\{U_x(\pi) || C_x(\pi)\} &: (s_x(\bm{k}),s_y(\bm{k}),s_z(\bm{k}))  \to (s_x(R_{2x}\bm{k}),-s_y(R_{2x}\bm{k}),-s_z(R_{2x}\bm{k})) \\
\{E||I\} &: (s_x(\bm{k}),s_y(\bm{k}),s_z(\bm{k}))  \to (s_x(-\bm{k}),s_y(-\bm{k}),s_z(-\bm{k})) \\
\{U_z(\pi){\cal T} || I\} &: (s_x(\bm{k}),s_y(\bm{k}),s_z(\bm{k}))  \to (s_x(\bm{k}),s_y(\bm{k}),-s_z(\bm{k})) 
\end{split}
\end{align}
Moreover, if we focus on the $\bm{k}$ around the $\Gamma$ point expand up to the $k^2$ order, all the aforementioned symmetry constraints can lead to 
\begin{align} \label{reply2-spin-texture}
(s_x(\bm{k}),s_y(\bm{k}),s_z(\bm{k})) \propto (k_x^2-k_y^2,2k_xk_y,0)     
\end{align}
This represents the spin-orbit-coupling-free spin texture enforced by the spin-space group $^{3_z}6/^1m^{2_x} m^{2_{xy}}m^{m_z}1$.

\item[(2)] {\bf Spin-space symmetry constraints on pairing correlations.} \\   
The spin-to-triplet conversion is not a result of net magnetization but rather due to non-collinear spins that break the SU(2) spin symmetry. This allows total $s_z=0$ Cooper pairs to be converted into total $s_z=\pm 1$ Cooper pairs, even in absence of spin-orbit coupling. For example, 
\begin{align}
\{U_x(\pi) || C_x(\pi)\}: \;  \{ c_{A,s}, c_{B,s}, c_{C,s} \} \to i \times \{ c_{B,\bar{s}}, c_{A,\bar{s}}, c_{C,\bar{s}} \}
\end{align}
with $s\in\{\uparrow,\downarrow\}$ and  $\bar{s}$ denotes the spin opposite to $s$. This symmetry gives rise to 
\begin{align}\label{reply-eq-fupup}
\begin{split}
& {\cal F}_{\uparrow\uparrow} = \langle c_{A,\uparrow}(\Vec{R})c_{A,\uparrow}(\Vec{R}) \rangle + \langle c_{B,\uparrow}(\Vec{R})c_{B,\uparrow}(\Vec{R}) \rangle + \langle c_{C,\uparrow}(\Vec{R})c_{C,\uparrow}(\Vec{R}) \rangle \\
\Rightarrow\; &  - \left\lbrack \langle c_{A,\downarrow}(\Vec{R})c_{A,\downarrow}(\Vec{R}) \rangle + \langle c_{B,\downarrow}(\Vec{R})c_{B,\downarrow}(\Vec{R}) \rangle + \langle c_{C,\downarrow}(\Vec{R})c_{C,\downarrow}(\Vec{R}) \rangle \right\rbrack = -{\cal F}_{\downarrow\downarrow} \\
\Rightarrow\; & {\cal F}_{\uparrow\uparrow} = -{\cal F}_{\downarrow\downarrow}.
\end{split}
\end{align}
Note that we are calculating the pairing for each magnetic unit cell. Thus we sum over all subllatices. Moreover, considering the total spin-zero triplet pairing:
\begin{align}
\begin{split}
& {\cal F}_{z}=\left\lbrack \langle c_{A,\uparrow}(\Vec{R})c_{A,\downarrow}(\Vec{R}) \rangle + \langle c_{B,\uparrow}(\Vec{R})c_{B,\downarrow}(\Vec{R}) \rangle + \langle c_{C,\uparrow}(\Vec{R})c_{C,\downarrow}(\Vec{R}) \rangle \right\rbrack \\
& + \left\lbrack \langle c_{A,\downarrow}(\Vec{R})c_{A,\uparrow}(\Vec{R}) \rangle + \langle c_{B,\downarrow}(\Vec{R})c_{B,\uparrow}(\Vec{R}) \rangle + \langle c_{C,\downarrow}(\Vec{R})c_{C,\uparrow}(\Vec{R}) \rangle \right\rbrack  \\ 
\Rightarrow\; & - \left\lbrack \langle c_{A,\downarrow}(\Vec{R})c_{A,\uparrow}(\Vec{R}) \rangle + \langle c_{B,\downarrow}(\Vec{R})c_{B,\uparrow}(\Vec{R}) \rangle + \langle c_{C,\downarrow}(\Vec{R})c_{C,\uparrow}(\Vec{R}) \rangle \right\rbrack \\
& -\left\lbrack \langle c_{A,\uparrow}(\Vec{R})c_{A,\downarrow}(\Vec{R}) \rangle + \langle c_{B,\uparrow}(\Vec{R})c_{B,\downarrow}(\Vec{R}) \rangle + \langle c_{C,\uparrow}(\Vec{R})c_{C,\downarrow}(\Vec{R}) \rangle \right\rbrack = - {\cal F}_{z} \\
\Rightarrow\; & {\cal F}_{z} = 0.
\end{split}
\end{align}
This symmetry analysis implies that the equal-spin triplet correlation must satisfy ${\cal F}_{\downarrow\downarrow} = -{\cal F}_{\uparrow\uparrow}$ and ${\cal F}_{z}=0$, as we discuss in the main text [below Eq.~(2) and also Fig.~2(a)]. 

Inversion symmetry $\{E||I\}$ is also important. It ensures that the spin texture, being even-parity as shown in Eq.~\eqref{reply2-spin-texture}, implies that the singlet-triplet mixed pairing must also preserve this parity. Consequently, the non-vanishing triplet correlations ${\cal F}_{\uparrow\uparrow/ \downarrow\downarrow}$ are even-parity spin-triplet correlations, which must be odd in frequency. 
\end{itemize}

\section{$0$-$\pi$ transitions in Josephson junctions}
\label{0-pi transition}

\begin{figure}[!htbp]
\includegraphics[width=0.7\textwidth]{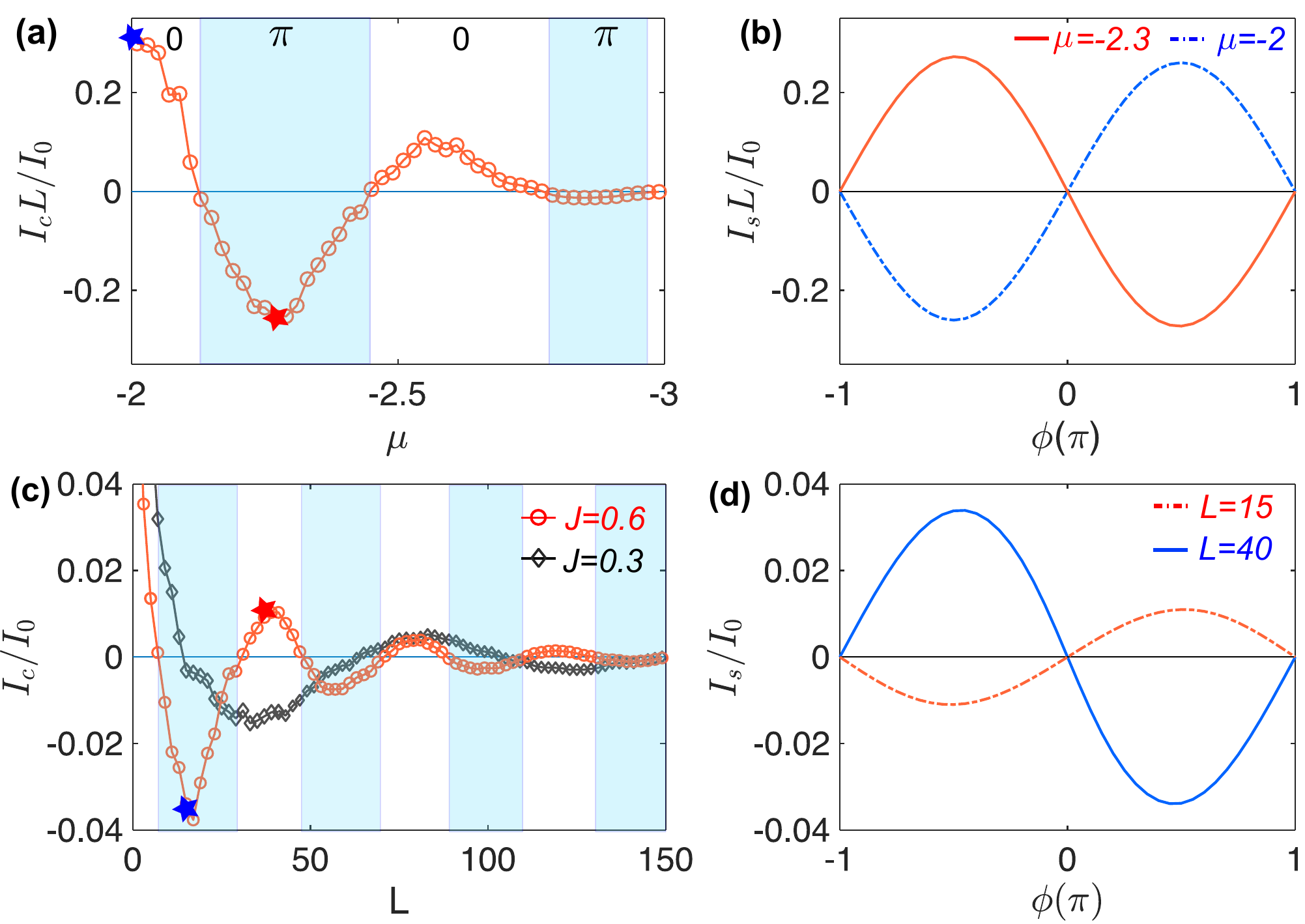}
	
\caption{(a) Critical Josephson current $I_c$ as a function of chemical potential $\mu$ in the cAFM for  $L=80$. The cyan shadow indicates the regions for the $\pi$-state, while the white regions correspond to the $0$-state of the Josephson junction. 
(b) Current-phase relation for $\mu=-2t$ and $-2.3t$ [corresponding to the two colored stars in (a)], respectively. 
(c) $I_c$ (orange) as a function of junction length $L$ for $\mu_N=-2t$. We plot the result (back curve) for a smaller AFM strength $J=0.3t$ for comparison. 
(d) Current-phase relation for $L=15$ and $40$ [corresponding to the two colored stars in (c)], respectively. 
Other parameters are $\mu_S=-2t$, $J=0.6t$, $\Delta=0.05t$ and $k_BT=0.02\Delta$.}
	
\label{fig:Josephson}
\end{figure}

The finite-momentum Cooper pairing in the cAFM manifests as $0$-$\pi$ transitions in Josephson supercurrent although the net magnetization of the system is zero.  The critical current $I_{c}$ exhibits pronounced oscillations around zero as a function of junction length $L$. The periodicity in $L$ is given by $\mathcal{P}_{L}=2\pi/Q_{\Gamma(K)}$. For the Sch\"odinger electrons, $I_c$ also oscillates with changing the chemical potential $\mu_\Gamma$. For weak AFM strength, the periodicity in $\mu_\Gamma$ is $\mathcal{P}_\mu=2\pi/(m_\Gamma L)$. These oscillations are similar to those in the pairing corrlation. Note that a positive $I_{c}$ corresponds to a $0$-junction, whereas a negative $I_{c}$ indicates a $\pi$-junction where the system has an intrinsic $\pi$ phase difference across the junction at the ground state. These oscillations indicate $0$-$\pi$ transitions of the Josephson junction when we vary $L$ or $\mu_\Gamma$. 
Similar to the pairing correlation, these results are also independent of the details of $s$-wave superconductors and SC/cAFM interfaces.

To confirm the above analytical results, we perform numerical calculations of the supercurrent on the full kagome lattice model, employing the lattice Green function with the recursive technique~\cite{Sancho1984JPFMP,Sancho1985JPFMP,SBZhang20PRB}. Similar to the calculations for the pairing correlation, we assume translation symmetry in $y$-direction, which allows us to compute the contribution of each $k_y$ independently. For concreteness, we choose the chemical potential $\mu_S=-2t$ and pairing potential $\Delta=0.05t$ of the superconducting leads. The contributions of 400 $k_y$ modes over the Brillouin zone are taken into account in the calculation. Furthermore, we set the hopping strength at the interface to be only 30\% of that in the bulk of the kagome system to mimic a weak interface coupling.

In Fig.~\ref{fig:Josephson}(a), we plot the critical current $I_c$ as a function of chemical potential $\mu$ of the cAFM near the band edge for $L=80$ layers of unit cells. We see that $I_c$ oscillates significantly about zero as $\mu$ varies. The amplitude of $I_c$ decays for decreasing $\mu_\Gamma$ mainly because of the increase of Fermi surface mismatch. For positive $I_c$, the current-phase relation is approximately given by $I_s\propto \sin \phi$, whereas for negative $I_c$, it is $I_s\propto \sin (\phi+\pi)$ [see Fig.~\ref{fig:Josephson}(b)]. This indicates  0-$\pi$ transitions by changing gate voltage on the cAFM region. In Fig.~\ref{fig:Josephson}(c), we plot $I_c$ as a function of $L$ for $\mu=\mu_S=-2t$ and two different cAFM strengths $J=0.3t$ and $0.6t$, respectively. Evidently, $I_c$ exhibits a damped oscillation around zero as $L$ increases. The oscillation periodicity is smaller for a stronger $J$. Two typical current-phase relations for positive and negative $I_c$ are displayed in Fig.~\ref{fig:Josephson}(d). Therefore, we can alternatively realize 0-$\pi$ transitions by changing $L$. The results are insensitive to the junction orientation when $\mu$ is close to the band edge $E_{\text{edge}}=-3t$.

\begin{figure}[!htbp]
\includegraphics[width=0.4\textwidth]{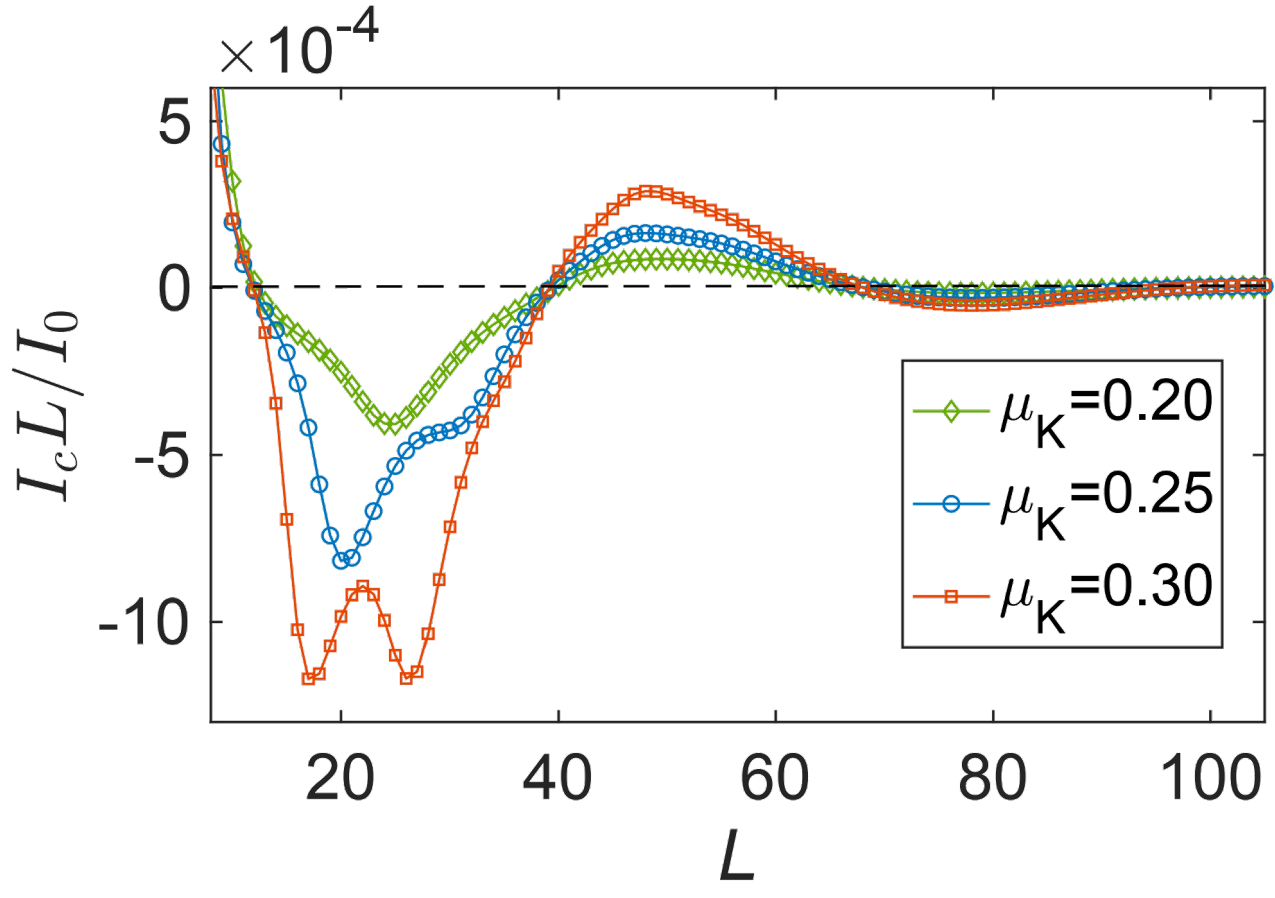}
	
\caption{Critical Josephson current $I_c$ as a function of junction length $L$ (in units of unit cells) for $\mu_K=0.2t$, $0.25t$ and $0.3t$, respectively. The chemical potential in the superconductor leads is set to be the same as $\mu_K$ in the cAFM. $I_c$ oscillates around 0. The oscillation periodicity is independent of $\mu_K$, consistent with our analytical result. The contributions from 400 $k_y$ points over the Brillouin zone are taken into account. Other parameters are $J=0.1t$, $\Delta=0.05t$ and $k_BT=0.01t$. }
\label{fig:Josephson-K}
\end{figure}

The Josephson $0$-$\pi$ transition as a function of junction length $L$ also occurs for the Dirac electrons near the band center. 
To show this, in Fig.~\ref{fig:Josephson-K}, we consider $J=0.1t$, $\Delta=0.05t$ and $k_BT=0.01t$ and calculate the critical current $I_c$ for different chemical potentials ($\mu_K=\mu=0.2t$, $0.25t$ and $0.3t$) near the band center. The contributions from 400 $k_y$ points over the Brillouin zone are taken into account in the calculation, as considered in Fig.~\ref{fig:Josephson}. Similar to the case of the Schr\"odinger electrons, $I_c$ oscillates around zero when increasing the junction length $L$, showing the $0$-$\pi$ transitions. However, the oscillation periodicity is insensitive to the change of chemical potential, in stark contrast to that for the Schr\"odinger electrons near the band edge. This can be attributed to the fact the finite momentum of the Dirac Cooper pairs is independent of Fermi energy close to the band center, as we discussed in the main text. This result indicates that it is hard to realize $0$-$\pi$ transitions by gating the cAFM in this energy regime. We also notice pronounced fluctuations in $I_c$, which are due to the Fabry-Perot resonance in the Josephson junction with partial interface transparency. These fluctuations also occur in the absence of magnetic order. They can be suppressed (smeared) by bringing the chemical potential closer to the band center or by increasing the temperature.  

Therefore, all the behaviors are consistent with our analytical results for both Schr\"odinger and Dirac electrons.

\section{Supplementary results for the mixed pairing and Josephson current}
\label{sm-sec-D}

Above, we have analytically demonstrated, using the Cooper pair propagation method, that the cAFM can accommodate spin-singlet Cooper pairs with finite momentum. The BCS-like singlet pairing correlation is an even function of frequency $\omega$ and typically attains its maximum in the static limit $\omega = 0$. Thus, focusing on the static limit is sufficient for understanding the singlet pairing. Here, $\omega$ is the frequency component of the Fourier transform corresponding to time. This property also serves as the foundation for analytically deriving the singlet pairing correlation [cf. Eq.~\eqref{eq-order-para-both}] in the SC/cAFM junction, implying that the singlet finite-momentum pairing state may require only a band gap between the spin-splitting bands. 
Below, we use numerical simulation to verify the analytical result and discuss the odd-frequency superconductivity that goes beyond the static limit.

To explore this intriguing possibility, we employ a standard Green function approach incorporating finite frequencies and consider similar SC/cAFM junctions as described before. 
Cooper pairing correlations are encoded in terms of the anomalous part of Green function in Nambu space. Thus, we further calculate the anomalous Green function of the junction based on the full kagome lattice model. It not only confirms our analytical findings obtained using the Cooper-pair propagator approach (shown below) but also enables the investigation of proximity-induced pairing beyond the static limit ($\omega \neq 0$). Moreover, we consider translation symmetry in $y$-direction such that the momentum $k_y$ remains a good quantum number. In the Matsubara frequency representation and Nambu space, the full Green function can be written as
\begin{align}
	{\bf \mathcal{ G}}(x,x',k_y,i\omega) 
	= \begin{pmatrix}
		{\bm G}_{ee} & {\bm G}_{eh} \\
		{\bm G}_{he}  & {\bm G}_{hh}
	\end{pmatrix}.
\end{align}
Here, $x$ and $x'$ denote positions of the triangle units of the kagome lattice in the $x$ direction. Dependencies inside the matrices are omitted for ease of notation. In the spin basis, the anomalous Green function can be decomposed into singlet and triplet components
\begin{equation}
	{\bm G}_{eh} 
	= -is_y \sum_{j={0,x,y,z}} {\bm f}_j(x,x',k_y,i\omega) s_j, 
\end{equation} 
where ${\bm f}_0$ is the singlet, and ${\bm f}_z$ as well as ${\bm f}_{\uparrow\uparrow/\downarrow\downarrow}= \mp {\bm f}_x - i {\bm f}_y$ are triplet amplitudes. They are all three-by-three matrix functions and diagonal in sublattice space. We focus on local pairings with $x=x'$ in this work. Averaging across the three sublattices within each unit cell, we obtain the local pairing amplitudes as 
\begin{align}
\mathcal{F}_j(x,k_y,i\omega) = \dfrac{1}{3} \text{Tr} [{\bm f}_j(x,x,k_y,i\omega)],\;\;  
\end{align}
where $j\in\{0,z,\uparrow\uparrow,\downarrow\downarrow\}$. In the numerics, we consider a self-infinite superconducting lead with the same kagome lattice and compute its self-energy contribution to the cAFM using a well-established recursive technique~\cite{Sancho1984JPFMP,Sancho1985JPFMP,Asano01PRB}. 
{The recursive Green’s functions are obtained using the following essential relations:
\begin{subequations}
\begin{align}
    G_j^L = [i\omega_n - H_{0,j}-T_j^+ G_{j-1}^LT_j^-]^{-1}, \\
    G_j^R = [i\omega_n - H_{0,j}-T_j^- G_{j+1}^LT_j^+]^{-1},
\end{align}
\end{subequations}
where we start with initial Green’s functions $G^L_{j_L}$ and $G^R_{j_R}$ for the left and right recursions,
respectively.
}

For the kagome lattice junction oriented along $x$-direction, the intra-layer local Hamiltonian matrix in the superconducting leads can be written as $H_{0,j}=H_\Delta+H_0$, where $H_\Delta$ is the $s$-wave pairing potential given by  
\begin{align}
    H_\Delta = \begin{pmatrix}
        0 & -i \mathbb{I}_3 s_y \\
        i \mathbb{I}_3 s_y & 0
    \end{pmatrix}.
\end{align}   
$\mathbb{I}_3$ is the $3\times 3$ unit matrix acting in the sublattice space. 
$H_0$ is intra-layer local Hamiltonian matrix 
\begin{equation}
    H_0 = -\tau_z \otimes \begin{pmatrix}
        \mu & t & 0\\
        t & \mu & t \\
        0 & t & \mu 
    \end{pmatrix}
    \otimes s_0,
\end{equation}
where $\tau_z$ is the Pauli matrix acting in Nambu space. 
The inter-layer coupling matrices acting throughout the system are
\begin{align}
    T_j^+ = (T^-_j)^\dagger =
    - \tau_z \otimes
    \begin{pmatrix}
    0 & 0 & t e^{-ik_y} \\
    t & 0 & t e^{-ik_y} \\
    t & 0 & 0
    \end{pmatrix}
    \otimes s_0.
\end{align}
In the cAFM region, the non-collinear antiferromagnetic order is introduced via onsite exchange coupling between local magnetic moments and electron spin. This is captured by $H_{\text{cAFM}}$ in Eq.~\eqref{eq:HJ}. 
In Nambu space, it takes the form
\begin{equation}
    H_{J,\text{BdG}} = \begin{pmatrix}
        H_{\text{cAFM}} & 0 \\
        0 & - H_{\text{cAFM}}^*
    \end{pmatrix} .
\end{equation}
Without loss of generality, we set the pairing potential phase of the superconductor to be zero. For concreteness, we consider chemical potential $\mu=-2.5t$ (near the band edge) and $J=0.5t$ and illustrate the results in the main text.

\begin{figure*}[!htbp]
\includegraphics[width=0.7\textwidth]{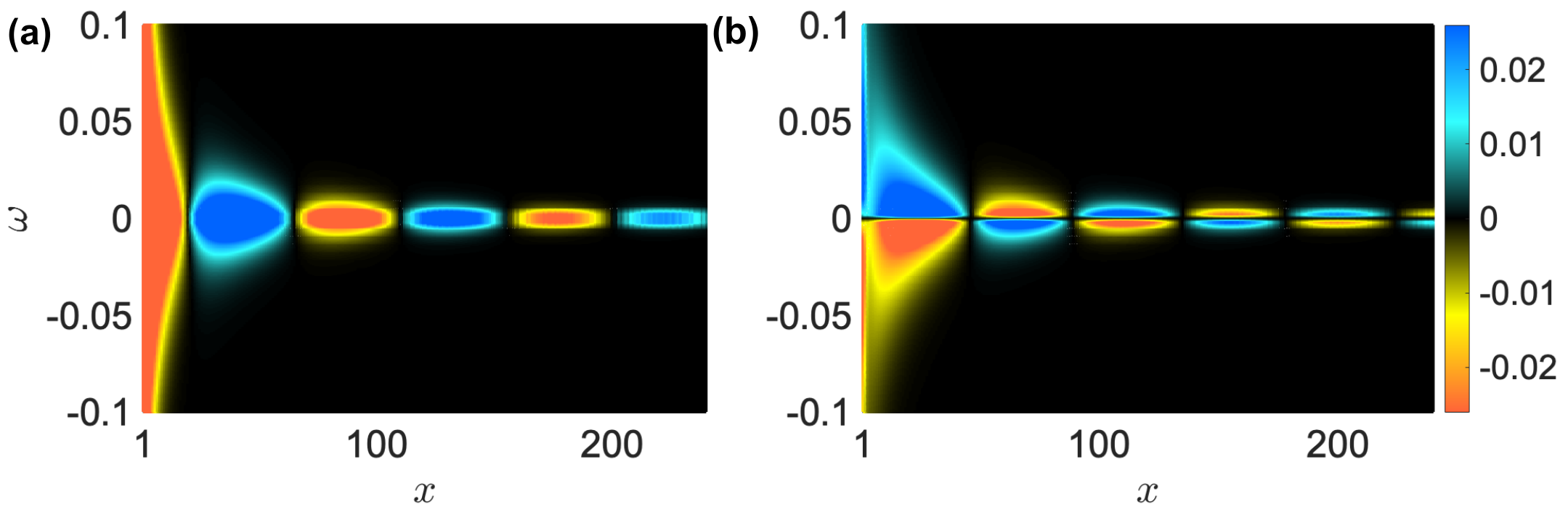}
	
\caption{(a) Contour plots of the singlet local pairing amplitude $F_0=({W}/{2\pi})\int_{-\pi/W}^{\pi/W}d{k_y}\mathcal{F}_0(k_y)$ as a function of Matsubara frequency $i \omega$ and position $x$ in the cAFM. 
(b) the same as (a) but for the triplet amplitudes, $F_t=({W}/{2\pi})\int_{-\pi/W}^{\pi/W}d{k_y}\mathcal{F}_t(k_y)$. The singlet amplitude is even in $\omega$, whereas the triplet amplitudes are odd in $\omega$. The results sum over the contributions from 400 discrete $k_y$ points over the Brillouin zone. Other parameters are the same as Fig.~2 in the main text.}
	
\label{fig:triplet-FFLO-sumk}
\end{figure*}

\subsection{Pairing amplitudes with considering all modes}

In the main text, we present the results of the pairing amplitudes for the $k_y=0$ modes to show the presence of the mixed spin singlet-triplet pairing. This exotic mixed pairing state persists after integrating over $k_y$ in the Brillouin zone, as shown in Fig.~\ref{fig:triplet-FFLO-sumk}. All the important features persist after the integration.

\begin{figure*}[!htbp]
\includegraphics[width=0.7\textwidth]{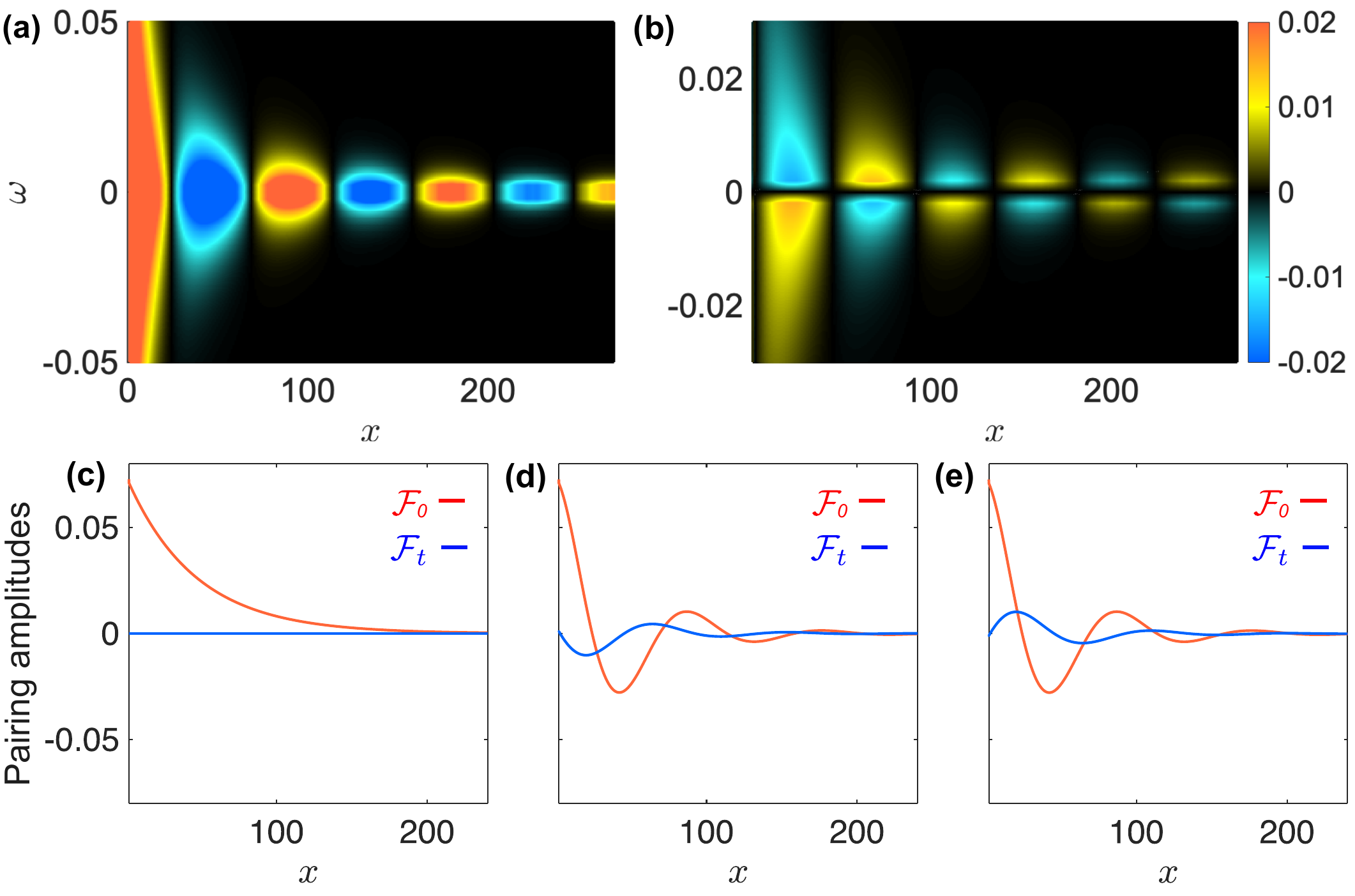}
	
\caption{(a) Contour plots of the singlet local pairing amplitude $\mathcal{F}_0$ (in units of $t^{-1}$) as a function of Matsubara frequency $\omega$ and position $x$ (in units of unit cells) in the cAFM. We consider $k_x=0.7\pi$, $\mu=0.5t$, $J=0.05t$ and $\Delta=0.1t$ for illustration. 
(b) the same as (a) but for the triplet component, $\mathcal{F}_t=-i\mathcal{F}_{\uparrow\uparrow}=i\mathcal{F}_{\downarrow\downarrow}$. 
(c) Singlet (red) and triplet (blue) amplitudes, $\mathcal{F}_0$ and $\mathcal{F}_t$, as functions of position $x$ in the cAFM in the absence of the cAFM order ($J=0$) and for $\omega=0.1\Delta$. $\mathcal{F}_0$ decays monotonically when away from the SC/cAFM interface, while $\mathcal{F}_t$ is absent anywhere, $\mathcal{F}_t=0$. 
(d) and (e) the same as (c) but for $J=0.05t$ and $\omega=+0.1\Delta$ and $-0.1\Delta$, respectively. A finite $\mathcal{F}_t$ emerges, as a result of the cAFM order $J\neq 0$. All pairing components decay with periodic oscillations about zero when away from the SC/cAFM interface.}
	
\label{fig:triplet-FFLO-SM}
\end{figure*}

\subsection{Mixed singlet-triplet pairings of Dirac electrons}

In the main text, we mainly use numerical results for the Schr\"odinger electrons near the band edge for illustration. The qualitatively same results occur for the Dirac electrons near the band center. To illustrate this, we consider instead $\mu=0.5t$ (i.e., $\mu_K=0.5t$), $J=0.05t$, and $k_y=0.7\pi$ (which is close to the $K/K'$ point) in the full kagome lattice model. Other parameters are the same as those used in Fig.~3 in the main text.  The numerical results are displayed in Fig.~\ref{fig:triplet-FFLO-SM}. A similar mixed singlet-triplet pairing state also emerges in this energy regime. This mixed pairing state only exists for the presence of cAFM order.

\subsection{Pairing amplitudes for negative canting and without AFM order }

\begin{figure}[!htbp]
\includegraphics[width=0.6\textwidth]{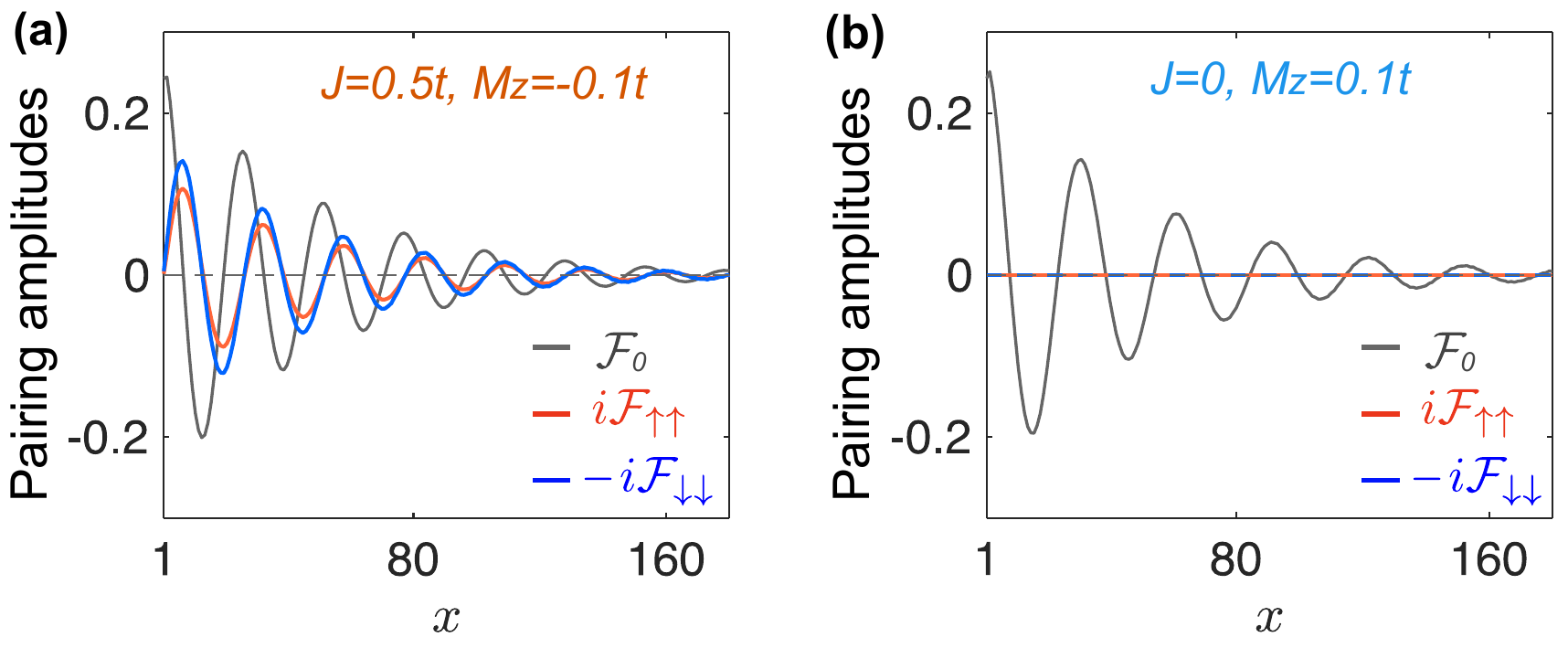}

\caption{(a) Spin singlet (back) and two equal-spin triplet amplitudes (red and blue), $\mathcal{F}_0$, $\mathcal{F}_{\uparrow\uparrow}$ and $\mathcal{F}_{\downarrow\downarrow}$, as functions of position $x$ in the AFM under the presence of a spin-canting in the $-z$ direction (i.e., $J=0.5t$ and $M_z=-0.1t$). (b) the same as (a) but for the absence of AFM order and the presence of spin canting (i.e., $J=0$ and $M_z=0.1t$). Other parameters are $\mu=\mu_S=-2t$, $\omega=0.01t$ and $\Delta=0.05t$.}

\label{figSM:Mz-canting}
\end{figure}

Figure~\ref{figSM:Mz-canting}(a) presents the pairing amplitudes for an opposite out-of-plane canting $M_z=-0.1t$. They are the same as those for $M_z=0.1t$ but with exchanging $\mathcal{F}_{\uparrow\uparrow}$ and $\mathcal{F}_{\downarrow\downarrow}$.
Figure~\ref{figSM:Mz-canting}(b) presents the pairing amplitudes for the case without cAFM order ($J=0$). We see that the triplet pairings disappear completely.

\subsection{Phases of the pairing correlations}
\begin{figure}[!htbp]
\includegraphics[width=0.7\textwidth]{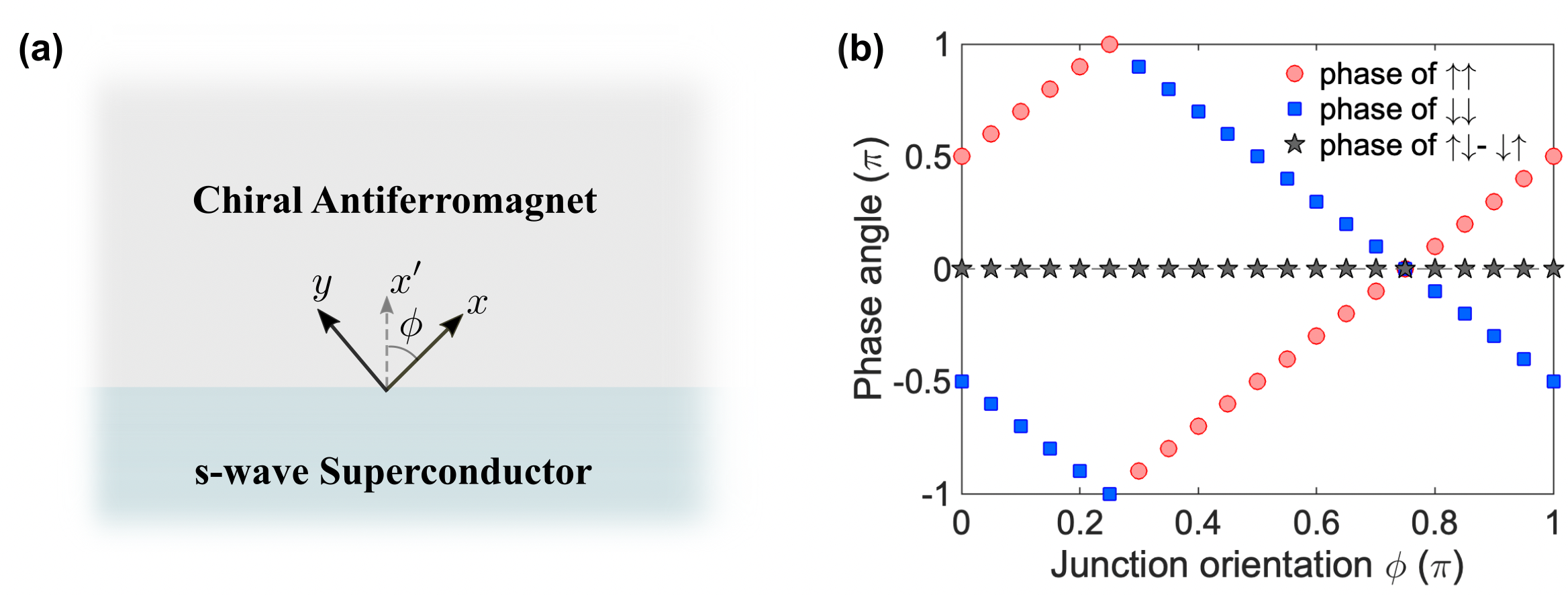}

\caption{(a) Schematic of the planar junction. (b) The phases of the pairing amplitudes, $\mathcal{F}_{\uparrow\uparrow}$, $\mathcal{F}_{\downarrow\downarrow}$ and $\mathcal{F}_0$, at $x'=30$ as functions of junction orientation $\phi$ with respect to $x$ axis. $\phi=0$ corresponds to the case where the junction is aligned with the $x$ axis. While the singlet amplitude keeps the same phase as that in the superconductor (i.e., real-valued), the phases of the two triplet amplitudes vary as the junction orientation changes.  Other parameters are the same as Fig.~2 in the main text.}

\label{figSM:phase}
\end{figure}

When the junction orientation changes, the phase of the triplet pairing changes accordingly, while the phase of the singlet pairing remains the same as that of the superconducting region. This is clearly supported by our numerical calculations in  Fig.~\ref{figSM:phase}.

\begin{thebibliography}{120}%
\makeatletter
\providecommand \@ifxundefined [1]{%
 \@ifx{#1\undefined}
}%
\providecommand \@ifnum [1]{%
 \ifnum #1\expandafter \@firstoftwo
 \else \expandafter \@secondoftwo
 \fi
}%
\providecommand \@ifx [1]{%
 \ifx #1\expandafter \@firstoftwo
 \else \expandafter \@secondoftwo
 \fi
}%
\providecommand \natexlab [1]{#1}%
\providecommand \enquote  [1]{``#1''}%
\providecommand \bibnamefont  [1]{#1}%
\providecommand \bibfnamefont [1]{#1}%
\providecommand \citenamefont [1]{#1}%
\providecommand \href@noop [0]{\@secondoftwo}%
\providecommand \href [0]{\begingroup \@sanitize@url \@href}%
\providecommand \@href[1]{\@@startlink{#1}\@@href}%
\providecommand \@@href[1]{\endgroup#1\@@endlink}%
\providecommand \@sanitize@url [0]{\catcode `\\12\catcode `\$12\catcode `\&12\catcode `\#12\catcode `\^12\catcode `\_12\catcode `\%12\relax}%
\providecommand \@@startlink[1]{}%
\providecommand \@@endlink[0]{}%
\providecommand \url  [0]{\begingroup\@sanitize@url \@url }%
\providecommand \@url [1]{\endgroup\@href {#1}{\urlprefix }}%
\providecommand \urlprefix  [0]{URL }%
\providecommand \Eprint [0]{\href }%
\providecommand \doibase [0]{https://doi.org/}%
\providecommand \selectlanguage [0]{\@gobble}%
\providecommand \bibinfo  [0]{\@secondoftwo}%
\providecommand \bibfield  [0]{\@secondoftwo}%
\providecommand \translation [1]{[#1]}%
\providecommand \BibitemOpen [0]{}%
\providecommand \bibitemStop [0]{}%
\providecommand \bibitemNoStop [0]{.\EOS\space}%
\providecommand \EOS [0]{\spacefactor3000\relax}%
\providecommand \BibitemShut  [1]{\csname bibitem#1\endcsname}%
\let\auto@bib@innerbib\@empty
\bibitem [{\citenamefont {Chen}\ \emph {et~al.}(2014)\citenamefont {Chen}, \citenamefont {Niu},\ and\ \citenamefont {MacDonald}}]{Chen2014Anomalous}%
  \BibitemOpen
  \bibfield  {author} {\bibinfo {author} {\bibfnamefont {H.}~\bibnamefont {Chen}}, \bibinfo {author} {\bibfnamefont {Q.}~\bibnamefont {Niu}},\ and\ \bibinfo {author} {\bibfnamefont {A.~H.}\ \bibnamefont {MacDonald}},\ }\bibfield  {title} {\bibinfo {title} {Anomalous hall effect arising from noncollinear antiferromagnetism},\ }\href {https://doi.org/10.1103/PhysRevLett.112.017205} {\bibfield  {journal} {\bibinfo  {journal} {Phys. Rev. Lett.}\ }\textbf {\bibinfo {volume} {112}},\ \bibinfo {pages} {017205} (\bibinfo {year} {2014})}\BibitemShut {NoStop}%
\bibitem [{\citenamefont {Nakatsuji}\ \emph {et~al.}(2015)\citenamefont {Nakatsuji}, \citenamefont {Kiyohara},\ and\ \citenamefont {Higo}}]{nakatsuji2015large}%
  \BibitemOpen
  \bibfield  {author} {\bibinfo {author} {\bibfnamefont {S.}~\bibnamefont {Nakatsuji}}, \bibinfo {author} {\bibfnamefont {N.}~\bibnamefont {Kiyohara}},\ and\ \bibinfo {author} {\bibfnamefont {T.}~\bibnamefont {Higo}},\ }\bibfield  {title} {\bibinfo {title} {Large anomalous hall effect in a non-collinear antiferromagnet at room temperature},\ }\href {https://doi.org/10.1038/nature15723} {\bibfield  {journal} {\bibinfo  {journal} {Nature}\ }\textbf {\bibinfo {volume} {527}},\ \bibinfo {pages} {212} (\bibinfo {year} {2015})}\BibitemShut {NoStop}%
\bibitem [{\citenamefont {Wu}\ and\ \citenamefont {Zhang}(2004)}]{wu2004prl}%
  \BibitemOpen
  \bibfield  {author} {\bibinfo {author} {\bibfnamefont {C.}~\bibnamefont {Wu}}\ and\ \bibinfo {author} {\bibfnamefont {S.-C.}\ \bibnamefont {Zhang}},\ }\bibfield  {title} {\bibinfo {title} {Dynamic generation of spin-orbit coupling},\ }\href {https://doi.org/10.1103/PhysRevLett.93.036403} {\bibfield  {journal} {\bibinfo  {journal} {Phys. Rev. Lett.}\ }\textbf {\bibinfo {volume} {93}},\ \bibinfo {pages} {036403} (\bibinfo {year} {2004})}\BibitemShut {NoStop}%
\bibitem [{\citenamefont {Wu}\ \emph {et~al.}(2007)\citenamefont {Wu}, \citenamefont {Sun}, \citenamefont {Fradkin},\ and\ \citenamefont {Zhang}}]{CJWu07PRB}%
  \BibitemOpen
  \bibfield  {author} {\bibinfo {author} {\bibfnamefont {C.}~\bibnamefont {Wu}}, \bibinfo {author} {\bibfnamefont {K.}~\bibnamefont {Sun}}, \bibinfo {author} {\bibfnamefont {E.}~\bibnamefont {Fradkin}},\ and\ \bibinfo {author} {\bibfnamefont {S.-C.}\ \bibnamefont {Zhang}},\ }\bibfield  {title} {\bibinfo {title} {Fermi liquid instabilities in the spin channel},\ }\href {https://doi.org/10.1103/PhysRevB.75.115103} {\bibfield  {journal} {\bibinfo  {journal} {Phys. Rev. B}\ }\textbf {\bibinfo {volume} {75}},\ \bibinfo {pages} {115103} (\bibinfo {year} {2007})}\BibitemShut {NoStop}%
\bibitem [{\citenamefont {{\v{S}}mejkal}\ \emph {et~al.}(2020)\citenamefont {{\v{S}}mejkal}, \citenamefont {Gonz{\'a}lez-Hern{\'a}ndez}, \citenamefont {Jungwirth},\ and\ \citenamefont {Sinova}}]{vsmejkal2020crystal}%
  \BibitemOpen
  \bibfield  {author} {\bibinfo {author} {\bibfnamefont {L.}~\bibnamefont {{\v{S}}mejkal}}, \bibinfo {author} {\bibfnamefont {R.}~\bibnamefont {Gonz{\'a}lez-Hern{\'a}ndez}}, \bibinfo {author} {\bibfnamefont {T.}~\bibnamefont {Jungwirth}},\ and\ \bibinfo {author} {\bibfnamefont {J.}~\bibnamefont {Sinova}},\ }\bibfield  {title} {\bibinfo {title} {{Crystal time-reversal symmetry breaking and spontaneous Hall effect in collinear antiferromagnets}},\ }\href {https://www.science.org/doi/full/10.1126/sciadv.aaz8809} {\bibfield  {journal} {\bibinfo  {journal} {Sci. Adv.}\ }\textbf {\bibinfo {volume} {6}},\ \bibinfo {pages} {eaaz8809} (\bibinfo {year} {2020})}\BibitemShut {NoStop}%
\bibitem [{\citenamefont {Naka}\ \emph {et~al.}(2019)\citenamefont {Naka}, \citenamefont {Hayami}, \citenamefont {Kusunose}, \citenamefont {Yanagi}, \citenamefont {Motome},\ and\ \citenamefont {Seo}}]{Naka19NC}%
  \BibitemOpen
  \bibfield  {author} {\bibinfo {author} {\bibfnamefont {M.}~\bibnamefont {Naka}}, \bibinfo {author} {\bibfnamefont {S.}~\bibnamefont {Hayami}}, \bibinfo {author} {\bibfnamefont {H.}~\bibnamefont {Kusunose}}, \bibinfo {author} {\bibfnamefont {Y.}~\bibnamefont {Yanagi}}, \bibinfo {author} {\bibfnamefont {Y.}~\bibnamefont {Motome}},\ and\ \bibinfo {author} {\bibfnamefont {H.}~\bibnamefont {Seo}},\ }\bibfield  {title} {\bibinfo {title} {Spin current generation in organic antiferromagnets},\ }\href {https://doi.org/10.1038/s41467-019-12229-y} {\bibfield  {journal} {\bibinfo  {journal} {Nat. Commun.}\ }\textbf {\bibinfo {volume} {10}},\ \bibinfo {pages} {4305} (\bibinfo {year} {2019})}\BibitemShut {NoStop}%
\bibitem [{\citenamefont {Hayami}\ \emph {et~al.}(2019)\citenamefont {Hayami}, \citenamefont {Yanagi},\ and\ \citenamefont {Kusunose}}]{hayami2019momentum}%
  \BibitemOpen
  \bibfield  {author} {\bibinfo {author} {\bibfnamefont {S.}~\bibnamefont {Hayami}}, \bibinfo {author} {\bibfnamefont {Y.}~\bibnamefont {Yanagi}},\ and\ \bibinfo {author} {\bibfnamefont {H.}~\bibnamefont {Kusunose}},\ }\bibfield  {title} {\bibinfo {title} {Momentum-dependent spin splitting by collinear antiferromagnetic ordering},\ }\href {https://doi.org/10.7566/JPSJ.88.123702} {\bibfield  {journal} {\bibinfo  {journal} {J. Phys. Soc. Jpn.}\ }\textbf {\bibinfo {volume} {88}},\ \bibinfo {pages} {123702} (\bibinfo {year} {2019})}\BibitemShut {NoStop}%
\bibitem [{\citenamefont {Yuan}\ \emph {et~al.}(2020)\citenamefont {Yuan}, \citenamefont {Wang}, \citenamefont {Luo}, \citenamefont {Rashba},\ and\ \citenamefont {Zunger}}]{Yuan2020Giant}%
  \BibitemOpen
  \bibfield  {author} {\bibinfo {author} {\bibfnamefont {L.-D.}\ \bibnamefont {Yuan}}, \bibinfo {author} {\bibfnamefont {Z.}~\bibnamefont {Wang}}, \bibinfo {author} {\bibfnamefont {J.-W.}\ \bibnamefont {Luo}}, \bibinfo {author} {\bibfnamefont {E.~I.}\ \bibnamefont {Rashba}},\ and\ \bibinfo {author} {\bibfnamefont {A.}~\bibnamefont {Zunger}},\ }\bibfield  {title} {\bibinfo {title} {Giant momentum-dependent spin splitting in centrosymmetric low-$z$ antiferromagnets},\ }\href {https://doi.org/10.1103/PhysRevB.102.014422} {\bibfield  {journal} {\bibinfo  {journal} {Phys. Rev. B}\ }\textbf {\bibinfo {volume} {102}},\ \bibinfo {pages} {014422} (\bibinfo {year} {2020})}\BibitemShut {NoStop}%
\bibitem [{\citenamefont {Mazin}\ \emph {et~al.}(2021)\citenamefont {Mazin}, \citenamefont {Koepernik}, \citenamefont {Johannes}, \citenamefont {Gonz{\'a}lez-Hern{\'a}ndez},\ and\ \citenamefont {{\v{S}}mejkal}}]{mazin2021prediction}%
  \BibitemOpen
  \bibfield  {author} {\bibinfo {author} {\bibfnamefont {I.~I.}\ \bibnamefont {Mazin}}, \bibinfo {author} {\bibfnamefont {K.}~\bibnamefont {Koepernik}}, \bibinfo {author} {\bibfnamefont {M.~D.}\ \bibnamefont {Johannes}}, \bibinfo {author} {\bibfnamefont {R.}~\bibnamefont {Gonz{\'a}lez-Hern{\'a}ndez}},\ and\ \bibinfo {author} {\bibfnamefont {L.}~\bibnamefont {{\v{S}}mejkal}},\ }\bibfield  {title} {\bibinfo {title} {{Prediction of unconventional magnetism in doped FeSb$_2$}},\ }\href {https://doi.org/10.1073/pnas.2108924118} {\bibfield  {journal} {\bibinfo  {journal} {Proc. Nat. Acad. Sci.}\ }\textbf {\bibinfo {volume} {118}},\ \bibinfo {pages} {e2108924118} (\bibinfo {year} {2021})}\BibitemShut {NoStop}%
\bibitem [{\citenamefont {Ma}\ \emph {et~al.}(2021)\citenamefont {Ma}, \citenamefont {Hu}, \citenamefont {Li}, \citenamefont {Liu}, \citenamefont {Yao}, \citenamefont {Jia},\ and\ \citenamefont {Liu}}]{ma2021multifunctional}%
  \BibitemOpen
  \bibfield  {author} {\bibinfo {author} {\bibfnamefont {H.-Y.}\ \bibnamefont {Ma}}, \bibinfo {author} {\bibfnamefont {M.}~\bibnamefont {Hu}}, \bibinfo {author} {\bibfnamefont {N.}~\bibnamefont {Li}}, \bibinfo {author} {\bibfnamefont {J.}~\bibnamefont {Liu}}, \bibinfo {author} {\bibfnamefont {W.}~\bibnamefont {Yao}}, \bibinfo {author} {\bibfnamefont {J.-F.}\ \bibnamefont {Jia}},\ and\ \bibinfo {author} {\bibfnamefont {J.}~\bibnamefont {Liu}},\ }\bibfield  {title} {\bibinfo {title} {Multifunctional antiferromagnetic materials with giant piezomagnetism and noncollinear spin current},\ }\href {https://doi.org/10.1038/s41467-021-23127-7} {\bibfield  {journal} {\bibinfo  {journal} {Nat. Commun.}\ }\textbf {\bibinfo {volume} {12}},\ \bibinfo {pages} {2846} (\bibinfo {year} {2021})}\BibitemShut {NoStop}%
\bibitem [{\citenamefont {\ifmmode~\check{S}\else \v{S}\fi{}mejkal}\ \emph {et~al.}(2022{\natexlab{a}})\citenamefont {\ifmmode~\check{S}\else \v{S}\fi{}mejkal}, \citenamefont {Sinova},\ and\ \citenamefont {Jungwirth}}]{ifmmode2022Beyond}%
  \BibitemOpen
  \bibfield  {author} {\bibinfo {author} {\bibfnamefont {L.}~\bibnamefont {\ifmmode~\check{S}\else \v{S}\fi{}mejkal}}, \bibinfo {author} {\bibfnamefont {J.}~\bibnamefont {Sinova}},\ and\ \bibinfo {author} {\bibfnamefont {T.}~\bibnamefont {Jungwirth}},\ }\bibfield  {title} {\bibinfo {title} {Beyond conventional ferromagnetism and antiferromagnetism: A phase with nonrelativistic spin and crystal rotation symmetry},\ }\href {https://doi.org/10.1103/PhysRevX.12.031042} {\bibfield  {journal} {\bibinfo  {journal} {Phys. Rev. X}\ }\textbf {\bibinfo {volume} {12}},\ \bibinfo {pages} {031042} (\bibinfo {year} {2022}{\natexlab{a}})}\BibitemShut {NoStop}%
\bibitem [{\citenamefont {\ifmmode~\check{S}\else \v{S}\fi{}mejkal}\ \emph {et~al.}(2022{\natexlab{b}})\citenamefont {\ifmmode~\check{S}\else \v{S}\fi{}mejkal}, \citenamefont {Sinova},\ and\ \citenamefont {Jungwirth}}]{ifmmode2022Emerging}%
  \BibitemOpen
  \bibfield  {author} {\bibinfo {author} {\bibfnamefont {L.}~\bibnamefont {\ifmmode~\check{S}\else \v{S}\fi{}mejkal}}, \bibinfo {author} {\bibfnamefont {J.}~\bibnamefont {Sinova}},\ and\ \bibinfo {author} {\bibfnamefont {T.}~\bibnamefont {Jungwirth}},\ }\bibfield  {title} {\bibinfo {title} {Emerging research landscape of altermagnetism},\ }\href {https://doi.org/10.1103/PhysRevX.12.040501} {\bibfield  {journal} {\bibinfo  {journal} {Phys. Rev. X}\ }\textbf {\bibinfo {volume} {12}},\ \bibinfo {pages} {040501} (\bibinfo {year} {2022}{\natexlab{b}})}\BibitemShut {NoStop}%
\bibitem [{\citenamefont {Ahn}\ \emph {et~al.}(2019)\citenamefont {Ahn}, \citenamefont {Hariki}, \citenamefont {Lee},\ and\ \citenamefont {Kune\ifmmode~\check{s}\else \v{s}\fi{}}}]{Ahn19PRB}%
  \BibitemOpen
  \bibfield  {author} {\bibinfo {author} {\bibfnamefont {K.-H.}\ \bibnamefont {Ahn}}, \bibinfo {author} {\bibfnamefont {A.}~\bibnamefont {Hariki}}, \bibinfo {author} {\bibfnamefont {K.-W.}\ \bibnamefont {Lee}},\ and\ \bibinfo {author} {\bibfnamefont {J.}~\bibnamefont {Kune\ifmmode~\check{s}\else \v{s}\fi{}}},\ }\bibfield  {title} {\bibinfo {title} {Antiferromagnetism in ${\mathrm{ruo}}_{2}$ as $d$-wave pomeranchuk instability},\ }\href {https://doi.org/10.1103/PhysRevB.99.184432} {\bibfield  {journal} {\bibinfo  {journal} {Phys. Rev. B}\ }\textbf {\bibinfo {volume} {99}},\ \bibinfo {pages} {184432} (\bibinfo {year} {2019})}\BibitemShut {NoStop}%
\bibitem [{\citenamefont {Shao}\ \emph {et~al.}(2021)\citenamefont {Shao}, \citenamefont {Zhang}, \citenamefont {Li}, \citenamefont {Eom},\ and\ \citenamefont {Tsymbal}}]{shao2021spin}%
  \BibitemOpen
  \bibfield  {author} {\bibinfo {author} {\bibfnamefont {D.-F.}\ \bibnamefont {Shao}}, \bibinfo {author} {\bibfnamefont {S.-H.}\ \bibnamefont {Zhang}}, \bibinfo {author} {\bibfnamefont {M.}~\bibnamefont {Li}}, \bibinfo {author} {\bibfnamefont {C.-B.}\ \bibnamefont {Eom}},\ and\ \bibinfo {author} {\bibfnamefont {E.~Y.}\ \bibnamefont {Tsymbal}},\ }\bibfield  {title} {\bibinfo {title} {Spin-neutral currents for spintronics},\ }\href {https://doi.org/10.1038/s41467-021-26915-3} {\bibfield  {journal} {\bibinfo  {journal} {Nat. Commun.}\ }\textbf {\bibinfo {volume} {12}},\ \bibinfo {pages} {7061} (\bibinfo {year} {2021})}\BibitemShut {NoStop}%
\bibitem [{\citenamefont {Feng}\ \emph {et~al.}(2022)\citenamefont {Feng}, \citenamefont {Zhou}, \citenamefont {{\v{S}}mejkal}, \citenamefont {Wu}, \citenamefont {Zhu}, \citenamefont {Guo}, \citenamefont {Gonz{\'a}lez-Hern{\'a}ndez}, \citenamefont {Wang}, \citenamefont {Yan}, \citenamefont {Qin} \emph {et~al.}}]{FengZX22NE}%
  \BibitemOpen
  \bibfield  {author} {\bibinfo {author} {\bibfnamefont {Z.}~\bibnamefont {Feng}}, \bibinfo {author} {\bibfnamefont {X.}~\bibnamefont {Zhou}}, \bibinfo {author} {\bibfnamefont {L.}~\bibnamefont {{\v{S}}mejkal}}, \bibinfo {author} {\bibfnamefont {L.}~\bibnamefont {Wu}}, \bibinfo {author} {\bibfnamefont {Z.}~\bibnamefont {Zhu}}, \bibinfo {author} {\bibfnamefont {H.}~\bibnamefont {Guo}}, \bibinfo {author} {\bibfnamefont {R.}~\bibnamefont {Gonz{\'a}lez-Hern{\'a}ndez}}, \bibinfo {author} {\bibfnamefont {X.}~\bibnamefont {Wang}}, \bibinfo {author} {\bibfnamefont {H.}~\bibnamefont {Yan}}, \bibinfo {author} {\bibfnamefont {P.}~\bibnamefont {Qin}}, \emph {et~al.},\ }\bibfield  {title} {\bibinfo {title} {{An anomalous Hall effect in altermagnetic ruthenium dioxide}},\ }\href {https://www.nature.com/articles/s41928-022-00866-z} {\bibfield  {journal} {\bibinfo  {journal} {Nat. Electron.}\ }\textbf {\bibinfo {volume} {5}},\ \bibinfo {pages} {735} (\bibinfo {year} {2022})}\BibitemShut {NoStop}%
\bibitem [{\citenamefont {\ifmmode~\check{S}\else \v{S}\fi{}mejkal}\ \emph {et~al.}(2022{\natexlab{c}})\citenamefont {\ifmmode~\check{S}\else \v{S}\fi{}mejkal}, \citenamefont {Hellenes}, \citenamefont {Gonz\'alez-Hern\'andez}, \citenamefont {Sinova},\ and\ \citenamefont {Jungwirth}}]{ifmmode2022Giant}%
  \BibitemOpen
  \bibfield  {author} {\bibinfo {author} {\bibfnamefont {L.}~\bibnamefont {\ifmmode~\check{S}\else \v{S}\fi{}mejkal}}, \bibinfo {author} {\bibfnamefont {A.~B.}\ \bibnamefont {Hellenes}}, \bibinfo {author} {\bibfnamefont {R.}~\bibnamefont {Gonz\'alez-Hern\'andez}}, \bibinfo {author} {\bibfnamefont {J.}~\bibnamefont {Sinova}},\ and\ \bibinfo {author} {\bibfnamefont {T.}~\bibnamefont {Jungwirth}},\ }\bibfield  {title} {\bibinfo {title} {Giant and tunneling magnetoresistance in unconventional collinear antiferromagnets with nonrelativistic spin-momentum coupling},\ }\href {https://doi.org/10.1103/PhysRevX.12.011028} {\bibfield  {journal} {\bibinfo  {journal} {Phys. Rev. X}\ }\textbf {\bibinfo {volume} {12}},\ \bibinfo {pages} {011028} (\bibinfo {year} {2022}{\natexlab{c}})}\BibitemShut {NoStop}%
\bibitem [{\citenamefont {Lin}\ \emph {et~al.}(2025)\citenamefont {Lin}, \citenamefont {Zhang}, \citenamefont {Lu},\ and\ \citenamefont {Xie}}]{HJLin25PRL}%
  \BibitemOpen
  \bibfield  {author} {\bibinfo {author} {\bibfnamefont {H.-J.}\ \bibnamefont {Lin}}, \bibinfo {author} {\bibfnamefont {S.-B.}\ \bibnamefont {Zhang}}, \bibinfo {author} {\bibfnamefont {H.-Z.}\ \bibnamefont {Lu}},\ and\ \bibinfo {author} {\bibfnamefont {X.~C.}\ \bibnamefont {Xie}},\ }\bibfield  {title} {\bibinfo {title} {Coulomb drag in altermagnets},\ }\href {https://doi.org/10.1103/PhysRevLett.134.136301} {\bibfield  {journal} {\bibinfo  {journal} {Phys. Rev. Lett.}\ }\textbf {\bibinfo {volume} {134}},\ \bibinfo {pages} {136301} (\bibinfo {year} {2025})}\BibitemShut {NoStop}%
\bibitem [{\citenamefont {Zhang}\ \emph {et~al.}(2024)\citenamefont {Zhang}, \citenamefont {Hu},\ and\ \citenamefont {Neupert}}]{SBZ2023arXiv}%
  \BibitemOpen
  \bibfield  {author} {\bibinfo {author} {\bibfnamefont {S.-B.}\ \bibnamefont {Zhang}}, \bibinfo {author} {\bibfnamefont {L.-H.}\ \bibnamefont {Hu}},\ and\ \bibinfo {author} {\bibfnamefont {T.}~\bibnamefont {Neupert}},\ }\bibfield  {title} {\bibinfo {title} {{Finite-momentum Cooper pairing in proximitized altermagnets}},\ }\href {https://doi.org/10.1038/s41467-024-45951-3} {\bibfield  {journal} {\bibinfo  {journal} {Nat. Commun.}\ }\textbf {\bibinfo {volume} {15}},\ \bibinfo {pages} {1801} (\bibinfo {year} {2024})}\BibitemShut {NoStop}%
\bibitem [{\citenamefont {Sumita}\ \emph {et~al.}(2023)\citenamefont {Sumita}, \citenamefont {Naka},\ and\ \citenamefont {Seo}}]{Sumita2023Fulde}%
  \BibitemOpen
  \bibfield  {author} {\bibinfo {author} {\bibfnamefont {S.}~\bibnamefont {Sumita}}, \bibinfo {author} {\bibfnamefont {M.}~\bibnamefont {Naka}},\ and\ \bibinfo {author} {\bibfnamefont {H.}~\bibnamefont {Seo}},\ }\bibfield  {title} {\bibinfo {title} {{Fulde-Ferrell-Larkin-Ovchinnikov state induced by antiferromagnetic order in $\ensuremath{\kappa}$-type organic conductors}},\ }\href {https://doi.org/10.1103/PhysRevResearch.5.043171} {\bibfield  {journal} {\bibinfo  {journal} {Phys. Rev. Res.}\ }\textbf {\bibinfo {volume} {5}},\ \bibinfo {pages} {043171} (\bibinfo {year} {2023})}\BibitemShut {NoStop}%
\bibitem [{\citenamefont {Chakraborty}\ and\ \citenamefont {Black-Schaffer}(2024)}]{chakraborty2023zero}%
  \BibitemOpen
  \bibfield  {author} {\bibinfo {author} {\bibfnamefont {D.}~\bibnamefont {Chakraborty}}\ and\ \bibinfo {author} {\bibfnamefont {A.~M.}\ \bibnamefont {Black-Schaffer}},\ }\bibfield  {title} {\bibinfo {title} {Zero-field finite-momentum and field-induced superconductivity in altermagnets},\ }\href {https://doi.org/10.1103/PhysRevB.110.L060508} {\bibfield  {journal} {\bibinfo  {journal} {Phys. Rev. B}\ }\textbf {\bibinfo {volume} {110}},\ \bibinfo {pages} {L060508} (\bibinfo {year} {2024})}\BibitemShut {NoStop}%
\bibitem [{\citenamefont {Ouassou}\ \emph {et~al.}(2023)\citenamefont {Ouassou}, \citenamefont {Brataas},\ and\ \citenamefont {Linder}}]{Ouassou2023dc}%
  \BibitemOpen
  \bibfield  {author} {\bibinfo {author} {\bibfnamefont {J.~A.}\ \bibnamefont {Ouassou}}, \bibinfo {author} {\bibfnamefont {A.}~\bibnamefont {Brataas}},\ and\ \bibinfo {author} {\bibfnamefont {J.}~\bibnamefont {Linder}},\ }\bibfield  {title} {\bibinfo {title} {{dc Josephson Effect in Altermagnets}},\ }\href {https://doi.org/10.1103/PhysRevLett.131.076003} {\bibfield  {journal} {\bibinfo  {journal} {Phys. Rev. Lett.}\ }\textbf {\bibinfo {volume} {131}},\ \bibinfo {pages} {076003} (\bibinfo {year} {2023})}\BibitemShut {NoStop}%
\bibitem [{\citenamefont {Beenakker}\ and\ \citenamefont {Vakhtel}(2023)}]{Beenakker2023Phase}%
  \BibitemOpen
  \bibfield  {author} {\bibinfo {author} {\bibfnamefont {C.~W.~J.}\ \bibnamefont {Beenakker}}\ and\ \bibinfo {author} {\bibfnamefont {T.}~\bibnamefont {Vakhtel}},\ }\bibfield  {title} {\bibinfo {title} {{Phase-shifted Andreev levels in an altermagnet Josephson junction}},\ }\href {https://doi.org/10.1103/PhysRevB.108.075425} {\bibfield  {journal} {\bibinfo  {journal} {Phys. Rev. B}\ }\textbf {\bibinfo {volume} {108}},\ \bibinfo {pages} {075425} (\bibinfo {year} {2023})}\BibitemShut {NoStop}%
\bibitem [{\citenamefont {Cheng}\ and\ \citenamefont {Sun}(2024)}]{Cheng2024Orientation}%
  \BibitemOpen
  \bibfield  {author} {\bibinfo {author} {\bibfnamefont {Q.}~\bibnamefont {Cheng}}\ and\ \bibinfo {author} {\bibfnamefont {Q.}~\bibnamefont {Sun}},\ }\bibfield  {title} {\bibinfo {title} {{Orientation-dependent Josephson effect in spin-singlet superconductor/altermagnet/spin-triplet superconductor junctions}},\ }\href {https://doi.org/10.1103/PhysRevB.109.024517} {\bibfield  {journal} {\bibinfo  {journal} {Phys. Rev. B}\ }\textbf {\bibinfo {volume} {109}},\ \bibinfo {pages} {024517} (\bibinfo {year} {2024})}\BibitemShut {NoStop}%
\bibitem [{\citenamefont {Sun}\ \emph {et~al.}(2023)\citenamefont {Sun}, \citenamefont {Brataas},\ and\ \citenamefont {Linder}}]{sun2023Andreev}%
  \BibitemOpen
  \bibfield  {author} {\bibinfo {author} {\bibfnamefont {C.}~\bibnamefont {Sun}}, \bibinfo {author} {\bibfnamefont {A.}~\bibnamefont {Brataas}},\ and\ \bibinfo {author} {\bibfnamefont {J.}~\bibnamefont {Linder}},\ }\bibfield  {title} {\bibinfo {title} {Andreev reflection in altermagnets},\ }\href {https://doi.org/10.1103/PhysRevB.108.054511} {\bibfield  {journal} {\bibinfo  {journal} {Phys. Rev. B}\ }\textbf {\bibinfo {volume} {108}},\ \bibinfo {pages} {054511} (\bibinfo {year} {2023})}\BibitemShut {NoStop}%
\bibitem [{\citenamefont {Papaj}(2023)}]{Papaj2023Andreev}%
  \BibitemOpen
  \bibfield  {author} {\bibinfo {author} {\bibfnamefont {M.}~\bibnamefont {Papaj}},\ }\bibfield  {title} {\bibinfo {title} {Andreev reflection at the altermagnet-superconductor interface},\ }\href {https://doi.org/10.1103/PhysRevB.108.L060508} {\bibfield  {journal} {\bibinfo  {journal} {Phys. Rev. B}\ }\textbf {\bibinfo {volume} {108}},\ \bibinfo {pages} {L060508} (\bibinfo {year} {2023})}\BibitemShut {NoStop}%
\bibitem [{\citenamefont {Wei}\ \emph {et~al.}(2024)\citenamefont {Wei}, \citenamefont {Xiang}, \citenamefont {Xu}, \citenamefont {Zhang}, \citenamefont {Tang},\ and\ \citenamefont {Wang}}]{wei2023gapless}%
  \BibitemOpen
  \bibfield  {author} {\bibinfo {author} {\bibfnamefont {M.}~\bibnamefont {Wei}}, \bibinfo {author} {\bibfnamefont {L.}~\bibnamefont {Xiang}}, \bibinfo {author} {\bibfnamefont {F.}~\bibnamefont {Xu}}, \bibinfo {author} {\bibfnamefont {L.}~\bibnamefont {Zhang}}, \bibinfo {author} {\bibfnamefont {G.}~\bibnamefont {Tang}},\ and\ \bibinfo {author} {\bibfnamefont {J.}~\bibnamefont {Wang}},\ }\bibfield  {title} {\bibinfo {title} {Gapless superconducting state and mirage gap in altermagnets},\ }\href {https://doi.org/10.1103/PhysRevB.109.L201404} {\bibfield  {journal} {\bibinfo  {journal} {Phys. Rev. B}\ }\textbf {\bibinfo {volume} {109}},\ \bibinfo {pages} {L201404} (\bibinfo {year} {2024})}\BibitemShut {NoStop}%
\bibitem [{\citenamefont {Zhu}\ \emph {et~al.}(2023)\citenamefont {Zhu}, \citenamefont {Zhuang}, \citenamefont {Wu},\ and\ \citenamefont {Yan}}]{Zhu2023Topological}%
  \BibitemOpen
  \bibfield  {author} {\bibinfo {author} {\bibfnamefont {D.}~\bibnamefont {Zhu}}, \bibinfo {author} {\bibfnamefont {Z.-Y.}\ \bibnamefont {Zhuang}}, \bibinfo {author} {\bibfnamefont {Z.}~\bibnamefont {Wu}},\ and\ \bibinfo {author} {\bibfnamefont {Z.}~\bibnamefont {Yan}},\ }\bibfield  {title} {\bibinfo {title} {Topological superconductivity in two-dimensional altermagnetic metals},\ }\href {https://doi.org/10.1103/PhysRevB.108.184505} {\bibfield  {journal} {\bibinfo  {journal} {Phys. Rev. B}\ }\textbf {\bibinfo {volume} {108}},\ \bibinfo {pages} {184505} (\bibinfo {year} {2023})}\BibitemShut {NoStop}%
\bibitem [{\citenamefont {Li}\ and\ \citenamefont {Liu}(2023)}]{Li2023Majorana}%
  \BibitemOpen
  \bibfield  {author} {\bibinfo {author} {\bibfnamefont {Y.-X.}\ \bibnamefont {Li}}\ and\ \bibinfo {author} {\bibfnamefont {C.-C.}\ \bibnamefont {Liu}},\ }\bibfield  {title} {\bibinfo {title} {Majorana corner modes and tunable patterns in an altermagnet heterostructure},\ }\href {https://doi.org/10.1103/PhysRevB.108.205410} {\bibfield  {journal} {\bibinfo  {journal} {Phys. Rev. B}\ }\textbf {\bibinfo {volume} {108}},\ \bibinfo {pages} {205410} (\bibinfo {year} {2023})}\BibitemShut {NoStop}%
\bibitem [{\citenamefont {Ghorashi}\ \emph {et~al.}(2024)\citenamefont {Ghorashi}, \citenamefont {Hughes},\ and\ \citenamefont {Cano}}]{ghorashi2023altermagnetic}%
  \BibitemOpen
  \bibfield  {author} {\bibinfo {author} {\bibfnamefont {S.~A.~A.}\ \bibnamefont {Ghorashi}}, \bibinfo {author} {\bibfnamefont {T.~L.}\ \bibnamefont {Hughes}},\ and\ \bibinfo {author} {\bibfnamefont {J.}~\bibnamefont {Cano}},\ }\bibfield  {title} {\bibinfo {title} {Altermagnetic routes to majorana modes in zero net magnetization},\ }\href {https://doi.org/10.1103/PhysRevLett.133.106601} {\bibfield  {journal} {\bibinfo  {journal} {Phys. Rev. Lett.}\ }\textbf {\bibinfo {volume} {133}},\ \bibinfo {pages} {106601} (\bibinfo {year} {2024})}\BibitemShut {NoStop}%
\bibitem [{\citenamefont {Banerjee}\ and\ \citenamefont {Scheurer}(2024)}]{banerjee2024altermagnetic}%
  \BibitemOpen
  \bibfield  {author} {\bibinfo {author} {\bibfnamefont {S.}~\bibnamefont {Banerjee}}\ and\ \bibinfo {author} {\bibfnamefont {M.~S.}\ \bibnamefont {Scheurer}},\ }\bibfield  {title} {\bibinfo {title} {Altermagnetic superconducting diode effect},\ }\href {https://doi.org/10.1103/PhysRevB.110.024503} {\bibfield  {journal} {\bibinfo  {journal} {Phys. Rev. B}\ }\textbf {\bibinfo {volume} {110}},\ \bibinfo {pages} {024503} (\bibinfo {year} {2024})}\BibitemShut {NoStop}%
\bibitem [{\citenamefont {Sun}\ \emph {et~al.}(2025)\citenamefont {Sun}, \citenamefont {Zhang}, \citenamefont {Li},\ and\ \citenamefont {Trauzettel}}]{HPSun2024arXiv}%
  \BibitemOpen
  \bibfield  {author} {\bibinfo {author} {\bibfnamefont {H.-P.}\ \bibnamefont {Sun}}, \bibinfo {author} {\bibfnamefont {S.-B.}\ \bibnamefont {Zhang}}, \bibinfo {author} {\bibfnamefont {C.-A.}\ \bibnamefont {Li}},\ and\ \bibinfo {author} {\bibfnamefont {B.}~\bibnamefont {Trauzettel}},\ }\bibfield  {title} {\bibinfo {title} {Tunable second harmonic in altermagnetic josephson junctions},\ }\href {https://doi.org/10.1103/PhysRevB.111.165406} {\bibfield  {journal} {\bibinfo  {journal} {Phys. Rev. B}\ }\textbf {\bibinfo {volume} {111}},\ \bibinfo {pages} {165406} (\bibinfo {year} {2025})}\BibitemShut {NoStop}%
\bibitem [{\citenamefont {Li}\ and\ \citenamefont {et~al.}()}]{CLi2025submit}%
  \BibitemOpen
  \bibfield  {author} {\bibinfo {author} {\bibfnamefont {C.}~\bibnamefont {Li}}\ and\ \bibinfo {author} {\bibnamefont {et~al.}},\ }\bibfield  {title} {\bibinfo {title} {{Spin-Polarized Josephson Supercurrent in Nodeless Altermagnets}},\ }\href@noop {} {\bibinfo  {journal} {Submitted}\ }\BibitemShut {NoStop}%
\bibitem [{\citenamefont {Fulde}\ and\ \citenamefont {Ferrell}(1964)}]{Fulde64PR}%
  \BibitemOpen
\bibfield  {journal} {  }\bibfield  {author} {\bibinfo {author} {\bibfnamefont {P.}~\bibnamefont {Fulde}}\ and\ \bibinfo {author} {\bibfnamefont {R.~A.}\ \bibnamefont {Ferrell}},\ }\bibfield  {title} {\bibinfo {title} {Superconductivity in a strong spin-exchange field},\ }\href {https://doi.org/10.1103/PhysRev.135.A550} {\bibfield  {journal} {\bibinfo  {journal} {Phys. Rev.}\ }\textbf {\bibinfo {volume} {135}},\ \bibinfo {pages} {A550} (\bibinfo {year} {1964})}\BibitemShut {NoStop}%
\bibitem [{\citenamefont {Larkin}\ and\ \citenamefont {Ovchinnikov}(1964)}]{Larkin64JETP}%
  \BibitemOpen
  \bibfield  {author} {\bibinfo {author} {\bibfnamefont {A.~I.}\ \bibnamefont {Larkin}}\ and\ \bibinfo {author} {\bibfnamefont {Y.~N.}\ \bibnamefont {Ovchinnikov}},\ }\bibfield  {title} {\bibinfo {title} {Nonuniform state of superconductors},\ }\href@noop {} {\bibfield  {journal} {\bibinfo  {journal} {Zh. Eksp. Teor. Fiz}\ }\textbf {\bibinfo {volume} {47}},\ \bibinfo {pages} {1136} (\bibinfo {year} {1964})}\BibitemShut {NoStop}%
\bibitem [{\citenamefont {Lee}\ \emph {et~al.}(2024{\natexlab{a}})\citenamefont {Lee}, \citenamefont {Lee}, \citenamefont {Jung}, \citenamefont {Jung}, \citenamefont {Kim}, \citenamefont {Lee}, \citenamefont {Seok}, \citenamefont {Kim}, \citenamefont {Park}, \citenamefont {\ifmmode~\check{S}\else \v{S}\fi{}mejkal}, \citenamefont {Kang},\ and\ \citenamefont {Kim}}]{Lee2024Broken}%
  \BibitemOpen
  \bibfield  {author} {\bibinfo {author} {\bibfnamefont {S.}~\bibnamefont {Lee}}, \bibinfo {author} {\bibfnamefont {S.}~\bibnamefont {Lee}}, \bibinfo {author} {\bibfnamefont {S.}~\bibnamefont {Jung}}, \bibinfo {author} {\bibfnamefont {J.}~\bibnamefont {Jung}}, \bibinfo {author} {\bibfnamefont {D.}~\bibnamefont {Kim}}, \bibinfo {author} {\bibfnamefont {Y.}~\bibnamefont {Lee}}, \bibinfo {author} {\bibfnamefont {B.}~\bibnamefont {Seok}}, \bibinfo {author} {\bibfnamefont {J.}~\bibnamefont {Kim}}, \bibinfo {author} {\bibfnamefont {B.~G.}\ \bibnamefont {Park}}, \bibinfo {author} {\bibfnamefont {L.}~\bibnamefont {\ifmmode~\check{S}\else \v{S}\fi{}mejkal}}, \bibinfo {author} {\bibfnamefont {C.-J.}\ \bibnamefont {Kang}},\ and\ \bibinfo {author} {\bibfnamefont {C.}~\bibnamefont {Kim}},\ }\bibfield  {title} {\bibinfo {title} {Broken kramers degeneracy in altermagnetic mnte},\ }\href {https://doi.org/10.1103/PhysRevLett.132.036702} {\bibfield  {journal} {\bibinfo  {journal} {Phys. Rev. Lett.}\ }\textbf {\bibinfo {volume}
  {132}},\ \bibinfo {pages} {036702} (\bibinfo {year} {2024}{\natexlab{a}})}\BibitemShut {NoStop}%
\bibitem [{\citenamefont {Zhou}\ \emph {et~al.}(2024)\citenamefont {Zhou}, \citenamefont {Feng}, \citenamefont {Zhang}, \citenamefont {\ifmmode~\check{S}\else \v{S}\fi{}mejkal}, \citenamefont {Sinova}, \citenamefont {Mokrousov},\ and\ \citenamefont {Yao}}]{Zhou2024Crystal}%
  \BibitemOpen
  \bibfield  {author} {\bibinfo {author} {\bibfnamefont {X.}~\bibnamefont {Zhou}}, \bibinfo {author} {\bibfnamefont {W.}~\bibnamefont {Feng}}, \bibinfo {author} {\bibfnamefont {R.-W.}\ \bibnamefont {Zhang}}, \bibinfo {author} {\bibfnamefont {L.}~\bibnamefont {\ifmmode~\check{S}\else \v{S}\fi{}mejkal}}, \bibinfo {author} {\bibfnamefont {J.}~\bibnamefont {Sinova}}, \bibinfo {author} {\bibfnamefont {Y.}~\bibnamefont {Mokrousov}},\ and\ \bibinfo {author} {\bibfnamefont {Y.}~\bibnamefont {Yao}},\ }\bibfield  {title} {\bibinfo {title} {Crystal thermal transport in altermagnetic ${\mathrm{ruo}}_{2}$},\ }\href {https://doi.org/10.1103/PhysRevLett.132.056701} {\bibfield  {journal} {\bibinfo  {journal} {Phys. Rev. Lett.}\ }\textbf {\bibinfo {volume} {132}},\ \bibinfo {pages} {056701} (\bibinfo {year} {2024})}\BibitemShut {NoStop}%
\bibitem [{\citenamefont {Fedchenko}\ \emph {et~al.}(2024)\citenamefont {Fedchenko}, \citenamefont {Min{\'a}r}, \citenamefont {Akashdeep}, \citenamefont {D’Souza}, \citenamefont {Vasilyev}, \citenamefont {Tkach}, \citenamefont {Odenbreit}, \citenamefont {Nguyen}, \citenamefont {Kutnyakhov}, \citenamefont {Wind} \emph {et~al.}}]{fedchenko2024observation}%
  \BibitemOpen
  \bibfield  {author} {\bibinfo {author} {\bibfnamefont {O.}~\bibnamefont {Fedchenko}}, \bibinfo {author} {\bibfnamefont {J.}~\bibnamefont {Min{\'a}r}}, \bibinfo {author} {\bibfnamefont {A.}~\bibnamefont {Akashdeep}}, \bibinfo {author} {\bibfnamefont {S.~W.}\ \bibnamefont {D’Souza}}, \bibinfo {author} {\bibfnamefont {D.}~\bibnamefont {Vasilyev}}, \bibinfo {author} {\bibfnamefont {O.}~\bibnamefont {Tkach}}, \bibinfo {author} {\bibfnamefont {L.}~\bibnamefont {Odenbreit}}, \bibinfo {author} {\bibfnamefont {Q.}~\bibnamefont {Nguyen}}, \bibinfo {author} {\bibfnamefont {D.}~\bibnamefont {Kutnyakhov}}, \bibinfo {author} {\bibfnamefont {N.}~\bibnamefont {Wind}}, \emph {et~al.},\ }\bibfield  {title} {\bibinfo {title} {{Observation of time-reversal symmetry breaking in the band structure of altermagnetic RuO$_2$}},\ }\href {https://www.science.org/doi/10.1126/sciadv.adj4883} {\bibfield  {journal} {\bibinfo  {journal} {Sci. Adv.}\ }\textbf {\bibinfo {volume} {10}},\ \bibinfo {pages} {eadj4883} (\bibinfo {year}
  {2024})}\BibitemShut {NoStop}%
\bibitem [{\citenamefont {Han}\ \emph {et~al.}(2024)\citenamefont {Han}, \citenamefont {Fu}, \citenamefont {Peng}, \citenamefont {Cheng}, \citenamefont {Dai}, \citenamefont {Liu}, \citenamefont {Li}, \citenamefont {Zhang}, \citenamefont {Zhu}, \citenamefont {Bai} \emph {et~al.}}]{han2024electrical}%
  \BibitemOpen
  \bibfield  {author} {\bibinfo {author} {\bibfnamefont {L.}~\bibnamefont {Han}}, \bibinfo {author} {\bibfnamefont {X.}~\bibnamefont {Fu}}, \bibinfo {author} {\bibfnamefont {R.}~\bibnamefont {Peng}}, \bibinfo {author} {\bibfnamefont {X.}~\bibnamefont {Cheng}}, \bibinfo {author} {\bibfnamefont {J.}~\bibnamefont {Dai}}, \bibinfo {author} {\bibfnamefont {L.}~\bibnamefont {Liu}}, \bibinfo {author} {\bibfnamefont {Y.}~\bibnamefont {Li}}, \bibinfo {author} {\bibfnamefont {Y.}~\bibnamefont {Zhang}}, \bibinfo {author} {\bibfnamefont {W.}~\bibnamefont {Zhu}}, \bibinfo {author} {\bibfnamefont {H.}~\bibnamefont {Bai}}, \emph {et~al.},\ }\bibfield  {title} {\bibinfo {title} {{Electrical 180° switching of N{\'e}el vector in spin-splitting antiferromagnet}},\ }\href {https://www.science.org/doi/10.1126/sciadv.adn0479} {\bibfield  {journal} {\bibinfo  {journal} {Sci. Adv.}\ }\textbf {\bibinfo {volume} {10}},\ \bibinfo {pages} {eadn0479} (\bibinfo {year} {2024})}\BibitemShut {NoStop}%
\bibitem [{\citenamefont {Feng}\ \emph {et~al.}(2024)\citenamefont {Feng}, \citenamefont {Bai}, \citenamefont {Fan}, \citenamefont {Guo}, \citenamefont {Zhang}, \citenamefont {Chai}, \citenamefont {Wang}, \citenamefont {Xue}, \citenamefont {Song},\ and\ \citenamefont {Fan}}]{Feng2024Incommensurate}%
  \BibitemOpen
  \bibfield  {author} {\bibinfo {author} {\bibfnamefont {X.}~\bibnamefont {Feng}}, \bibinfo {author} {\bibfnamefont {H.}~\bibnamefont {Bai}}, \bibinfo {author} {\bibfnamefont {X.}~\bibnamefont {Fan}}, \bibinfo {author} {\bibfnamefont {M.}~\bibnamefont {Guo}}, \bibinfo {author} {\bibfnamefont {Z.}~\bibnamefont {Zhang}}, \bibinfo {author} {\bibfnamefont {G.}~\bibnamefont {Chai}}, \bibinfo {author} {\bibfnamefont {T.}~\bibnamefont {Wang}}, \bibinfo {author} {\bibfnamefont {D.}~\bibnamefont {Xue}}, \bibinfo {author} {\bibfnamefont {C.}~\bibnamefont {Song}},\ and\ \bibinfo {author} {\bibfnamefont {X.}~\bibnamefont {Fan}},\ }\bibfield  {title} {\bibinfo {title} {{Incommensurate Spin Density Wave in Antiferromagnetic ${\mathrm{RuO}}_{2}$ Evinced by Abnormal Spin Splitting Torque}},\ }\href {https://doi.org/10.1103/PhysRevLett.132.086701} {\bibfield  {journal} {\bibinfo  {journal} {Phys. Rev. Lett.}\ }\textbf {\bibinfo {volume} {132}},\ \bibinfo {pages} {086701} (\bibinfo {year} {2024})}\BibitemShut {NoStop}%
\bibitem [{\citenamefont {Osumi}\ \emph {et~al.}(2024)\citenamefont {Osumi}, \citenamefont {Souma}, \citenamefont {Aoyama}, \citenamefont {Yamauchi}, \citenamefont {Honma}, \citenamefont {Nakayama}, \citenamefont {Takahashi}, \citenamefont {Ohgushi},\ and\ \citenamefont {Sato}}]{Osumi2024Observation}%
  \BibitemOpen
  \bibfield  {author} {\bibinfo {author} {\bibfnamefont {T.}~\bibnamefont {Osumi}}, \bibinfo {author} {\bibfnamefont {S.}~\bibnamefont {Souma}}, \bibinfo {author} {\bibfnamefont {T.}~\bibnamefont {Aoyama}}, \bibinfo {author} {\bibfnamefont {K.}~\bibnamefont {Yamauchi}}, \bibinfo {author} {\bibfnamefont {A.}~\bibnamefont {Honma}}, \bibinfo {author} {\bibfnamefont {K.}~\bibnamefont {Nakayama}}, \bibinfo {author} {\bibfnamefont {T.}~\bibnamefont {Takahashi}}, \bibinfo {author} {\bibfnamefont {K.}~\bibnamefont {Ohgushi}},\ and\ \bibinfo {author} {\bibfnamefont {T.}~\bibnamefont {Sato}},\ }\bibfield  {title} {\bibinfo {title} {Observation of a giant band splitting in altermagnetic mnte},\ }\href {https://doi.org/10.1103/PhysRevB.109.115102} {\bibfield  {journal} {\bibinfo  {journal} {Phys. Rev. B}\ }\textbf {\bibinfo {volume} {109}},\ \bibinfo {pages} {115102} (\bibinfo {year} {2024})}\BibitemShut {NoStop}%
\bibitem [{\citenamefont {Chen}\ \emph {et~al.}(2024{\natexlab{a}})\citenamefont {Chen}, \citenamefont {Liu}, \citenamefont {Zhou}, \citenamefont {Meng}, \citenamefont {Wang}, \citenamefont {Duan}, \citenamefont {Zhao}, \citenamefont {Yan}, \citenamefont {Qin},\ and\ \citenamefont {Liu}}]{chen2024emerging}%
  \BibitemOpen
  \bibfield  {author} {\bibinfo {author} {\bibfnamefont {H.}~\bibnamefont {Chen}}, \bibinfo {author} {\bibfnamefont {L.}~\bibnamefont {Liu}}, \bibinfo {author} {\bibfnamefont {X.}~\bibnamefont {Zhou}}, \bibinfo {author} {\bibfnamefont {Z.}~\bibnamefont {Meng}}, \bibinfo {author} {\bibfnamefont {X.}~\bibnamefont {Wang}}, \bibinfo {author} {\bibfnamefont {Z.}~\bibnamefont {Duan}}, \bibinfo {author} {\bibfnamefont {G.}~\bibnamefont {Zhao}}, \bibinfo {author} {\bibfnamefont {H.}~\bibnamefont {Yan}}, \bibinfo {author} {\bibfnamefont {P.}~\bibnamefont {Qin}},\ and\ \bibinfo {author} {\bibfnamefont {Z.}~\bibnamefont {Liu}},\ }\bibfield  {title} {\bibinfo {title} {{Emerging Antiferromagnets for Spintronics}},\ }\href {https://doi.org/10.1002/adma.202310379} {\bibfield  {journal} {\bibinfo  {journal} {Adv. Mater.}\ ,\ \bibinfo {pages} {2310379}} (\bibinfo {year} {2024}{\natexlab{a}})}\BibitemShut {NoStop}%
\bibitem [{\citenamefont {Bhowal}\ and\ \citenamefont {Spaldin}(2024)}]{Bhowal2024Ferroically}%
  \BibitemOpen
  \bibfield  {author} {\bibinfo {author} {\bibfnamefont {S.}~\bibnamefont {Bhowal}}\ and\ \bibinfo {author} {\bibfnamefont {N.~A.}\ \bibnamefont {Spaldin}},\ }\bibfield  {title} {\bibinfo {title} {Ferroically ordered magnetic octupoles in $d$-wave altermagnets},\ }\href {https://doi.org/10.1103/PhysRevX.14.011019} {\bibfield  {journal} {\bibinfo  {journal} {Phys. Rev. X}\ }\textbf {\bibinfo {volume} {14}},\ \bibinfo {pages} {011019} (\bibinfo {year} {2024})}\BibitemShut {NoStop}%
\bibitem [{\citenamefont {Cheong}\ and\ \citenamefont {Huang}(2024)}]{cheong2024altermagnetism}%
  \BibitemOpen
  \bibfield  {author} {\bibinfo {author} {\bibfnamefont {S.-W.}\ \bibnamefont {Cheong}}\ and\ \bibinfo {author} {\bibfnamefont {F.-T.}\ \bibnamefont {Huang}},\ }\bibfield  {title} {\bibinfo {title} {Altermagnetism with non-collinear spins},\ }\href {https://doi.org/10.1038/s41535-024-00626-6} {\bibfield  {journal} {\bibinfo  {journal} {npj Quantum Materials}\ }\textbf {\bibinfo {volume} {9}},\ \bibinfo {pages} {13} (\bibinfo {year} {2024})}\BibitemShut {NoStop}%
\bibitem [{\citenamefont {Roig}\ \emph {et~al.}(2024)\citenamefont {Roig}, \citenamefont {Kreisel}, \citenamefont {Yu}, \citenamefont {Andersen},\ and\ \citenamefont {Agterberg}}]{roig2024minimal}%
  \BibitemOpen
  \bibfield  {author} {\bibinfo {author} {\bibfnamefont {M.}~\bibnamefont {Roig}}, \bibinfo {author} {\bibfnamefont {A.}~\bibnamefont {Kreisel}}, \bibinfo {author} {\bibfnamefont {Y.}~\bibnamefont {Yu}}, \bibinfo {author} {\bibfnamefont {B.~M.}\ \bibnamefont {Andersen}},\ and\ \bibinfo {author} {\bibfnamefont {D.~F.}\ \bibnamefont {Agterberg}},\ }\bibfield  {title} {\bibinfo {title} {Minimal models for altermagnetism},\ }\href {https://doi.org/10.1103/PhysRevB.110.144412} {\bibfield  {journal} {\bibinfo  {journal} {Phys. Rev. B}\ }\textbf {\bibinfo {volume} {110}},\ \bibinfo {pages} {144412} (\bibinfo {year} {2024})}\BibitemShut {NoStop}%
\bibitem [{\citenamefont {Yuan}\ \emph {et~al.}(2024)\citenamefont {Yuan}, \citenamefont {Georgescu},\ and\ \citenamefont {Rondinelli}}]{yuan2024non}%
  \BibitemOpen
  \bibfield  {author} {\bibinfo {author} {\bibfnamefont {L.-D.}\ \bibnamefont {Yuan}}, \bibinfo {author} {\bibfnamefont {A.~B.}\ \bibnamefont {Georgescu}},\ and\ \bibinfo {author} {\bibfnamefont {J.~M.}\ \bibnamefont {Rondinelli}},\ }\bibfield  {title} {\bibinfo {title} {Nonrelativistic spin splitting at the brillouin zone center in compensated magnets},\ }\href {https://doi.org/10.1103/PhysRevLett.133.216701} {\bibfield  {journal} {\bibinfo  {journal} {Phys. Rev. Lett.}\ }\textbf {\bibinfo {volume} {133}},\ \bibinfo {pages} {216701} (\bibinfo {year} {2024})}\BibitemShut {NoStop}%
\bibitem [{\citenamefont {He}\ \emph {et~al.}(2023)\citenamefont {He}, \citenamefont {Wang}, \citenamefont {Luo}, \citenamefont {Zeng}, \citenamefont {Chen},\ and\ \citenamefont {Tang}}]{He2023Nonrelativistic}%
  \BibitemOpen
  \bibfield  {author} {\bibinfo {author} {\bibfnamefont {R.}~\bibnamefont {He}}, \bibinfo {author} {\bibfnamefont {D.}~\bibnamefont {Wang}}, \bibinfo {author} {\bibfnamefont {N.}~\bibnamefont {Luo}}, \bibinfo {author} {\bibfnamefont {J.}~\bibnamefont {Zeng}}, \bibinfo {author} {\bibfnamefont {K.-Q.}\ \bibnamefont {Chen}},\ and\ \bibinfo {author} {\bibfnamefont {L.-M.}\ \bibnamefont {Tang}},\ }\bibfield  {title} {\bibinfo {title} {Nonrelativistic spin-momentum coupling in antiferromagnetic twisted bilayers},\ }\href {https://doi.org/10.1103/PhysRevLett.130.046401} {\bibfield  {journal} {\bibinfo  {journal} {Phys. Rev. Lett.}\ }\textbf {\bibinfo {volume} {130}},\ \bibinfo {pages} {046401} (\bibinfo {year} {2023})}\BibitemShut {NoStop}%
\bibitem [{\citenamefont {Zhu}\ \emph {et~al.}(2024)\citenamefont {Zhu}, \citenamefont {Chen}, \citenamefont {Liu}, \citenamefont {Liu}, \citenamefont {Liu}, \citenamefont {Zha}, \citenamefont {Qu}, \citenamefont {Hong}, \citenamefont {Li}, \citenamefont {Jiang} \emph {et~al.}}]{zhu2024Nature}%
  \BibitemOpen
  \bibfield  {author} {\bibinfo {author} {\bibfnamefont {Y.-P.}\ \bibnamefont {Zhu}}, \bibinfo {author} {\bibfnamefont {X.}~\bibnamefont {Chen}}, \bibinfo {author} {\bibfnamefont {X.-R.}\ \bibnamefont {Liu}}, \bibinfo {author} {\bibfnamefont {Y.}~\bibnamefont {Liu}}, \bibinfo {author} {\bibfnamefont {P.}~\bibnamefont {Liu}}, \bibinfo {author} {\bibfnamefont {H.}~\bibnamefont {Zha}}, \bibinfo {author} {\bibfnamefont {G.}~\bibnamefont {Qu}}, \bibinfo {author} {\bibfnamefont {C.}~\bibnamefont {Hong}}, \bibinfo {author} {\bibfnamefont {J.}~\bibnamefont {Li}}, \bibinfo {author} {\bibfnamefont {Z.}~\bibnamefont {Jiang}}, \emph {et~al.},\ }\bibfield  {title} {\bibinfo {title} {Observation of plaid-like spin splitting in a noncoplanar antiferromagnet},\ }\href {https://doi.org/10.1038/s41586-024-07023-w} {\bibfield  {journal} {\bibinfo  {journal} {Nature}\ }\textbf {\bibinfo {volume} {626}},\ \bibinfo {pages} {523} (\bibinfo {year} {2024})}\BibitemShut {NoStop}%
\bibitem [{\citenamefont {Krempask{\`y}}\ \emph {et~al.}(2024)\citenamefont {Krempask{\`y}}, \citenamefont {{\v{S}}mejkal}, \citenamefont {D’Souza}, \citenamefont {Hajlaoui}, \citenamefont {Springholz}, \citenamefont {Uhl{\'\i}{\v{r}}ov{\'a}}, \citenamefont {Alarab}, \citenamefont {Constantinou}, \citenamefont {Strocov}, \citenamefont {Usanov} \emph {et~al.}}]{krempasky2024Nature}%
  \BibitemOpen
  \bibfield  {author} {\bibinfo {author} {\bibfnamefont {J.}~\bibnamefont {Krempask{\`y}}}, \bibinfo {author} {\bibfnamefont {L.}~\bibnamefont {{\v{S}}mejkal}}, \bibinfo {author} {\bibfnamefont {S.}~\bibnamefont {D’Souza}}, \bibinfo {author} {\bibfnamefont {M.}~\bibnamefont {Hajlaoui}}, \bibinfo {author} {\bibfnamefont {G.}~\bibnamefont {Springholz}}, \bibinfo {author} {\bibfnamefont {K.}~\bibnamefont {Uhl{\'\i}{\v{r}}ov{\'a}}}, \bibinfo {author} {\bibfnamefont {F.}~\bibnamefont {Alarab}}, \bibinfo {author} {\bibfnamefont {P.}~\bibnamefont {Constantinou}}, \bibinfo {author} {\bibfnamefont {V.}~\bibnamefont {Strocov}}, \bibinfo {author} {\bibfnamefont {D.}~\bibnamefont {Usanov}}, \emph {et~al.},\ }\bibfield  {title} {\bibinfo {title} {Altermagnetic lifting of kramers spin degeneracy},\ }\href {https://doi.org/10.1038/s41586-023-06907-7} {\bibfield  {journal} {\bibinfo  {journal} {Nature}\ }\textbf {\bibinfo {volume} {626}},\ \bibinfo {pages} {517} (\bibinfo {year} {2024})}\BibitemShut {NoStop}%
\bibitem [{\citenamefont {Reimers}\ \emph {et~al.}(2024)\citenamefont {Reimers}, \citenamefont {Odenbreit}, \citenamefont {{\v{S}}mejkal}, \citenamefont {Strocov}, \citenamefont {Constantinou}, \citenamefont {Hellenes}, \citenamefont {Jaeschke~Ubiergo}, \citenamefont {Campos}, \citenamefont {Bharadwaj}, \citenamefont {Chakraborty} \emph {et~al.}}]{reimers2024NatComm}%
  \BibitemOpen
  \bibfield  {author} {\bibinfo {author} {\bibfnamefont {S.}~\bibnamefont {Reimers}}, \bibinfo {author} {\bibfnamefont {L.}~\bibnamefont {Odenbreit}}, \bibinfo {author} {\bibfnamefont {L.}~\bibnamefont {{\v{S}}mejkal}}, \bibinfo {author} {\bibfnamefont {V.~N.}\ \bibnamefont {Strocov}}, \bibinfo {author} {\bibfnamefont {P.}~\bibnamefont {Constantinou}}, \bibinfo {author} {\bibfnamefont {A.~B.}\ \bibnamefont {Hellenes}}, \bibinfo {author} {\bibfnamefont {R.}~\bibnamefont {Jaeschke~Ubiergo}}, \bibinfo {author} {\bibfnamefont {W.~H.}\ \bibnamefont {Campos}}, \bibinfo {author} {\bibfnamefont {V.~K.}\ \bibnamefont {Bharadwaj}}, \bibinfo {author} {\bibfnamefont {A.}~\bibnamefont {Chakraborty}}, \emph {et~al.},\ }\bibfield  {title} {\bibinfo {title} {Direct observation of altermagnetic band splitting in crsb thin films},\ }\href {https://doi.org/10.1038/s41467-024-46476-5} {\bibfield  {journal} {\bibinfo  {journal} {Nat. Commun.}\ }\textbf {\bibinfo {volume} {15}},\ \bibinfo {pages} {2116} (\bibinfo {year}
  {2024})}\BibitemShut {NoStop}%
\bibitem [{\citenamefont {Hu}\ \emph {et~al.}(2025)\citenamefont {Hu}, \citenamefont {Matsyshyn},\ and\ \citenamefont {Song}}]{JXHu25PRL}%
  \BibitemOpen
  \bibfield  {author} {\bibinfo {author} {\bibfnamefont {J.-X.}\ \bibnamefont {Hu}}, \bibinfo {author} {\bibfnamefont {O.}~\bibnamefont {Matsyshyn}},\ and\ \bibinfo {author} {\bibfnamefont {J.~C.~W.}\ \bibnamefont {Song}},\ }\bibfield  {title} {\bibinfo {title} {Nonlinear superconducting magnetoelectric effect},\ }\href {https://doi.org/10.1103/PhysRevLett.134.026001} {\bibfield  {journal} {\bibinfo  {journal} {Phys. Rev. Lett.}\ }\textbf {\bibinfo {volume} {134}},\ \bibinfo {pages} {026001} (\bibinfo {year} {2025})}\BibitemShut {NoStop}%
\bibitem [{\citenamefont {Zyuzin}(2024)}]{zyuzin2024magnetoelectric}%
  \BibitemOpen
  \bibfield  {author} {\bibinfo {author} {\bibfnamefont {A.~A.}\ \bibnamefont {Zyuzin}},\ }\bibfield  {title} {\bibinfo {title} {Magnetoelectric effect in superconductors with d-wave magnetization},\ }\href {https://doi.org/10.48550/arXiv.2402.15459} {\bibfield  {journal} {\bibinfo  {journal} {arXiv:2402.15459}\ } (\bibinfo {year} {2024})}\BibitemShut {NoStop}%
\bibitem [{\citenamefont {Zhu}\ \emph {et~al.}(2025)\citenamefont {Zhu}, \citenamefont {Huo}, \citenamefont {Feng}, \citenamefont {Zhang}, \citenamefont {Yang},\ and\ \citenamefont {Guo}}]{XCZhu25PRL}%
  \BibitemOpen
  \bibfield  {author} {\bibinfo {author} {\bibfnamefont {X.}~\bibnamefont {Zhu}}, \bibinfo {author} {\bibfnamefont {X.}~\bibnamefont {Huo}}, \bibinfo {author} {\bibfnamefont {S.}~\bibnamefont {Feng}}, \bibinfo {author} {\bibfnamefont {S.-B.}\ \bibnamefont {Zhang}}, \bibinfo {author} {\bibfnamefont {S.~A.}\ \bibnamefont {Yang}},\ and\ \bibinfo {author} {\bibfnamefont {H.}~\bibnamefont {Guo}},\ }\bibfield  {title} {\bibinfo {title} {{Design of Altermagnetic Models from Spin Clusters}},\ }\href {https://doi.org/10.1103/PhysRevLett.134.166701} {\bibfield  {journal} {\bibinfo  {journal} {Phys. Rev. Lett.}\ }\textbf {\bibinfo {volume} {134}},\ \bibinfo {pages} {166701} (\bibinfo {year} {2025})}\BibitemShut {NoStop}%
\bibitem [{\citenamefont {Duan}\ \emph {et~al.}(2025)\citenamefont {Duan}, \citenamefont {Zhang}, \citenamefont {Zhu}, \citenamefont {Liu}, \citenamefont {Zhang}, \citenamefont {{\v{Z}}uti{\'c}},\ and\ \citenamefont {Zhou}}]{Duan2025prl}%
  \BibitemOpen
  \bibfield  {author} {\bibinfo {author} {\bibfnamefont {X.}~\bibnamefont {Duan}}, \bibinfo {author} {\bibfnamefont {J.}~\bibnamefont {Zhang}}, \bibinfo {author} {\bibfnamefont {Z.}~\bibnamefont {Zhu}}, \bibinfo {author} {\bibfnamefont {Y.}~\bibnamefont {Liu}}, \bibinfo {author} {\bibfnamefont {Z.}~\bibnamefont {Zhang}}, \bibinfo {author} {\bibfnamefont {I.}~\bibnamefont {{\v{Z}}uti{\'c}}},\ and\ \bibinfo {author} {\bibfnamefont {T.}~\bibnamefont {Zhou}},\ }\bibfield  {title} {\bibinfo {title} {{Antiferroelectric altermagnets: Antiferroelectricity alters magnets}},\ }\href {https://journals.aps.org/prl/abstract/10.1103/PhysRevLett.134.106801} {\bibfield  {journal} {\bibinfo  {journal} {Phys. Rev. Lett.}\ }\textbf {\bibinfo {volume} {134}},\ \bibinfo {pages} {106801} (\bibinfo {year} {2025})}\BibitemShut {NoStop}%
\bibitem [{\citenamefont {Gu}\ \emph {et~al.}(2025)\citenamefont {Gu}, \citenamefont {Liu}, \citenamefont {Zhu}, \citenamefont {Yananose}, \citenamefont {Chen}, \citenamefont {Hu}, \citenamefont {Stroppa},\ and\ \citenamefont {Liu}}]{MQGu25PRL}%
  \BibitemOpen
  \bibfield  {author} {\bibinfo {author} {\bibfnamefont {M.}~\bibnamefont {Gu}}, \bibinfo {author} {\bibfnamefont {Y.}~\bibnamefont {Liu}}, \bibinfo {author} {\bibfnamefont {H.}~\bibnamefont {Zhu}}, \bibinfo {author} {\bibfnamefont {K.}~\bibnamefont {Yananose}}, \bibinfo {author} {\bibfnamefont {X.}~\bibnamefont {Chen}}, \bibinfo {author} {\bibfnamefont {Y.}~\bibnamefont {Hu}}, \bibinfo {author} {\bibfnamefont {A.}~\bibnamefont {Stroppa}},\ and\ \bibinfo {author} {\bibfnamefont {Q.}~\bibnamefont {Liu}},\ }\bibfield  {title} {\bibinfo {title} {{Ferroelectric switchable altermagnetism}},\ }\href {https://journals.aps.org/prl/abstract/10.1103/PhysRevLett.134.106802} {\bibfield  {journal} {\bibinfo  {journal} {Phys. Rev. Lett.}\ }\textbf {\bibinfo {volume} {134}},\ \bibinfo {pages} {106802} (\bibinfo {year} {2025})}\BibitemShut {NoStop}%
\bibitem [{\citenamefont {Chen}\ \emph {et~al.}(2025)\citenamefont {Chen}, \citenamefont {Liu}, \citenamefont {Lu},\ and\ \citenamefont {Xie}}]{YYChen25PRL}%
  \BibitemOpen
  \bibfield  {author} {\bibinfo {author} {\bibfnamefont {Y.}~\bibnamefont {Chen}}, \bibinfo {author} {\bibfnamefont {X.}~\bibnamefont {Liu}}, \bibinfo {author} {\bibfnamefont {H.-Z.}\ \bibnamefont {Lu}},\ and\ \bibinfo {author} {\bibfnamefont {X.~C.}\ \bibnamefont {Xie}},\ }\bibfield  {title} {\bibinfo {title} {{Electrical Switching of Altermagnetism}},\ }\href {https://doi.org/10.1103/zm5y-vy41} {\bibfield  {journal} {\bibinfo  {journal} {Phys. Rev. Lett.}\ }\textbf {\bibinfo {volume} {135}},\ \bibinfo {pages} {016701} (\bibinfo {year} {2025})}\BibitemShut {NoStop}%
\bibitem [{\citenamefont {Wang}\ and\ \citenamefont {et~al.}(2025)}]{ZMWang25PRL}%
  \BibitemOpen
  \bibfield  {author} {\bibinfo {author} {\bibfnamefont {Z.-M.}\ \bibnamefont {Wang}}\ and\ \bibinfo {author} {\bibnamefont {et~al.}},\ }\bibfield  {title} {\bibinfo {title} {{Spin-Orbital Altermagnetism}},\ }\href {https://journals.aps.org/prl/accepted/10.1103/cjzw-j4v7} {\bibfield  {journal} {\bibinfo  {journal} {Phys. Rew. Lett.}\ } (\bibinfo {year} {2025})}\BibitemShut {NoStop}%
\bibitem [{\citenamefont {Jungwirth}\ \emph {et~al.}(2016)\citenamefont {Jungwirth}, \citenamefont {Marti}, \citenamefont {Wadley},\ and\ \citenamefont {Wunderlich}}]{Jungwirth2016NNanoTech}%
  \BibitemOpen
  \bibfield  {author} {\bibinfo {author} {\bibfnamefont {T.}~\bibnamefont {Jungwirth}}, \bibinfo {author} {\bibfnamefont {X.}~\bibnamefont {Marti}}, \bibinfo {author} {\bibfnamefont {P.}~\bibnamefont {Wadley}},\ and\ \bibinfo {author} {\bibfnamefont {J.}~\bibnamefont {Wunderlich}},\ }\bibfield  {title} {\bibinfo {title} {Antiferromagnetic spintronics},\ }\href {https://doi.org/10.1038/nnano.2016.18} {\bibfield  {journal} {\bibinfo  {journal} {Nat. Nanotech.}\ }\textbf {\bibinfo {volume} {11}},\ \bibinfo {pages} {231} (\bibinfo {year} {2016})}\BibitemShut {NoStop}%
\bibitem [{\citenamefont {{\v{S}}mejkal}\ \emph {et~al.}(2022)\citenamefont {{\v{S}}mejkal}, \citenamefont {MacDonald}, \citenamefont {Sinova}, \citenamefont {Nakatsuji},\ and\ \citenamefont {Jungwirth}}]{vsmejkal2022anomalous}%
  \BibitemOpen
  \bibfield  {author} {\bibinfo {author} {\bibfnamefont {L.}~\bibnamefont {{\v{S}}mejkal}}, \bibinfo {author} {\bibfnamefont {A.~H.}\ \bibnamefont {MacDonald}}, \bibinfo {author} {\bibfnamefont {J.}~\bibnamefont {Sinova}}, \bibinfo {author} {\bibfnamefont {S.}~\bibnamefont {Nakatsuji}},\ and\ \bibinfo {author} {\bibfnamefont {T.}~\bibnamefont {Jungwirth}},\ }\bibfield  {title} {\bibinfo {title} {Anomalous hall antiferromagnets},\ }\href {https://doi.org/10.1038/s41578-022-00430-3} {\bibfield  {journal} {\bibinfo  {journal} {Nat. Rev. Mater.}\ }\textbf {\bibinfo {volume} {7}},\ \bibinfo {pages} {482} (\bibinfo {year} {2022})}\BibitemShut {NoStop}%
\bibitem [{\citenamefont {Baltz}\ \emph {et~al.}(2018)\citenamefont {Baltz}, \citenamefont {Manchon}, \citenamefont {Tsoi}, \citenamefont {Moriyama}, \citenamefont {Ono},\ and\ \citenamefont {Tserkovnyak}}]{Baltz18RMP}%
  \BibitemOpen
  \bibfield  {author} {\bibinfo {author} {\bibfnamefont {V.}~\bibnamefont {Baltz}}, \bibinfo {author} {\bibfnamefont {A.}~\bibnamefont {Manchon}}, \bibinfo {author} {\bibfnamefont {M.}~\bibnamefont {Tsoi}}, \bibinfo {author} {\bibfnamefont {T.}~\bibnamefont {Moriyama}}, \bibinfo {author} {\bibfnamefont {T.}~\bibnamefont {Ono}},\ and\ \bibinfo {author} {\bibfnamefont {Y.}~\bibnamefont {Tserkovnyak}},\ }\bibfield  {title} {\bibinfo {title} {Antiferromagnetic spintronics},\ }\href {https://doi.org/10.1103/RevModPhys.90.015005} {\bibfield  {journal} {\bibinfo  {journal} {Rev. Mod. Phys.}\ }\textbf {\bibinfo {volume} {90}},\ \bibinfo {pages} {015005} (\bibinfo {year} {2018})}\BibitemShut {NoStop}%
\bibitem [{\citenamefont {Manchon}\ \emph {et~al.}(2019)\citenamefont {Manchon}, \citenamefont {\ifmmode~\check{Z}\else \v{Z}\fi{}elezn\'y}, \citenamefont {Miron}, \citenamefont {Jungwirth}, \citenamefont {Sinova}, \citenamefont {Thiaville}, \citenamefont {Garello},\ and\ \citenamefont {Gambardella}}]{manchon2019RMP}%
  \BibitemOpen
  \bibfield  {author} {\bibinfo {author} {\bibfnamefont {A.}~\bibnamefont {Manchon}}, \bibinfo {author} {\bibfnamefont {J.}~\bibnamefont {\ifmmode~\check{Z}\else \v{Z}\fi{}elezn\'y}}, \bibinfo {author} {\bibfnamefont {I.~M.}\ \bibnamefont {Miron}}, \bibinfo {author} {\bibfnamefont {T.}~\bibnamefont {Jungwirth}}, \bibinfo {author} {\bibfnamefont {J.}~\bibnamefont {Sinova}}, \bibinfo {author} {\bibfnamefont {A.}~\bibnamefont {Thiaville}}, \bibinfo {author} {\bibfnamefont {K.}~\bibnamefont {Garello}},\ and\ \bibinfo {author} {\bibfnamefont {P.}~\bibnamefont {Gambardella}},\ }\bibfield  {title} {\bibinfo {title} {Current-induced spin-orbit torques in ferromagnetic and antiferromagnetic systems},\ }\href {https://doi.org/10.1103/RevModPhys.91.035004} {\bibfield  {journal} {\bibinfo  {journal} {Rev. Mod. Phys.}\ }\textbf {\bibinfo {volume} {91}},\ \bibinfo {pages} {035004} (\bibinfo {year} {2019})}\BibitemShut {NoStop}%
\bibitem [{\citenamefont {Jiang}\ \emph {et~al.}(2024)\citenamefont {Jiang}, \citenamefont {Song}, \citenamefont {Zhu}, \citenamefont {Fang}, \citenamefont {Weng}, \citenamefont {Liu}, \citenamefont {Yang},\ and\ \citenamefont {Fang}}]{jiang2023enumeration}%
  \BibitemOpen
  \bibfield  {author} {\bibinfo {author} {\bibfnamefont {Y.}~\bibnamefont {Jiang}}, \bibinfo {author} {\bibfnamefont {Z.}~\bibnamefont {Song}}, \bibinfo {author} {\bibfnamefont {T.}~\bibnamefont {Zhu}}, \bibinfo {author} {\bibfnamefont {Z.}~\bibnamefont {Fang}}, \bibinfo {author} {\bibfnamefont {H.}~\bibnamefont {Weng}}, \bibinfo {author} {\bibfnamefont {Z.-X.}\ \bibnamefont {Liu}}, \bibinfo {author} {\bibfnamefont {J.}~\bibnamefont {Yang}},\ and\ \bibinfo {author} {\bibfnamefont {C.}~\bibnamefont {Fang}},\ }\bibfield  {title} {\bibinfo {title} {Enumeration of spin-space groups: Toward a complete description of symmetries of magnetic orders},\ }\href {https://doi.org/10.1103/PhysRevX.14.031039} {\bibfield  {journal} {\bibinfo  {journal} {Phys. Rev. X}\ }\textbf {\bibinfo {volume} {14}},\ \bibinfo {pages} {031039} (\bibinfo {year} {2024})}\BibitemShut {NoStop}%
\bibitem [{\citenamefont {Chen}\ \emph {et~al.}(2024{\natexlab{b}})\citenamefont {Chen}, \citenamefont {Ren}, \citenamefont {Zhu}, \citenamefont {Yu}, \citenamefont {Zhang}, \citenamefont {Liu}, \citenamefont {Li}, \citenamefont {Liu}, \citenamefont {Li},\ and\ \citenamefont {Liu}}]{ren2023enumeration}%
  \BibitemOpen
  \bibfield  {author} {\bibinfo {author} {\bibfnamefont {X.}~\bibnamefont {Chen}}, \bibinfo {author} {\bibfnamefont {J.}~\bibnamefont {Ren}}, \bibinfo {author} {\bibfnamefont {Y.}~\bibnamefont {Zhu}}, \bibinfo {author} {\bibfnamefont {Y.}~\bibnamefont {Yu}}, \bibinfo {author} {\bibfnamefont {A.}~\bibnamefont {Zhang}}, \bibinfo {author} {\bibfnamefont {P.}~\bibnamefont {Liu}}, \bibinfo {author} {\bibfnamefont {J.}~\bibnamefont {Li}}, \bibinfo {author} {\bibfnamefont {Y.}~\bibnamefont {Liu}}, \bibinfo {author} {\bibfnamefont {C.}~\bibnamefont {Li}},\ and\ \bibinfo {author} {\bibfnamefont {Q.}~\bibnamefont {Liu}},\ }\bibfield  {title} {\bibinfo {title} {Enumeration and representation theory of spin space groups},\ }\href {https://doi.org/10.1103/PhysRevX.14.031038} {\bibfield  {journal} {\bibinfo  {journal} {Phys. Rev. X}\ }\textbf {\bibinfo {volume} {14}},\ \bibinfo {pages} {031038} (\bibinfo {year} {2024}{\natexlab{b}})}\BibitemShut {NoStop}%
\bibitem [{\citenamefont {Xiao}\ \emph {et~al.}(2024)\citenamefont {Xiao}, \citenamefont {Zhao}, \citenamefont {Li}, \citenamefont {Shindou},\ and\ \citenamefont {Song}}]{xiao2023spin}%
  \BibitemOpen
  \bibfield  {author} {\bibinfo {author} {\bibfnamefont {Z.}~\bibnamefont {Xiao}}, \bibinfo {author} {\bibfnamefont {J.}~\bibnamefont {Zhao}}, \bibinfo {author} {\bibfnamefont {Y.}~\bibnamefont {Li}}, \bibinfo {author} {\bibfnamefont {R.}~\bibnamefont {Shindou}},\ and\ \bibinfo {author} {\bibfnamefont {Z.-D.}\ \bibnamefont {Song}},\ }\bibfield  {title} {\bibinfo {title} {Spin space groups: Full classification and applications},\ }\href {https://doi.org/10.1103/PhysRevX.14.031037} {\bibfield  {journal} {\bibinfo  {journal} {Phys. Rev. X}\ }\textbf {\bibinfo {volume} {14}},\ \bibinfo {pages} {031037} (\bibinfo {year} {2024})}\BibitemShut {NoStop}%
\bibitem [{\citenamefont {Liu}\ \emph {et~al.}(2018)\citenamefont {Liu}, \citenamefont {Chen}, \citenamefont {Wang}, \citenamefont {Liu}, \citenamefont {Wang}, \citenamefont {Feng}, \citenamefont {Yan}, \citenamefont {Wang}, \citenamefont {Jiang}, \citenamefont {Coey} \emph {et~al.}}]{liu2018electrical}%
  \BibitemOpen
  \bibfield  {author} {\bibinfo {author} {\bibfnamefont {Z.}~\bibnamefont {Liu}}, \bibinfo {author} {\bibfnamefont {H.}~\bibnamefont {Chen}}, \bibinfo {author} {\bibfnamefont {J.}~\bibnamefont {Wang}}, \bibinfo {author} {\bibfnamefont {J.}~\bibnamefont {Liu}}, \bibinfo {author} {\bibfnamefont {K.}~\bibnamefont {Wang}}, \bibinfo {author} {\bibfnamefont {Z.}~\bibnamefont {Feng}}, \bibinfo {author} {\bibfnamefont {H.}~\bibnamefont {Yan}}, \bibinfo {author} {\bibfnamefont {X.}~\bibnamefont {Wang}}, \bibinfo {author} {\bibfnamefont {C.}~\bibnamefont {Jiang}}, \bibinfo {author} {\bibfnamefont {J.}~\bibnamefont {Coey}}, \emph {et~al.},\ }\bibfield  {title} {\bibinfo {title} {{Electrical switching of the topological anomalous Hall effect in a non-collinear antiferromagnet above room temperature}},\ }\href {https://doi.org/10.1038/s41928-018-0040-1} {\bibfield  {journal} {\bibinfo  {journal} {Nat. Electron.}\ }\textbf {\bibinfo {volume} {1}},\ \bibinfo {pages} {172} (\bibinfo {year} {2018})}\BibitemShut {NoStop}%
\bibitem [{\citenamefont {Kiyohara}\ \emph {et~al.}(2016)\citenamefont {Kiyohara}, \citenamefont {Tomita},\ and\ \citenamefont {Nakatsuji}}]{Kiyohara2016Giant}%
  \BibitemOpen
  \bibfield  {author} {\bibinfo {author} {\bibfnamefont {N.}~\bibnamefont {Kiyohara}}, \bibinfo {author} {\bibfnamefont {T.}~\bibnamefont {Tomita}},\ and\ \bibinfo {author} {\bibfnamefont {S.}~\bibnamefont {Nakatsuji}},\ }\bibfield  {title} {\bibinfo {title} {{Giant Anomalous Hall Effect in the Chiral Antiferromagnet Mn$_{3}\mathrm{Ge}$}},\ }\href {https://doi.org/10.1103/PhysRevApplied.5.064009} {\bibfield  {journal} {\bibinfo  {journal} {Phys. Rev. Appl.}\ }\textbf {\bibinfo {volume} {5}},\ \bibinfo {pages} {064009} (\bibinfo {year} {2016})}\BibitemShut {NoStop}%
\bibitem [{\citenamefont {Nayak}\ \emph {et~al.}(2016)\citenamefont {Nayak}, \citenamefont {Fischer}, \citenamefont {Sun}, \citenamefont {Yan}, \citenamefont {Karel}, \citenamefont {Komarek}, \citenamefont {Shekhar}, \citenamefont {Kumar}, \citenamefont {Schnelle}, \citenamefont {K{\"u}bler} \emph {et~al.}}]{nayak2016large}%
  \BibitemOpen
  \bibfield  {author} {\bibinfo {author} {\bibfnamefont {A.~K.}\ \bibnamefont {Nayak}}, \bibinfo {author} {\bibfnamefont {J.~E.}\ \bibnamefont {Fischer}}, \bibinfo {author} {\bibfnamefont {Y.}~\bibnamefont {Sun}}, \bibinfo {author} {\bibfnamefont {B.}~\bibnamefont {Yan}}, \bibinfo {author} {\bibfnamefont {J.}~\bibnamefont {Karel}}, \bibinfo {author} {\bibfnamefont {A.~C.}\ \bibnamefont {Komarek}}, \bibinfo {author} {\bibfnamefont {C.}~\bibnamefont {Shekhar}}, \bibinfo {author} {\bibfnamefont {N.}~\bibnamefont {Kumar}}, \bibinfo {author} {\bibfnamefont {W.}~\bibnamefont {Schnelle}}, \bibinfo {author} {\bibfnamefont {J.}~\bibnamefont {K{\"u}bler}}, \emph {et~al.},\ }\bibfield  {title} {\bibinfo {title} {{Large anomalous Hall effect driven by a nonvanishing Berry curvature in the noncolinear antiferromagnet Mn$_3$Ge}},\ }\href {https://www.science.org/doi/10.1126/sciadv.1501870} {\bibfield  {journal} {\bibinfo  {journal} {Sci. Adv.}\ }\textbf {\bibinfo {volume} {2}},\ \bibinfo {pages} {e1501870} (\bibinfo {year}
  {2016})}\BibitemShut {NoStop}%
\bibitem [{\citenamefont {Liu}\ \emph {et~al.}(2017)\citenamefont {Liu}, \citenamefont {Zhang}, \citenamefont {Liu}, \citenamefont {Ding}, \citenamefont {Liu}, \citenamefont {Jafri}, \citenamefont {Hou}, \citenamefont {Wang}, \citenamefont {Ma},\ and\ \citenamefont {Wu}}]{liu2017SciRep}%
  \BibitemOpen
  \bibfield  {author} {\bibinfo {author} {\bibfnamefont {Z.}~\bibnamefont {Liu}}, \bibinfo {author} {\bibfnamefont {Y.}~\bibnamefont {Zhang}}, \bibinfo {author} {\bibfnamefont {G.}~\bibnamefont {Liu}}, \bibinfo {author} {\bibfnamefont {B.}~\bibnamefont {Ding}}, \bibinfo {author} {\bibfnamefont {E.}~\bibnamefont {Liu}}, \bibinfo {author} {\bibfnamefont {H.~M.}\ \bibnamefont {Jafri}}, \bibinfo {author} {\bibfnamefont {Z.}~\bibnamefont {Hou}}, \bibinfo {author} {\bibfnamefont {W.}~\bibnamefont {Wang}}, \bibinfo {author} {\bibfnamefont {X.}~\bibnamefont {Ma}},\ and\ \bibinfo {author} {\bibfnamefont {G.}~\bibnamefont {Wu}},\ }\bibfield  {title} {\bibinfo {title} {Transition from anomalous hall effect to topological hall effect in hexagonal non-collinear magnet mn3ga},\ }\href {https://doi.org/10.1038/s41598-017-00621-x} {\bibfield  {journal} {\bibinfo  {journal} {Scientific Reports}\ }\textbf {\bibinfo {volume} {7}},\ \bibinfo {pages} {515} (\bibinfo {year} {2017})}\BibitemShut {NoStop}%
\bibitem [{\citenamefont {Song}\ \emph {et~al.}(2024)\citenamefont {Song}, \citenamefont {Zhou}, \citenamefont {Li}, \citenamefont {Ding}, \citenamefont {Li}, \citenamefont {Xi}, \citenamefont {Yao}, \citenamefont {Lau},\ and\ \citenamefont {Wang}}]{Song2024AFM}%
  \BibitemOpen
  \bibfield  {author} {\bibinfo {author} {\bibfnamefont {L.}~\bibnamefont {Song}}, \bibinfo {author} {\bibfnamefont {F.}~\bibnamefont {Zhou}}, \bibinfo {author} {\bibfnamefont {H.}~\bibnamefont {Li}}, \bibinfo {author} {\bibfnamefont {B.}~\bibnamefont {Ding}}, \bibinfo {author} {\bibfnamefont {X.}~\bibnamefont {Li}}, \bibinfo {author} {\bibfnamefont {X.}~\bibnamefont {Xi}}, \bibinfo {author} {\bibfnamefont {Y.}~\bibnamefont {Yao}}, \bibinfo {author} {\bibfnamefont {Y.-C.}\ \bibnamefont {Lau}},\ and\ \bibinfo {author} {\bibfnamefont {W.}~\bibnamefont {Wang}},\ }\bibfield  {title} {\bibinfo {title} {{Large Anomalous Hall Effect at Room Temperature in a Fermi-Level-Tuned Kagome Antiferromagnet}},\ }\href {https://doi.org/10.1002/adfm.202316588} {\bibfield  {journal} {\bibinfo  {journal} {Adv. Funct. Mater.}\ }\textbf {\bibinfo {volume} {2024}},\ \bibinfo {pages} {2316588} (\bibinfo {year} {2024})}\BibitemShut {NoStop}%
\bibitem [{\citenamefont {Zhang}\ \emph {et~al.}(2016)\citenamefont {Zhang}, \citenamefont {Han}, \citenamefont {Yang}, \citenamefont {Sun}, \citenamefont {Zhang}, \citenamefont {Yan},\ and\ \citenamefont {Parkin}}]{zhang2016giant}%
  \BibitemOpen
  \bibfield  {author} {\bibinfo {author} {\bibfnamefont {W.}~\bibnamefont {Zhang}}, \bibinfo {author} {\bibfnamefont {W.}~\bibnamefont {Han}}, \bibinfo {author} {\bibfnamefont {S.-H.}\ \bibnamefont {Yang}}, \bibinfo {author} {\bibfnamefont {Y.}~\bibnamefont {Sun}}, \bibinfo {author} {\bibfnamefont {Y.}~\bibnamefont {Zhang}}, \bibinfo {author} {\bibfnamefont {B.}~\bibnamefont {Yan}},\ and\ \bibinfo {author} {\bibfnamefont {S.~S.}\ \bibnamefont {Parkin}},\ }\bibfield  {title} {\bibinfo {title} {{Giant facet-dependent spin-orbit torque and spin Hall conductivity in the triangular antiferromagnet IrMn$_3$}},\ }\href {https://www.science.org/doi/10.1126/sciadv.1600759} {\bibfield  {journal} {\bibinfo  {journal} {Sci. Adv.}\ }\textbf {\bibinfo {volume} {2}},\ \bibinfo {pages} {e1600759} (\bibinfo {year} {2016})}\BibitemShut {NoStop}%
\bibitem [{\citenamefont {Jeon}\ \emph {et~al.}(2021)\citenamefont {Jeon}, \citenamefont {Hazra}, \citenamefont {Cho}, \citenamefont {Chakraborty}, \citenamefont {Jeon}, \citenamefont {Han}, \citenamefont {Meyerheim}, \citenamefont {Kontos},\ and\ \citenamefont {Parkin}}]{jeon2021long}%
  \BibitemOpen
  \bibfield  {author} {\bibinfo {author} {\bibfnamefont {K.-R.}\ \bibnamefont {Jeon}}, \bibinfo {author} {\bibfnamefont {B.~K.}\ \bibnamefont {Hazra}}, \bibinfo {author} {\bibfnamefont {K.}~\bibnamefont {Cho}}, \bibinfo {author} {\bibfnamefont {A.}~\bibnamefont {Chakraborty}}, \bibinfo {author} {\bibfnamefont {J.-C.}\ \bibnamefont {Jeon}}, \bibinfo {author} {\bibfnamefont {H.}~\bibnamefont {Han}}, \bibinfo {author} {\bibfnamefont {H.~L.}\ \bibnamefont {Meyerheim}}, \bibinfo {author} {\bibfnamefont {T.}~\bibnamefont {Kontos}},\ and\ \bibinfo {author} {\bibfnamefont {S.~S.}\ \bibnamefont {Parkin}},\ }\bibfield  {title} {\bibinfo {title} {{Long-range supercurrents through a chiral non-collinear antiferromagnet in lateral Josephson junctions}},\ }\href {https://doi.org/10.1038/s41563-021-01061-9} {\bibfield  {journal} {\bibinfo  {journal} {Nat. Mater.}\ }\textbf {\bibinfo {volume} {20}},\ \bibinfo {pages} {1358} (\bibinfo {year} {2021})}\BibitemShut {NoStop}%
\bibitem [{\citenamefont {Jeon}\ \emph {et~al.}(2023)\citenamefont {Jeon}, \citenamefont {Hazra}, \citenamefont {Kim}, \citenamefont {Jeon}, \citenamefont {Han}, \citenamefont {Meyerheim}, \citenamefont {Kontos}, \citenamefont {Cottet},\ and\ \citenamefont {Parkin}}]{jeon2023chiral}%
  \BibitemOpen
  \bibfield  {author} {\bibinfo {author} {\bibfnamefont {K.-R.}\ \bibnamefont {Jeon}}, \bibinfo {author} {\bibfnamefont {B.~K.}\ \bibnamefont {Hazra}}, \bibinfo {author} {\bibfnamefont {J.-K.}\ \bibnamefont {Kim}}, \bibinfo {author} {\bibfnamefont {J.-C.}\ \bibnamefont {Jeon}}, \bibinfo {author} {\bibfnamefont {H.}~\bibnamefont {Han}}, \bibinfo {author} {\bibfnamefont {H.~L.}\ \bibnamefont {Meyerheim}}, \bibinfo {author} {\bibfnamefont {T.}~\bibnamefont {Kontos}}, \bibinfo {author} {\bibfnamefont {A.}~\bibnamefont {Cottet}},\ and\ \bibinfo {author} {\bibfnamefont {S.~S.}\ \bibnamefont {Parkin}},\ }\bibfield  {title} {\bibinfo {title} {{Chiral antiferromagnetic Josephson junctions as spin-triplet supercurrent spin valves and dc SQUIDs}},\ }\href {https://doi.org/10.1038/s41565-023-01336-z} {\bibfield  {journal} {\bibinfo  {journal} {Nat. Nanotech.}\ }\textbf {\bibinfo {volume} {18}},\ \bibinfo {pages} {747} (\bibinfo {year} {2023})}\BibitemShut {NoStop}%
\bibitem [{\citenamefont {Bergeret}\ \emph {et~al.}(2001)\citenamefont {Bergeret}, \citenamefont {Volkov},\ and\ \citenamefont {Efetov}}]{Bergeret2001LongRange}%
  \BibitemOpen
  \bibfield  {author} {\bibinfo {author} {\bibfnamefont {F.~S.}\ \bibnamefont {Bergeret}}, \bibinfo {author} {\bibfnamefont {A.~F.}\ \bibnamefont {Volkov}},\ and\ \bibinfo {author} {\bibfnamefont {K.~B.}\ \bibnamefont {Efetov}},\ }\bibfield  {title} {\bibinfo {title} {Long-range proximity effects in superconductor-ferromagnet structures},\ }\href {https://doi.org/10.1103/PhysRevLett.86.4096} {\bibfield  {journal} {\bibinfo  {journal} {Phys. Rev. Lett.}\ }\textbf {\bibinfo {volume} {86}},\ \bibinfo {pages} {4096} (\bibinfo {year} {2001})}\BibitemShut {NoStop}%
\bibitem [{\citenamefont {Volkov}\ \emph {et~al.}(2003)\citenamefont {Volkov}, \citenamefont {Bergeret},\ and\ \citenamefont {Efetov}}]{Volkov03PRL}%
  \BibitemOpen
  \bibfield  {author} {\bibinfo {author} {\bibfnamefont {A.~F.}\ \bibnamefont {Volkov}}, \bibinfo {author} {\bibfnamefont {F.~S.}\ \bibnamefont {Bergeret}},\ and\ \bibinfo {author} {\bibfnamefont {K.~B.}\ \bibnamefont {Efetov}},\ }\bibfield  {title} {\bibinfo {title} {Odd triplet superconductivity in superconductor-ferromagnet multilayered structures},\ }\href {https://doi.org/10.1103/PhysRevLett.90.117006} {\bibfield  {journal} {\bibinfo  {journal} {Phys. Rev. Lett.}\ }\textbf {\bibinfo {volume} {90}},\ \bibinfo {pages} {117006} (\bibinfo {year} {2003})}\BibitemShut {NoStop}%
\bibitem [{\citenamefont {Bergeret}\ \emph {et~al.}(2005)\citenamefont {Bergeret}, \citenamefont {Volkov},\ and\ \citenamefont {Efetov}}]{Bergeret05RMP}%
  \BibitemOpen
  \bibfield  {author} {\bibinfo {author} {\bibfnamefont {F.~S.}\ \bibnamefont {Bergeret}}, \bibinfo {author} {\bibfnamefont {A.~F.}\ \bibnamefont {Volkov}},\ and\ \bibinfo {author} {\bibfnamefont {K.~B.}\ \bibnamefont {Efetov}},\ }\bibfield  {title} {\bibinfo {title} {Odd triplet superconductivity and related phenomena in superconductor-ferromagnet structures},\ }\href {https://doi.org/10.1103/RevModPhys.77.1321} {\bibfield  {journal} {\bibinfo  {journal} {Rev. Mod. Phys.}\ }\textbf {\bibinfo {volume} {77}},\ \bibinfo {pages} {1321} (\bibinfo {year} {2005})}\BibitemShut {NoStop}%
\bibitem [{\citenamefont {Buzdin}(2005)}]{Buzdin05RMP}%
  \BibitemOpen
  \bibfield  {author} {\bibinfo {author} {\bibfnamefont {A.~I.}\ \bibnamefont {Buzdin}},\ }\bibfield  {title} {\bibinfo {title} {Proximity effects in superconductor-ferromagnet heterostructures},\ }\href {https://doi.org/10.1103/RevModPhys.77.935} {\bibfield  {journal} {\bibinfo  {journal} {Rev. Mod. Phys.}\ }\textbf {\bibinfo {volume} {77}},\ \bibinfo {pages} {935} (\bibinfo {year} {2005})}\BibitemShut {NoStop}%
\bibitem [{\citenamefont {Linder}\ \emph {et~al.}(2009)\citenamefont {Linder}, \citenamefont {Yokoyama}, \citenamefont {Sudb\o{}},\ and\ \citenamefont {Eschrig}}]{Linder09PRL}%
  \BibitemOpen
  \bibfield  {author} {\bibinfo {author} {\bibfnamefont {J.}~\bibnamefont {Linder}}, \bibinfo {author} {\bibfnamefont {T.}~\bibnamefont {Yokoyama}}, \bibinfo {author} {\bibfnamefont {A.}~\bibnamefont {Sudb\o{}}},\ and\ \bibinfo {author} {\bibfnamefont {M.}~\bibnamefont {Eschrig}},\ }\bibfield  {title} {\bibinfo {title} {Pairing symmetry conversion by spin-active interfaces in magnetic normal-metal--superconductor junctions},\ }\href {https://doi.org/10.1103/PhysRevLett.102.107008} {\bibfield  {journal} {\bibinfo  {journal} {Phys. Rev. Lett.}\ }\textbf {\bibinfo {volume} {102}},\ \bibinfo {pages} {107008} (\bibinfo {year} {2009})}\BibitemShut {NoStop}%
\bibitem [{\citenamefont {Cottet}(2011)}]{Cottet11PRL}%
  \BibitemOpen
  \bibfield  {author} {\bibinfo {author} {\bibfnamefont {A.}~\bibnamefont {Cottet}},\ }\bibfield  {title} {\bibinfo {title} {{Inducing Odd-Frequency Triplet Superconducting Correlations in a Normal Metal}},\ }\href {https://doi.org/10.1103/PhysRevLett.107.177001} {\bibfield  {journal} {\bibinfo  {journal} {Phys. Rev. Lett.}\ }\textbf {\bibinfo {volume} {107}},\ \bibinfo {pages} {177001} (\bibinfo {year} {2011})}\BibitemShut {NoStop}%
\bibitem [{\citenamefont {Eschrig}(2015)}]{Eschrig2015review}%
  \BibitemOpen
  \bibfield  {author} {\bibinfo {author} {\bibfnamefont {M.}~\bibnamefont {Eschrig}},\ }\bibfield  {title} {\bibinfo {title} {Spin-polarized supercurrents for spintronics: a review of current progress},\ }\href {https://doi.org/10.1088/0034-4885/78/10/104501} {\bibfield  {journal} {\bibinfo  {journal} {Rep. Prog. Phys.}\ }\textbf {\bibinfo {volume} {78}},\ \bibinfo {pages} {104501} (\bibinfo {year} {2015})}\BibitemShut {NoStop}%
\bibitem [{\citenamefont {Demler}\ \emph {et~al.}(1997)\citenamefont {Demler}, \citenamefont {Arnold},\ and\ \citenamefont {Beasley}}]{Demler97prb}%
  \BibitemOpen
  \bibfield  {author} {\bibinfo {author} {\bibfnamefont {E.~A.}\ \bibnamefont {Demler}}, \bibinfo {author} {\bibfnamefont {G.~B.}\ \bibnamefont {Arnold}},\ and\ \bibinfo {author} {\bibfnamefont {M.~R.}\ \bibnamefont {Beasley}},\ }\bibfield  {title} {\bibinfo {title} {Superconducting proximity effects in magnetic metals},\ }\href {https://doi.org/10.1103/PhysRevB.55.15174} {\bibfield  {journal} {\bibinfo  {journal} {Phys. Rev. B}\ }\textbf {\bibinfo {volume} {55}},\ \bibinfo {pages} {15174} (\bibinfo {year} {1997})}\BibitemShut {NoStop}%
\bibitem [{\citenamefont {Gor'kov}\ and\ \citenamefont {Rashba}(2001)}]{Gorkov01PRL}%
  \BibitemOpen
  \bibfield  {author} {\bibinfo {author} {\bibfnamefont {L.~P.}\ \bibnamefont {Gor'kov}}\ and\ \bibinfo {author} {\bibfnamefont {E.~I.}\ \bibnamefont {Rashba}},\ }\bibfield  {title} {\bibinfo {title} {{Superconducting 2D System with Lifted Spin Degeneracy: Mixed Singlet-Triplet State}},\ }\href {https://doi.org/10.1103/PhysRevLett.87.037004} {\bibfield  {journal} {\bibinfo  {journal} {Phys. Rev. Lett.}\ }\textbf {\bibinfo {volume} {87}},\ \bibinfo {pages} {037004} (\bibinfo {year} {2001})}\BibitemShut {NoStop}%
\bibitem [{\citenamefont {Frigeri}\ \emph {et~al.}(2004)\citenamefont {Frigeri}, \citenamefont {Agterberg}, \citenamefont {Koga},\ and\ \citenamefont {Sigrist}}]{Frigeri04PRL}%
  \BibitemOpen
  \bibfield  {author} {\bibinfo {author} {\bibfnamefont {P.~A.}\ \bibnamefont {Frigeri}}, \bibinfo {author} {\bibfnamefont {D.~F.}\ \bibnamefont {Agterberg}}, \bibinfo {author} {\bibfnamefont {A.}~\bibnamefont {Koga}},\ and\ \bibinfo {author} {\bibfnamefont {M.}~\bibnamefont {Sigrist}},\ }\bibfield  {title} {\bibinfo {title} {{Superconductivity without Inversion Symmetry: MnSi versus ${\mathrm{C}\mathrm{e}\mathrm{P}\mathrm{t}}_{3}\mathrm{S}\mathrm{i}$}},\ }\href {https://doi.org/10.1103/PhysRevLett.92.097001} {\bibfield  {journal} {\bibinfo  {journal} {Phys. Rev. Lett.}\ }\textbf {\bibinfo {volume} {92}},\ \bibinfo {pages} {097001} (\bibinfo {year} {2004})}\BibitemShut {NoStop}%
\bibitem [{\citenamefont {Bergeret}\ and\ \citenamefont {Tokatly}(2013)}]{Bergeret13PRL}%
  \BibitemOpen
  \bibfield  {author} {\bibinfo {author} {\bibfnamefont {F.~S.}\ \bibnamefont {Bergeret}}\ and\ \bibinfo {author} {\bibfnamefont {I.~V.}\ \bibnamefont {Tokatly}},\ }\bibfield  {title} {\bibinfo {title} {{Singlet-Triplet Conversion and the Long-Range Proximity Effect in Superconductor-Ferromagnet Structures with Generic Spin Dependent Fields}},\ }\href {https://doi.org/10.1103/PhysRevLett.110.117003} {\bibfield  {journal} {\bibinfo  {journal} {Phys. Rev. Lett.}\ }\textbf {\bibinfo {volume} {110}},\ \bibinfo {pages} {117003} (\bibinfo {year} {2013})}\BibitemShut {NoStop}%
\bibitem [{\citenamefont {Bergeret}\ and\ \citenamefont {Tokatly}(2014)}]{Bergeret14PRB}%
  \BibitemOpen
  \bibfield  {author} {\bibinfo {author} {\bibfnamefont {F.~S.}\ \bibnamefont {Bergeret}}\ and\ \bibinfo {author} {\bibfnamefont {I.~V.}\ \bibnamefont {Tokatly}},\ }\bibfield  {title} {\bibinfo {title} {Spin-orbit coupling as a source of long-range triplet proximity effect in superconductor-ferromagnet hybrid structures},\ }\href {https://doi.org/10.1103/PhysRevB.89.134517} {\bibfield  {journal} {\bibinfo  {journal} {Phys. Rev. B}\ }\textbf {\bibinfo {volume} {89}},\ \bibinfo {pages} {134517} (\bibinfo {year} {2014})}\BibitemShut {NoStop}%
\bibitem [{\citenamefont {Cr\'epin}\ \emph {et~al.}(2015)\citenamefont {Cr\'epin}, \citenamefont {Burset},\ and\ \citenamefont {Trauzettel}}]{Crepin15prl}%
  \BibitemOpen
  \bibfield  {author} {\bibinfo {author} {\bibfnamefont {F.}~\bibnamefont {Cr\'epin}}, \bibinfo {author} {\bibfnamefont {P.}~\bibnamefont {Burset}},\ and\ \bibinfo {author} {\bibfnamefont {B.}~\bibnamefont {Trauzettel}},\ }\bibfield  {title} {\bibinfo {title} {Odd-frequency triplet superconductivity at the helical edge of a topological insulator},\ }\href {https://doi.org/10.1103/PhysRevB.92.100507} {\bibfield  {journal} {\bibinfo  {journal} {Phys. Rev. B}\ }\textbf {\bibinfo {volume} {92}},\ \bibinfo {pages} {100507} (\bibinfo {year} {2015})}\BibitemShut {NoStop}%
\bibitem [{\citenamefont {Smidman}\ \emph {et~al.}(2017)\citenamefont {Smidman}, \citenamefont {Salamon}, \citenamefont {Yuan},\ and\ \citenamefont {Agterberg}}]{smidman2017superconductivity}%
  \BibitemOpen
  \bibfield  {author} {\bibinfo {author} {\bibfnamefont {M.}~\bibnamefont {Smidman}}, \bibinfo {author} {\bibfnamefont {M.}~\bibnamefont {Salamon}}, \bibinfo {author} {\bibfnamefont {H.}~\bibnamefont {Yuan}},\ and\ \bibinfo {author} {\bibfnamefont {D.}~\bibnamefont {Agterberg}},\ }\bibfield  {title} {\bibinfo {title} {Superconductivity and spin--orbit coupling in non-centrosymmetric materials: a review},\ }\href {https://iopscience.iop.org/article/10.1088/1361-6633/80/3/036501} {\bibfield  {journal} {\bibinfo  {journal} {Rep. Prog. Phys.}\ }\textbf {\bibinfo {volume} {80}},\ \bibinfo {pages} {036501} (\bibinfo {year} {2017})}\BibitemShut {NoStop}%
\bibitem [{\citenamefont {Cayao}\ and\ \citenamefont {Black-Schaffer}(2017)}]{Cayao17PRB}%
  \BibitemOpen
  \bibfield  {author} {\bibinfo {author} {\bibfnamefont {J.}~\bibnamefont {Cayao}}\ and\ \bibinfo {author} {\bibfnamefont {A.~M.}\ \bibnamefont {Black-Schaffer}},\ }\bibfield  {title} {\bibinfo {title} {Odd-frequency superconducting pairing and subgap density of states at the edge of a two-dimensional topological insulator without magnetism},\ }\href {https://doi.org/10.1103/PhysRevB.96.155426} {\bibfield  {journal} {\bibinfo  {journal} {Phys. Rev. B}\ }\textbf {\bibinfo {volume} {96}},\ \bibinfo {pages} {155426} (\bibinfo {year} {2017})}\BibitemShut {NoStop}%
\bibitem [{\citenamefont {Cayao}\ and\ \citenamefont {Black-Schaffer}(2018)}]{Cayao18prb}%
  \BibitemOpen
  \bibfield  {author} {\bibinfo {author} {\bibfnamefont {J.}~\bibnamefont {Cayao}}\ and\ \bibinfo {author} {\bibfnamefont {A.~M.}\ \bibnamefont {Black-Schaffer}},\ }\bibfield  {title} {\bibinfo {title} {Odd-frequency superconducting pairing in junctions with rashba spin-orbit coupling},\ }\href {https://doi.org/10.1103/PhysRevB.98.075425} {\bibfield  {journal} {\bibinfo  {journal} {Phys. Rev. B}\ }\textbf {\bibinfo {volume} {98}},\ \bibinfo {pages} {075425} (\bibinfo {year} {2018})}\BibitemShut {NoStop}%
\bibitem [{\citenamefont {Fleckenstein}\ \emph {et~al.}(2018)\citenamefont {Fleckenstein}, \citenamefont {Ziani},\ and\ \citenamefont {Trauzettel}}]{Fleckenstein18prb}%
  \BibitemOpen
  \bibfield  {author} {\bibinfo {author} {\bibfnamefont {C.}~\bibnamefont {Fleckenstein}}, \bibinfo {author} {\bibfnamefont {N.~T.}\ \bibnamefont {Ziani}},\ and\ \bibinfo {author} {\bibfnamefont {B.}~\bibnamefont {Trauzettel}},\ }\bibfield  {title} {\bibinfo {title} {{Conductance signatures of odd-frequency superconductivity in quantum spin Hall systems using a quantum point contact}},\ }\href {https://doi.org/10.1103/PhysRevB.97.134523} {\bibfield  {journal} {\bibinfo  {journal} {Phys. Rev. B}\ }\textbf {\bibinfo {volume} {97}},\ \bibinfo {pages} {134523} (\bibinfo {year} {2018})}\BibitemShut {NoStop}%
\bibitem [{\citenamefont {Bobkova}\ \emph {et~al.}(2005)\citenamefont {Bobkova}, \citenamefont {Hirschfeld},\ and\ \citenamefont {Barash}}]{Bobkova05PRL}%
  \BibitemOpen
  \bibfield  {author} {\bibinfo {author} {\bibfnamefont {I.~V.}\ \bibnamefont {Bobkova}}, \bibinfo {author} {\bibfnamefont {P.~J.}\ \bibnamefont {Hirschfeld}},\ and\ \bibinfo {author} {\bibfnamefont {Y.~S.}\ \bibnamefont {Barash}},\ }\bibfield  {title} {\bibinfo {title} {Spin-dependent quasiparticle reflection and bound states at interfaces with itinerant antiferromagnets},\ }\href {https://doi.org/10.1103/PhysRevLett.94.037005} {\bibfield  {journal} {\bibinfo  {journal} {Phys. Rev. Lett.}\ }\textbf {\bibinfo {volume} {94}},\ \bibinfo {pages} {037005} (\bibinfo {year} {2005})}\BibitemShut {NoStop}%
\bibitem [{\citenamefont {Kamra}\ \emph {et~al.}(2018)\citenamefont {Kamra}, \citenamefont {Rezaei},\ and\ \citenamefont {Belzig}}]{Kamra18PRL}%
  \BibitemOpen
  \bibfield  {author} {\bibinfo {author} {\bibfnamefont {A.}~\bibnamefont {Kamra}}, \bibinfo {author} {\bibfnamefont {A.}~\bibnamefont {Rezaei}},\ and\ \bibinfo {author} {\bibfnamefont {W.}~\bibnamefont {Belzig}},\ }\bibfield  {title} {\bibinfo {title} {Spin splitting induced in a superconductor by an antiferromagnetic insulator},\ }\href {https://doi.org/10.1103/PhysRevLett.121.247702} {\bibfield  {journal} {\bibinfo  {journal} {Phys. Rev. Lett.}\ }\textbf {\bibinfo {volume} {121}},\ \bibinfo {pages} {247702} (\bibinfo {year} {2018})}\BibitemShut {NoStop}%
\bibitem [{\citenamefont {Rabinovich}\ \emph {et~al.}(2019)\citenamefont {Rabinovich}, \citenamefont {Bobkova},\ and\ \citenamefont {Bobkov}}]{Rabinovich19PRR}%
  \BibitemOpen
  \bibfield  {author} {\bibinfo {author} {\bibfnamefont {D.~S.}\ \bibnamefont {Rabinovich}}, \bibinfo {author} {\bibfnamefont {I.~V.}\ \bibnamefont {Bobkova}},\ and\ \bibinfo {author} {\bibfnamefont {A.~M.}\ \bibnamefont {Bobkov}},\ }\bibfield  {title} {\bibinfo {title} {{Anomalous phase shift in a Josephson junction via an antiferromagnetic interlayer}},\ }\href {https://doi.org/10.1103/PhysRevResearch.1.033095} {\bibfield  {journal} {\bibinfo  {journal} {Phys. Rev. Res.}\ }\textbf {\bibinfo {volume} {1}},\ \bibinfo {pages} {033095} (\bibinfo {year} {2019})}\BibitemShut {NoStop}%
\bibitem [{\citenamefont {Jakobsen}\ \emph {et~al.}(2020)\citenamefont {Jakobsen}, \citenamefont {Naess}, \citenamefont {Dutta}, \citenamefont {Brataas},\ and\ \citenamefont {Qaiumzadeh}}]{JakobsenPRB20}%
  \BibitemOpen
  \bibfield  {author} {\bibinfo {author} {\bibfnamefont {M.~F.}\ \bibnamefont {Jakobsen}}, \bibinfo {author} {\bibfnamefont {K.~B.}\ \bibnamefont {Naess}}, \bibinfo {author} {\bibfnamefont {P.}~\bibnamefont {Dutta}}, \bibinfo {author} {\bibfnamefont {A.}~\bibnamefont {Brataas}},\ and\ \bibinfo {author} {\bibfnamefont {A.}~\bibnamefont {Qaiumzadeh}},\ }\bibfield  {title} {\bibinfo {title} {Electrical and thermal transport in antiferromagnet-superconductor junctions},\ }\href {https://doi.org/10.1103/PhysRevB.102.140504} {\bibfield  {journal} {\bibinfo  {journal} {Phys. Rev. B}\ }\textbf {\bibinfo {volume} {102}},\ \bibinfo {pages} {140504} (\bibinfo {year} {2020})}\BibitemShut {NoStop}%
\bibitem [{\citenamefont {Bobkov}\ \emph {et~al.}(2022)\citenamefont {Bobkov}, \citenamefont {Bobkova}, \citenamefont {Bobkov},\ and\ \citenamefont {Kamra}}]{Bobkov22prb}%
  \BibitemOpen
  \bibfield  {author} {\bibinfo {author} {\bibfnamefont {G.~A.}\ \bibnamefont {Bobkov}}, \bibinfo {author} {\bibfnamefont {I.~V.}\ \bibnamefont {Bobkova}}, \bibinfo {author} {\bibfnamefont {A.~M.}\ \bibnamefont {Bobkov}},\ and\ \bibinfo {author} {\bibfnamefont {A.}~\bibnamefont {Kamra}},\ }\bibfield  {title} {\bibinfo {title} {N\'eel proximity effect at antiferromagnet/superconductor interfaces},\ }\href {https://doi.org/10.1103/PhysRevB.106.144512} {\bibfield  {journal} {\bibinfo  {journal} {Phys. Rev. B}\ }\textbf {\bibinfo {volume} {106}},\ \bibinfo {pages} {144512} (\bibinfo {year} {2022})}\BibitemShut {NoStop}%
\bibitem [{\citenamefont {Bobkov}\ \emph {et~al.}(2023)\citenamefont {Bobkov}, \citenamefont {Gordeeva}, \citenamefont {Bobkov},\ and\ \citenamefont {Bobkova}}]{Bobkov23PRB}%
  \BibitemOpen
  \bibfield  {author} {\bibinfo {author} {\bibfnamefont {G.~A.}\ \bibnamefont {Bobkov}}, \bibinfo {author} {\bibfnamefont {V.~M.}\ \bibnamefont {Gordeeva}}, \bibinfo {author} {\bibfnamefont {A.~M.}\ \bibnamefont {Bobkov}},\ and\ \bibinfo {author} {\bibfnamefont {I.~V.}\ \bibnamefont {Bobkova}},\ }\bibfield  {title} {\bibinfo {title} {Oscillatory superconducting transition temperature in superconductor/antiferromagnet heterostructures},\ }\href {https://doi.org/10.1103/PhysRevB.108.184509} {\bibfield  {journal} {\bibinfo  {journal} {Phys. Rev. B}\ }\textbf {\bibinfo {volume} {108}},\ \bibinfo {pages} {184509} (\bibinfo {year} {2023})}\BibitemShut {NoStop}%
\bibitem [{\citenamefont {Fyhn}\ \emph {et~al.}(2023)\citenamefont {Fyhn}, \citenamefont {Brataas}, \citenamefont {Qaiumzadeh},\ and\ \citenamefont {Linder}}]{Fyhn23PRL}%
  \BibitemOpen
  \bibfield  {author} {\bibinfo {author} {\bibfnamefont {E.~H.}\ \bibnamefont {Fyhn}}, \bibinfo {author} {\bibfnamefont {A.}~\bibnamefont {Brataas}}, \bibinfo {author} {\bibfnamefont {A.}~\bibnamefont {Qaiumzadeh}},\ and\ \bibinfo {author} {\bibfnamefont {J.}~\bibnamefont {Linder}},\ }\bibfield  {title} {\bibinfo {title} {Superconducting proximity effect and long-ranged triplets in dirty metallic antiferromagnets},\ }\href {https://doi.org/10.1103/PhysRevLett.131.076001} {\bibfield  {journal} {\bibinfo  {journal} {Phys. Rev. Lett.}\ }\textbf {\bibinfo {volume} {131}},\ \bibinfo {pages} {076001} (\bibinfo {year} {2023})}\BibitemShut {NoStop}%
\bibitem [{\citenamefont {Liu}\ \emph {et~al.}(2022)\citenamefont {Liu}, \citenamefont {Li}, \citenamefont {Han}, \citenamefont {Wan},\ and\ \citenamefont {Liu}}]{Liu2022spin}%
  \BibitemOpen
  \bibfield  {author} {\bibinfo {author} {\bibfnamefont {P.}~\bibnamefont {Liu}}, \bibinfo {author} {\bibfnamefont {J.}~\bibnamefont {Li}}, \bibinfo {author} {\bibfnamefont {J.}~\bibnamefont {Han}}, \bibinfo {author} {\bibfnamefont {X.}~\bibnamefont {Wan}},\ and\ \bibinfo {author} {\bibfnamefont {Q.}~\bibnamefont {Liu}},\ }\bibfield  {title} {\bibinfo {title} {Spin-group symmetry in magnetic materials with negligible spin-orbit coupling},\ }\href {https://doi.org/10.1103/PhysRevX.12.021016} {\bibfield  {journal} {\bibinfo  {journal} {Phys. Rev. X}\ }\textbf {\bibinfo {volume} {12}},\ \bibinfo {pages} {021016} (\bibinfo {year} {2022})}\BibitemShut {NoStop}%
\bibitem [{\citenamefont {Schiff}\ \emph {et~al.}(2025)\citenamefont {Schiff}, \citenamefont {Corticelli}, \citenamefont {Guerreiro}, \citenamefont {Romh{\'a}nyi},\ and\ \citenamefont {McClarty}}]{schiff2023spin}%
  \BibitemOpen
  \bibfield  {author} {\bibinfo {author} {\bibfnamefont {H.}~\bibnamefont {Schiff}}, \bibinfo {author} {\bibfnamefont {A.}~\bibnamefont {Corticelli}}, \bibinfo {author} {\bibfnamefont {A.}~\bibnamefont {Guerreiro}}, \bibinfo {author} {\bibfnamefont {J.}~\bibnamefont {Romh{\'a}nyi}},\ and\ \bibinfo {author} {\bibfnamefont {P.~A.}\ \bibnamefont {McClarty}},\ }\bibfield  {title} {\bibinfo {title} {The crystallographic spin point groups and their representations},\ }\href {https://scipost.org/10.21468/SciPostPhys.18.3.109} {\bibfield  {journal} {\bibinfo  {journal} {SciPost Physics}\ }\textbf {\bibinfo {volume} {18}},\ \bibinfo {pages} {109} (\bibinfo {year} {2025})}\BibitemShut {NoStop}%
\bibitem [{SM-()}]{SM-AFM2024}%
  \BibitemOpen
  \href@noop {} {\bibinfo  {journal} {See Supplemental Material at [?] for the details of the construction of the ${\bf k}\cdot{\bf p}$ models, the derivation of pairing correlations and Josephson currents for Schr\"odinger and Dirac electrons, a comparison of superconducting proximity effects for the two types of electrons, symmetry constraints on the spin texture and induced Cooper pairs, $0$-$\pi$ transitions in Josephson junctions, and supplementary results for other choices of parameters. The Supplemental Material also contains Refs.~\cite{Fernandes2024Topological,BHYan17PRL,zhang2018spin,SBZ2023arXiv,Sancho1984JPFMP,Sancho1985JPFMP,SBZhang20PRB,Asano01PRB}}\ }\BibitemShut {NoStop}%
\bibitem [{\citenamefont {Tanaka}\ and\ \citenamefont {Golubov}(2007)}]{Tanaka07PRL}%
  \BibitemOpen
\bibfield  {journal} {  }\bibfield  {author} {\bibinfo {author} {\bibfnamefont {Y.}~\bibnamefont {Tanaka}}\ and\ \bibinfo {author} {\bibfnamefont {A.~A.}\ \bibnamefont {Golubov}},\ }\bibfield  {title} {\bibinfo {title} {Theory of the proximity effect in junctions with unconventional superconductors},\ }\href {https://doi.org/10.1103/PhysRevLett.98.037003} {\bibfield  {journal} {\bibinfo  {journal} {Phys. Rev. Lett.}\ }\textbf {\bibinfo {volume} {98}},\ \bibinfo {pages} {037003} (\bibinfo {year} {2007})}\BibitemShut {NoStop}%
\bibitem [{\citenamefont {Linder}\ and\ \citenamefont {Balatsky}(2019)}]{Linder19RMP}%
  \BibitemOpen
  \bibfield  {author} {\bibinfo {author} {\bibfnamefont {J.}~\bibnamefont {Linder}}\ and\ \bibinfo {author} {\bibfnamefont {A.~V.}\ \bibnamefont {Balatsky}},\ }\bibfield  {title} {\bibinfo {title} {Odd-frequency superconductivity},\ }\href {https://doi.org/10.1103/RevModPhys.91.045005} {\bibfield  {journal} {\bibinfo  {journal} {Rev. Mod. Phys.}\ }\textbf {\bibinfo {volume} {91}},\ \bibinfo {pages} {045005} (\bibinfo {year} {2019})}\BibitemShut {NoStop}%
\bibitem [{\citenamefont {Andersen}\ \emph {et~al.}(2006)\citenamefont {Andersen}, \citenamefont {Bobkova}, \citenamefont {Hirschfeld},\ and\ \citenamefont {Barash}}]{Andersen06PRL}%
  \BibitemOpen
  \bibfield  {author} {\bibinfo {author} {\bibfnamefont {B.~M.}\ \bibnamefont {Andersen}}, \bibinfo {author} {\bibfnamefont {I.~V.}\ \bibnamefont {Bobkova}}, \bibinfo {author} {\bibfnamefont {P.~J.}\ \bibnamefont {Hirschfeld}},\ and\ \bibinfo {author} {\bibfnamefont {Y.~S.}\ \bibnamefont {Barash}},\ }\bibfield  {title} {\bibinfo {title} {{$0\ensuremath{-}\ensuremath{\pi}$ Transitions in Josephson Junctions with Antiferromagnetic Interlayers}},\ }\href {https://doi.org/10.1103/PhysRevLett.96.117005} {\bibfield  {journal} {\bibinfo  {journal} {Phys. Rev. Lett.}\ }\textbf {\bibinfo {volume} {96}},\ \bibinfo {pages} {117005} (\bibinfo {year} {2006})}\BibitemShut {NoStop}%
\bibitem [{\citenamefont {Tanaka}\ \emph {et~al.}(2007)\citenamefont {Tanaka}, \citenamefont {Tanuma},\ and\ \citenamefont {Golubov}}]{Tanaka07PRB}%
  \BibitemOpen
  \bibfield  {author} {\bibinfo {author} {\bibfnamefont {Y.}~\bibnamefont {Tanaka}}, \bibinfo {author} {\bibfnamefont {Y.}~\bibnamefont {Tanuma}},\ and\ \bibinfo {author} {\bibfnamefont {A.~A.}\ \bibnamefont {Golubov}},\ }\bibfield  {title} {\bibinfo {title} {Odd-frequency pairing in normal-metal/superconductor junctions},\ }\href {https://doi.org/10.1103/PhysRevB.76.054522} {\bibfield  {journal} {\bibinfo  {journal} {Phys. Rev. B}\ }\textbf {\bibinfo {volume} {76}},\ \bibinfo {pages} {054522} (\bibinfo {year} {2007})}\BibitemShut {NoStop}%
\bibitem [{\citenamefont {Black-Schaffer}\ and\ \citenamefont {Balatsky}(2013)}]{Black-Schaffer13PRB}%
  \BibitemOpen
  \bibfield  {author} {\bibinfo {author} {\bibfnamefont {A.~M.}\ \bibnamefont {Black-Schaffer}}\ and\ \bibinfo {author} {\bibfnamefont {A.~V.}\ \bibnamefont {Balatsky}},\ }\bibfield  {title} {\bibinfo {title} {Odd-frequency superconducting pairing in multiband superconductors},\ }\href {https://doi.org/10.1103/PhysRevB.88.104514} {\bibfield  {journal} {\bibinfo  {journal} {Phys. Rev. B}\ }\textbf {\bibinfo {volume} {88}},\ \bibinfo {pages} {104514} (\bibinfo {year} {2013})}\BibitemShut {NoStop}%
\bibitem [{\citenamefont {Sancho}\ \emph {et~al.}(1984)\citenamefont {Sancho}, \citenamefont {Sancho},\ and\ \citenamefont {Rubio}}]{Sancho1984JPFMP}%
  \BibitemOpen
  \bibfield  {author} {\bibinfo {author} {\bibfnamefont {M.~P.~L.}\ \bibnamefont {Sancho}}, \bibinfo {author} {\bibfnamefont {J.~M.~L.}\ \bibnamefont {Sancho}},\ and\ \bibinfo {author} {\bibfnamefont {J.}~\bibnamefont {Rubio}},\ }\bibfield  {title} {\bibinfo {title} {{Quick iterative scheme for the calculation of transfer matrices: application to Mo (100)}},\ }\href {https://doi.org/10.1088/0305-4608/14/5/016} {\bibfield  {journal} {\bibinfo  {journal} {J. Phys. F: Met. Phys.}\ }\textbf {\bibinfo {volume} {14}},\ \bibinfo {pages} {1205} (\bibinfo {year} {1984})}\BibitemShut {NoStop}%
\bibitem [{\citenamefont {Sancho}\ \emph {et~al.}(1985)\citenamefont {Sancho}, \citenamefont {Sancho}, \citenamefont {Sancho},\ and\ \citenamefont {Rubio}}]{Sancho1985JPFMP}%
  \BibitemOpen
  \bibfield  {author} {\bibinfo {author} {\bibfnamefont {M.~P.~L.}\ \bibnamefont {Sancho}}, \bibinfo {author} {\bibfnamefont {J.~M.~L.}\ \bibnamefont {Sancho}}, \bibinfo {author} {\bibfnamefont {J.~M.~L.}\ \bibnamefont {Sancho}},\ and\ \bibinfo {author} {\bibfnamefont {J.}~\bibnamefont {Rubio}},\ }\bibfield  {title} {\bibinfo {title} {{Highly convergent schemes for the calculation of bulk and surface Green functions}},\ }\href {https://doi.org/10.1088/0305-4608/15/4/009} {\bibfield  {journal} {\bibinfo  {journal} {J. Phys. F: Met. Phys.}\ }\textbf {\bibinfo {volume} {15}},\ \bibinfo {pages} {851} (\bibinfo {year} {1985})}\BibitemShut {NoStop}%
\bibitem [{\citenamefont {Breunig}\ \emph {et~al.}(2019)\citenamefont {Breunig}, \citenamefont {Zhang}, \citenamefont {Stehno},\ and\ \citenamefont {Trauzettel}}]{Breunig19prb}%
  \BibitemOpen
  \bibfield  {author} {\bibinfo {author} {\bibfnamefont {D.}~\bibnamefont {Breunig}}, \bibinfo {author} {\bibfnamefont {S.-B.}\ \bibnamefont {Zhang}}, \bibinfo {author} {\bibfnamefont {M.}~\bibnamefont {Stehno}},\ and\ \bibinfo {author} {\bibfnamefont {B.}~\bibnamefont {Trauzettel}},\ }\bibfield  {title} {\bibinfo {title} {Influence of a chiral chemical potential on weyl hybrid junctions},\ }\href {https://doi.org/10.1103/PhysRevB.99.174501} {\bibfield  {journal} {\bibinfo  {journal} {Phys. Rev. B}\ }\textbf {\bibinfo {volume} {99}},\ \bibinfo {pages} {174501} (\bibinfo {year} {2019})}\BibitemShut {NoStop}%
\bibitem [{\citenamefont {Lee}\ \emph {et~al.}(2024{\natexlab{b}})\citenamefont {Lee}, \citenamefont {Qian},\ and\ \citenamefont {Yang}}]{lee2024prl}%
  \BibitemOpen
  \bibfield  {author} {\bibinfo {author} {\bibfnamefont {S.~H.}\ \bibnamefont {Lee}}, \bibinfo {author} {\bibfnamefont {Y.}~\bibnamefont {Qian}},\ and\ \bibinfo {author} {\bibfnamefont {B.-J.}\ \bibnamefont {Yang}},\ }\bibfield  {title} {\bibinfo {title} {Fermi surface spin texture and topological superconductivity in spin-orbit free noncollinear antiferromagnets},\ }\href {https://doi.org/10.1103/PhysRevLett.132.196602} {\bibfield  {journal} {\bibinfo  {journal} {Phys. Rev. Lett.}\ }\textbf {\bibinfo {volume} {132}},\ \bibinfo {pages} {196602} (\bibinfo {year} {2024}{\natexlab{b}})}\BibitemShut {NoStop}%
\bibitem [{\citenamefont {Hu}\ and\ \citenamefont {Zhang}(2025)}]{LHHu25SCPMA}%
  \BibitemOpen
  \bibfield  {author} {\bibinfo {author} {\bibfnamefont {L.-H.}\ \bibnamefont {Hu}}\ and\ \bibinfo {author} {\bibfnamefont {S.-B.}\ \bibnamefont {Zhang}},\ }\bibfield  {title} {\bibinfo {title} {{Spin magnetization in unconventional antiferromagnets with collinear and non-collinear spins}},\ }\href {https://doi.org/10.1007/s11433-024-2567-6} {\bibfield  {journal} {\bibinfo  {journal} {Sci. China Phys. Mech. Astron.}\ }\textbf {\bibinfo {volume} {68}},\ \bibinfo {pages} {247211} (\bibinfo {year} {2025})}\BibitemShut {NoStop}%
\bibitem [{\citenamefont {{Zhang}}\ \emph {et~al.}()\citenamefont {{Zhang}}, \citenamefont {{Hu}}, \citenamefont {{Niu}},\ and\ \citenamefont {{Zhang}}}]{SBZhang2025RRP}%
  \BibitemOpen
  \bibfield  {author} {\bibinfo {author} {\bibfnamefont {S.-B.}\ \bibnamefont {{Zhang}}}, \bibinfo {author} {\bibfnamefont {L.-H.}\ \bibnamefont {{Hu}}}, \bibinfo {author} {\bibfnamefont {Q.}~\bibnamefont {{Niu}}},\ and\ \bibinfo {author} {\bibfnamefont {Z.}~\bibnamefont {{Zhang}}},\ }\bibfield  {title} {\bibinfo {title} {{Spin-Valley Locking and Pure Spin-Triplet Superconductivity in Noncollinear Antiferromagnets Proximitized to Conventional Superconductors}},\ }\href {https://doi.org/10.48550/arXiv.2507.11921} {\bibinfo  {journal} {e-prints}\ ,\ \bibinfo {pages} {arXiv:2507.11921}}\BibitemShut {NoStop}%
\bibitem [{\citenamefont {{Hou}}\ \emph {et~al.}()\citenamefont {{Hou}}, \citenamefont {{Sun}}, \citenamefont {{Trauzettel}},\ and\ \citenamefont {{Zhang}}}]{JXHou25PRB}%
  \BibitemOpen
\bibfield  {journal} {  }\bibfield  {author} {\bibinfo {author} {\bibfnamefont {J.-X.}\ \bibnamefont {{Hou}}}, \bibinfo {author} {\bibfnamefont {H.-P.}\ \bibnamefont {{Sun}}}, \bibinfo {author} {\bibfnamefont {B.}~\bibnamefont {{Trauzettel}}},\ and\ \bibinfo {author} {\bibfnamefont {S.-B.}\ \bibnamefont {{Zhang}}},\ }\bibfield  {title} {\bibinfo {title} {{Enhancement of Josephson Supercurrent and $\pi$-Junction by Chiral Antiferromagnetism}},\ }\href {https://doi.org/10.48550/arXiv.2507.12842} {\bibinfo  {journal} {arXiv e-prints}\ ,\ \bibinfo {eid} {arXiv:2507.12842}}\BibitemShut {NoStop}%
\bibitem [{Note1()}]{Note1}%
  \BibitemOpen
\bibfield  {journal} {  }\bibinfo {note} {The ratio of between $\protect \mathcal {F}_{\uparrow \uparrow }/\protect \mathcal {F}_{\downarrow \downarrow }$ is approximately independent of position $x$.}\BibitemShut {Stop}%
\bibitem [{\citenamefont {Chen}\ \emph {et~al.}(2020)\citenamefont {Chen}, \citenamefont {Wang}, \citenamefont {Xiao}, \citenamefont {Guo}, \citenamefont {Niu},\ and\ \citenamefont {MacDonald}}]{chen2020prb}%
  \BibitemOpen
  \bibfield  {author} {\bibinfo {author} {\bibfnamefont {H.}~\bibnamefont {Chen}}, \bibinfo {author} {\bibfnamefont {T.-C.}\ \bibnamefont {Wang}}, \bibinfo {author} {\bibfnamefont {D.}~\bibnamefont {Xiao}}, \bibinfo {author} {\bibfnamefont {G.-Y.}\ \bibnamefont {Guo}}, \bibinfo {author} {\bibfnamefont {Q.}~\bibnamefont {Niu}},\ and\ \bibinfo {author} {\bibfnamefont {A.~H.}\ \bibnamefont {MacDonald}},\ }\bibfield  {title} {\bibinfo {title} {Manipulating anomalous hall antiferromagnets with magnetic fields},\ }\href {https://doi.org/10.1103/PhysRevB.101.104418} {\bibfield  {journal} {\bibinfo  {journal} {Phys. Rev. B}\ }\textbf {\bibinfo {volume} {101}},\ \bibinfo {pages} {104418} (\bibinfo {year} {2020})}\BibitemShut {NoStop}%
\bibitem [{\citenamefont {Zhang}\ \emph {et~al.}(2017)\citenamefont {Zhang}, \citenamefont {Sun}, \citenamefont {Yang}, \citenamefont {\ifmmode~\check{Z}\else \v{Z}\fi{}elezn\'y}, \citenamefont {Parkin}, \citenamefont {Felser},\ and\ \citenamefont {Yan}}]{zhang2017Strong}%
  \BibitemOpen
  \bibfield  {author} {\bibinfo {author} {\bibfnamefont {Y.}~\bibnamefont {Zhang}}, \bibinfo {author} {\bibfnamefont {Y.}~\bibnamefont {Sun}}, \bibinfo {author} {\bibfnamefont {H.}~\bibnamefont {Yang}}, \bibinfo {author} {\bibfnamefont {J.}~\bibnamefont {\ifmmode~\check{Z}\else \v{Z}\fi{}elezn\'y}}, \bibinfo {author} {\bibfnamefont {S.~P.~P.}\ \bibnamefont {Parkin}}, \bibinfo {author} {\bibfnamefont {C.}~\bibnamefont {Felser}},\ and\ \bibinfo {author} {\bibfnamefont {B.}~\bibnamefont {Yan}},\ }\bibfield  {title} {\bibinfo {title} {{Strong anisotropic anomalous Hall effect and spin Hall effect in the chiral antiferromagnetic compounds ${\mathrm{Mn}}_{3}X$ ($X=\mathrm{Ge}$, Sn, Ga, Ir, Rh, and Pt)}},\ }\href {https://doi.org/10.1103/PhysRevB.95.075128} {\bibfield  {journal} {\bibinfo  {journal} {Phys. Rev. B}\ }\textbf {\bibinfo {volume} {95}},\ \bibinfo {pages} {075128} (\bibinfo {year} {2017})}\BibitemShut {NoStop}%
\bibitem [{Note2()}]{Note2}%
  \BibitemOpen
  \bibinfo {note} {We neglect changes in the chemical potential $\mu $ when the materials are made into thin films.}\BibitemShut {Stop}%
\bibitem [{\citenamefont {Martin}\ and\ \citenamefont {Batista}(2008)}]{Martin08PRL}%
  \BibitemOpen
  \bibfield  {author} {\bibinfo {author} {\bibfnamefont {I.}~\bibnamefont {Martin}}\ and\ \bibinfo {author} {\bibfnamefont {C.~D.}\ \bibnamefont {Batista}},\ }\bibfield  {title} {\bibinfo {title} {{Itinerant Electron-Driven Chiral Magnetic Ordering and Spontaneous Quantum Hall Effect in Triangular Lattice Models}},\ }\href {https://doi.org/10.1103/PhysRevLett.101.156402} {\bibfield  {journal} {\bibinfo  {journal} {Phys. Rev. Lett.}\ }\textbf {\bibinfo {volume} {101}},\ \bibinfo {pages} {156402} (\bibinfo {year} {2008})}\BibitemShut {NoStop}%
\bibitem [{\citenamefont {Fernandes}\ \emph {et~al.}(2024)\citenamefont {Fernandes}, \citenamefont {de~Carvalho}, \citenamefont {Birol},\ and\ \citenamefont {Pereira}}]{Fernandes2024Topological}%
  \BibitemOpen
  \bibfield  {author} {\bibinfo {author} {\bibfnamefont {R.~M.}\ \bibnamefont {Fernandes}}, \bibinfo {author} {\bibfnamefont {V.~S.}\ \bibnamefont {de~Carvalho}}, \bibinfo {author} {\bibfnamefont {T.}~\bibnamefont {Birol}},\ and\ \bibinfo {author} {\bibfnamefont {R.~G.}\ \bibnamefont {Pereira}},\ }\bibfield  {title} {\bibinfo {title} {Topological transition from nodal to nodeless zeeman splitting in altermagnets},\ }\href {https://doi.org/10.1103/PhysRevB.109.024404} {\bibfield  {journal} {\bibinfo  {journal} {Phys. Rev. B}\ }\textbf {\bibinfo {volume} {109}},\ \bibinfo {pages} {024404} (\bibinfo {year} {2024})}\BibitemShut {NoStop}%
\bibitem [{\citenamefont {\ifmmode~\check{Z}\else \v{Z}\fi{}elezn\'y}\ \emph {et~al.}(2017)\citenamefont {\ifmmode~\check{Z}\else \v{Z}\fi{}elezn\'y}, \citenamefont {Zhang}, \citenamefont {Felser},\ and\ \citenamefont {Yan}}]{BHYan17PRL}%
  \BibitemOpen
  \bibfield  {author} {\bibinfo {author} {\bibfnamefont {J.}~\bibnamefont {\ifmmode~\check{Z}\else \v{Z}\fi{}elezn\'y}}, \bibinfo {author} {\bibfnamefont {Y.}~\bibnamefont {Zhang}}, \bibinfo {author} {\bibfnamefont {C.}~\bibnamefont {Felser}},\ and\ \bibinfo {author} {\bibfnamefont {B.}~\bibnamefont {Yan}},\ }\bibfield  {title} {\bibinfo {title} {Spin-polarized current in noncollinear antiferromagnets},\ }\href {https://doi.org/10.1103/PhysRevLett.119.187204} {\bibfield  {journal} {\bibinfo  {journal} {Phys. Rev. Lett.}\ }\textbf {\bibinfo {volume} {119}},\ \bibinfo {pages} {187204} (\bibinfo {year} {2017})}\BibitemShut {NoStop}%
\bibitem [{\citenamefont {Zhang}\ \emph {et~al.}(2018)\citenamefont {Zhang}, \citenamefont {{\v{Z}}elezn{\`y}}, \citenamefont {Sun}, \citenamefont {Van Den~Brink},\ and\ \citenamefont {Yan}}]{zhang2018spin}%
  \BibitemOpen
  \bibfield  {author} {\bibinfo {author} {\bibfnamefont {Y.}~\bibnamefont {Zhang}}, \bibinfo {author} {\bibfnamefont {J.}~\bibnamefont {{\v{Z}}elezn{\`y}}}, \bibinfo {author} {\bibfnamefont {Y.}~\bibnamefont {Sun}}, \bibinfo {author} {\bibfnamefont {J.}~\bibnamefont {Van Den~Brink}},\ and\ \bibinfo {author} {\bibfnamefont {B.}~\bibnamefont {Yan}},\ }\bibfield  {title} {\bibinfo {title} {{Spin Hall effect emerging from a noncollinear magnetic lattice without spin--orbit coupling}},\ }\href {https://iopscience.iop.org/article/10.1088/1367-2630/aad1eb} {\bibfield  {journal} {\bibinfo  {journal} {New J. Phys.}\ }\textbf {\bibinfo {volume} {20}},\ \bibinfo {pages} {073028} (\bibinfo {year} {2018})}\BibitemShut {NoStop}%
\bibitem [{\citenamefont {Zhang}\ and\ \citenamefont {Trauzettel}(2020)}]{SBZhang20PRB}%
  \BibitemOpen
  \bibfield  {author} {\bibinfo {author} {\bibfnamefont {S.-B.}\ \bibnamefont {Zhang}}\ and\ \bibinfo {author} {\bibfnamefont {B.}~\bibnamefont {Trauzettel}},\ }\bibfield  {title} {\bibinfo {title} {{Detection of second-order topological superconductors by Josephson junctions}},\ }\href {https://doi.org/10.1103/PhysRevResearch.2.012018} {\bibfield  {journal} {\bibinfo  {journal} {Phys. Rev. Res.}\ }\textbf {\bibinfo {volume} {2}},\ \bibinfo {pages} {012018} (\bibinfo {year} {2020})}\BibitemShut {NoStop}%
\bibitem [{\citenamefont {Asano}(2001)}]{Asano01PRB}%
  \BibitemOpen
  \bibfield  {author} {\bibinfo {author} {\bibfnamefont {Y.}~\bibnamefont {Asano}},\ }\bibfield  {title} {\bibinfo {title} {{Numerical method for dc Josephson current between d-wave superconductors}},\ }\href {https://doi.org/10.1103/PhysRevB.63.052512} {\bibfield  {journal} {\bibinfo  {journal} {Phys. Rev. B}\ }\textbf {\bibinfo {volume} {63}},\ \bibinfo {pages} {052512} (\bibinfo {year} {2001})}\BibitemShut {NoStop}%
\end{thebibliography}
\end{document}